\documentclass[review]{elsarticle}
\usepackage{lineno,hyperref}
\modulolinenumbers[5]

\journal{Computers \& Fluids}

\usepackage{rotating}
\usepackage{graphicx}
\usepackage{amsmath}
\usepackage{amssymb}

\usepackage{combelow}
\usepackage{bm}
\usepackage{slashed}
\usepackage{braket}
\usepackage{color}
\usepackage{tikz}
\def\eqref#1{(\ref{#1})}

\def\feq{\ensuremath{f^{\rm{(eq)}}}}
\def\hh{\mathfrak{h}}

\begin{document}

\begin{frontmatter}

\title{Comparison of the Shakhov and ellipsoidal models for 
the Boltzmann equation and DSMC for {\it ab initio}-based particle 
interactions}

\author[wut,odu]{Victor E. Ambru\cb{s}}
\ead{victor.ambrus@e-uvt.ro}

\author[ufp]{Felix Sharipov}
\ead{sharipov@fisica.ufpr.br}

\author[arft]{Victor Sofonea}
\ead{sofonea@gmail.com, sofonea@acad-tim.tm.edu.ro}

\address[wut]{%
Department of Physics, West University of Timi\cb{s}oara, \\
Bd.~Vasile P\^arvan 4, Timi\cb{s}oara 300223, Romania${}^\dagger$\footnote{${}^\dagger$ Permanent address.}}
\address[odu]{%
Department of Mathematics and Statistics, 
Old Dominion University, \\ Norfolk, VA 23529, USA}
\address[ufp]{%
Departamento de F\'isica, 
Universidade Federal do Paran\'a, Curitiba, 81531-980
Brazil}
\address[arft]{%
Center for Fundamental and Advanced Technical Research, 
Romanian Academy, Bd.~Mihai Viteazu 24, Timi\cb{s}oara 300223, Romania}

\begin{abstract}
In this paper, we consider the capabilities of the
Boltzmann equation with the Shakhov and ellipsoidal models for the collision
term to capture the characteristics of rarefied gas flows.
The benchmark is performed by comparing the results obtained using 
these kinetic model equations with direct simulation
Monte Carlo (DSMC) results for particles interacting via {\it ab initio}
potentials. The analysis is restricted to channel flows between
parallel plates and we consider
three flow problems, namely: the
heat transfer between stationary plates, the Couette flow and
the heat transfer under shear. The simulations are performed in the
non-linear regime for the ${}^3{\rm He}$, ${}^4{\rm He}$, and ${\rm Ne}$ 
gases. The reference temperature ranges between $1\ {\rm K}$ and
$3000\ {\rm K}$ for ${}^3{\rm He}$ and ${}^4{\rm He}$ and 
between $20 \ {\rm K}$ and $5000\ {\rm K}$ for ${\rm Ne}$.
While good agreement is seen up to the transition
regime for the direct phenomena (shear stress, heat flux driven by
temperature gradient), the relative errors in the cross phenomena
(heat flux perpendicular to the temperature gradient) exceed $10\%$ even
in the slip-flow regime. The kinetic model equations are solved
using the finite difference lattice Boltzmann algorithm based on 
half-range Gauss-Hermite quadratures with the third order upwind 
method used for the implementation of the advection.
\end{abstract}

\begin{keyword}
Ab initio\sep DSMC\sep Ellipsoidal model\sep Shakhov model\sep
Half-range Gauss-Hermite quadrature
\end{keyword}

\end{frontmatter}

\linenumbers

\section{Introduction} \label{sec:intro}

Finding accurate solutions of the kinetic equations
governing rarefied gas flows is a challenging task because
of their complexity \cite{cercignani00, sone07}. In
the case of channel flows, it has been shown
under quite general assumptions that the velocity 
field in the vicinity of solid boundaries is non-analytic, its normal 
derivative presenting a logarithmic singularity with respect to the 
distance to the wall \cite{takata13}. Understanding the main 
properties of such flows is crucial when devising
micro/nano-electro-mechanical systems 
(MEMS/NEMS) \cite{gadelhaq06hbook}.

Since the kinetic equation is difficult to solve analytically, numerical methods remain the primary 
tools available for its investigation. It has been established 
in the research community that the direct simulation Monte Carlo (DSMC) method 
\cite{bird94} can provide solutions to realistic systems in a wide range of 
flow regimes. The main ingredient controlling the relevance of the DSMC 
formulation lies in specifying the interparticle interactions.
Recently, {\it ab initio} potentials have been implemented into the DSMC method \cite{sharipov13,Sha106,Sha115,Sha116,zhu19}. A quantum consideration of interatomic collisions \cite{sharipov18,Sha125} allowed to extend an application of {\it ab initio} potentials to low temperatures. To reduce the computational effort, lookup tables for the deflection angle of binary collisions of helium-3 (${}^3{\rm He}$), helium-4 (${}^4{\rm He}$), and neon (${\rm Ne}$) atoms have been calculated and reported in the Supplementary material to Ref.~\cite{Sha125}. The lookup tables can be used for any flow of these gases over a wide range of temperature.
Due to the stochastic nature of the DSMC method, 
its results often exhibit 
steady-state fluctuations, which are especially significant in the 
slip-flow regime and at small Mach numbers. Filtering out these 
fluctuations is a computationally demanding part of the algorithm,
making this method computationally convenient only in the transition 
and free molecular flow regimes.

Another approach for the description 
of rarefied gas flows starts from the Boltzmann equation, where the 
collision integral takes into account the details of the interparticle 
interactions. While recent years have seen significant progress in 
the development of numerical methods for evaluating the Boltzmann 
collision integral \cite{mouhot06,filbet12,wu13,wu15,gamba17}, this 
operation still remains the most expensive part of the solver, 
making the application of such methods for complex systems
computationally prohibitive.

As argued in the early '50s, the features of the collision integral
can be preserved, at least for small Knudsen numbers and mildly non-linear 
systems, by replacing the collision term through a relaxation time 
approach. The BGK model, introduced by Bhatnagar, Gross and 
Krook \cite{bhatnagar54}, employed a single relaxation time 
$\tau$ to control the departure of the Boltzmann distribution
function $f$ from local thermal equilibrium. This parameter could 
be used to match realistic flows by ensuring the correct recovery 
of the dynamic viscosity $\mu$ in the hydrodynamic regime, however 
it could not allow the heat conductivity $\kappa$ to be controlled 
independently. This difficulty was later alleviated through two 
extensions, known as the ellipsoidal-BGK (ES) and Shakhov (S) 
models, proposed in the late '60s by Holway \cite{holway66} and 
Shakhov \cite{shakhov68a,shakhov68b}, respectively. The accuracy of 
these models has been tested by considering the comparison to experimental 
\cite{sharipov02,sharipov03,graur09} or 
DSMC \cite{ambrus12pre,meng13,ambrus18pre} results.
In the following, we refer to these two models (the ES and S models) 
as {\it the model equations}.

Various methods have been developed over the years to solve 
the model equations and their variations. 
Amongst these, we mention the discrete velocity method (DVM) 
\cite{sone07,broadwell64,sharipov16,ho15}, the discrete unified 
gas kinetic scheme (DUGKS) \cite{guo13pre,guo15pre,zhu17jcp}, 
the discrete Boltzmann method (DBM) 
\cite{he98pre,lin17,zhang18} (generally 
restricted to the Navier-Stokes regime due to the small 
velocity set size)
and the lattice Boltzmann (LB) method 
\cite{yudistiawan08,yudistiawan10,feuchter16,atif18} 
with its finite difference (FDLB) version 
\cite{ambrus12pre,aidun10,meng11pre,shi15,ambrus16jcp}.

In the LB approach, the kinetic equation is employed
to obtain an accurate account of the evolution of the macroscopic 
moments of $f$ \cite{succi01,fede15,kruger17,succi18}.
Less attention is directed to the distribution $f$ itself. This allows the 
momentum space to be sampled in a manner optimized for the recovery of 
the moments of $f$ \cite{shan06}. Since the moments are defined as 
integrals of $f$, the momentum space discretization can be viewed 
as a quadrature method \cite{meng11pre}.
Our implementation is based on the idea of Gauss quadratures
\cite{hildebrand87,shizgal15}, which provide a prescription of 
choosing optimal quadrature points for the recovery of polynomial 
integrals, given a certain domain and integration weight. 

In this paper, we consider the systematic comparison between 
the numerical solutions of the Boltzmann equation with the 
S and ES models for the collision term, obtained using 
the FDLB algorithm,
and the numerical results obtained using DSMC.
The comparison is made in the frame of channel flows between 
parallel plates, where the fluid is assumed to be homogeneous 
with respect to the directions parallel to the plates.
Specifically, we address three flow problems. 
The first one is the heat transfer between stationary plates
at differing temperatures. The second is the Couette flow 
between parallel plates at equal temperatures. The third 
problem refers to the heat transfer between plates at 
differing temperatures undergoing parallel motion.
In future studies, it may be interesting to perform the 
comparison in more complex configurations, such as the 
thermal transpiration through a long channel attached 
to two vessels with different temperatures considered 
in Refs.~\cite{sharipov96} and \cite{wu14}, 
or the pressure-driven flow through a long rectangular 
channel setup considered in Ref.~\cite{sharipov99}; 
as well as in the highly nonlinear context of shock wave 
structure \cite{Sha115}. Since the focus in this paper is on 
introducing kinetic models for the simulation of various 
kinds of gas particles interacting via {\it ab initio} 
interparticle potentials, the present study 
is restricted only to the channel flows mentioned above.

In channel flows, it is known that the particle-wall interaction 
induces a discontinuity in the distribution function \cite{takata13,gross57}.
This discontinuity is responsible for microfluidics effects, such 
as the development of a slip velocity and temperature jump near the 
walls. Another important consequence of the discontinuity of 
$f$ is that the velocity profile becomes non-analytic
in the vicinity of the wall, where its derivative diverges logarithmically 
with respect to the distance to the wall \cite{takata13,sone64,jiang16}. 

As highlighted already in the late '50s by Gross and 
his collaborators 
\cite{gross57,gross58,ziering60,bhatnagar69}, taking into account 
the  discontinuity of the distribution function by considering 
separately its moments
with respect to the vectors pointing towards and away from the wall
($p_x > 0$ and $p_x < 0$, respectively) can give a dramatic 
increase in
the accuracy of the Knudsen layer representation, compared to the
full momentum space projection approach. Recent works have 
focused on employing half-range quadratures \cite{frezzotti09,gibelli12,ghiroldi14}
for the (semi-)analytical analysis of the solutions of the (linearised or 
non-linear) Boltzmann equation in the relaxation time approximation,

An important step in employing the idea of treating separately 
the distribution function for incoming and outgoing particles with 
respect to solid walls in the numerical simulation of rarefied gas flows 
was taken in the '60s by Huang and Giddens \cite{huang68}, who 
computed the quadrature points and weights for the one-dimensional
half-range Gauss-Hermite quadrature 
with the weight function $\omega(x) = e^{-x^2}$,
up to 8th order. The extension of the procedure to higher orders 
through a recurrence relation was discussed by Ball in Ref.~\cite{ball03}
and the algorithm was adapted in Ref.~\cite{ambrus16jcp} to the case 
of the weight function $\omega(x) = e^{-x^2/2} / \sqrt{2\pi}$.
A half-range (or modified) Gauss-Hermite quadrature was used
in the early 2000's by Li and his collaborators \cite{li03,li04} 
for kinetic theory simulations in the context of unbounded flows.
Recently, the half-range Gauss-Hermite quadrature was shown 
to offer significantly more accurate solutions of the kinetic model
equations than the full-range 
Gauss-Hermite quadrature with the same number of quadrature points
for the moderate and highly rarefied regimes 
\cite{shi15,ambrus16jcp,ghiroldi15}. As a side note, 
similarly accurate results can be obtained when the Gauss-Laguerre 
quadrature is used on the semi-axis, instead of the Gauss-Hermite 
quadrature \cite{ambrus14ijmpc,ambrus14pre}.

In order to take advantage of the geometry of the channel flows 
considered in this paper, we 
solve the kinetic model equations by employing
the mixed quadratures concept, according 
to which the quadrature is controlled separately on each axis
\cite{ambrus16jcp,gibelli12}. This approach allows 
the half-range Gauss-Hermite quadrature to be employed on the 
$x$ axis, which is perpendicular to the channel walls. On the axes 
parallel to the walls, the full-range Gauss-Hermite 
quadrature can be employed. Details regarding Gauss quadratures 
can be found in various textbooks, of which we remind 
Refs.~\cite{hildebrand87,shizgal15}. 

Furthermore, in the channel flows considered in this paper, the dynamics 
is non-trivial only along $d < D$ degrees of freedom (DOFs), where 
$D = 3$ is the number of DOFs for an ideal monatomic gas. In particular, 
we consider $d = 1$ when the walls are stationary and $d = 2$ 
when the plates are in motion. We then introduce two reduced distributions, 
$\phi$ and $\chi$, which are obtained by integrating
the distribution function $f$ multiplied by $1$ and 
$[p_{d+1}^2 + \dots p_D^2]/m$ with respect to $dp^{d+1} \cdots dp^D$ \cite{graur09}.
Thus, $\phi$ can be seen to describe the mass and momentum evolution and
$\chi$ contributes to the energy evolution \cite{ambrus18pre,ambrus19COST}. When 
$d = 2$, we employ the mixed quadrature paradigm \cite{ambrus16jcp,frezzotti09}
and discretize the momentum along the direction parallel to the wall using 
the full range Gauss-Hermite quadrature. Furthemore, the homogeneity of the 
fluid along these directions allows the system to be exactly described (i.e.,
without introducing any errors) using a relatively low order quadrature 
\cite{ambrus18pre,ambrus16jcp}. In this paper, we introduce a novel expansion 
of the Shakhov and ellipsoidal collision terms with 
respect to the full-range (standard) Hermite polynomials which allows the quadrature 
orders along the $y$ axis (which is parallel to the walls) to 
be set to $Q^\phi_y = 4$ and $Q^\chi_y = 2$ for the $\phi$ and $\chi$ distributions,
respectively. The resulting expansion coefficients remain 
dependent on the momentum component $p_x$ which is perpendicular to the 
boundary. While replacing the distributions with their truncated polynomial 
expansions is inherited from the standard LB algorithm \cite{shan06}, the 
expansion coefficients are evaluated directly (without resorting to polynomial 
expansions), which is closer to the standard DVM practice \cite{broadwell64}. 
In this sense, our approach is a hybrid FDLB-DVM method (we refer to it as 
{\it the hybrid method}), combining the 
advantages of both LB and DVM. For brevity, we use the notation FDLB 
to refer to the numerical scheme that we employ to solve the kinetic 
model equations. For completeness, the subsequent 
projection with respect to $p_x$ onto the half-range Hermite polynomials 
is also discussed (we refer to this latter approach as {\it the projection 
method}). The latter approach is more efficient than the hybrid approach in 
the hydrodynamic (small ${\rm Kn}$) regime.

For the analysis presented in this paper, only the stationary 
state is of interest. 
Since the transient solution is not important, iterative 
schemes can be employed to solve the kinetic model equation, 
as described, e.g., in Refs.~\cite{valougeorgis03,wu17,su19,zhu20}. 
However, since the computations in the one-dimensional 
settings that we consider in this paper are not 
very demanding, we compute the stationary solution using 
explicit time marching, implemented using the 
third order total variation diminishing Runge-Kutta (RK-3) method 
introduced in Refs.~\cite{shu88,gottlieb98,trangenstein07}.
For the advection operator, we introduce a third order 
upwind scheme which preserves the order of accuracy in the 
presence of diffuse reflecting boundaries which extends 
the one considered in Ref.~\cite{ambrus19luo} for the 
linearised Boltzmann-BGK equation. We further increase the resolution
inside the Knudsen layer by employing a grid stretching 
procedure \cite{mei98jcph,guo03,busuioc19pre}.
The upwind method is known to introduce numerical 
dissipation \cite{sofonea03jcp}. The numerical errors due to 
this spurious dissipation can be controlled by refining the 
grid. This becomes especially important in the inviscid regime,
where the numerical viscosity can dominate over the physical 
one \cite{sofonea03jcp,sofonea04pre}. In the slip-flow and transition regimes,
we find that $2S = 64$ points per channel width are sufficient 
to obtain results which have errors less than $0.1\%$ (for more 
details, see Sec.~\ref{sec:meth:FDLB}),
which is acceptable from a computational cost point of view,
e.g., compared to $100$ \cite{graur09}, $101$ \cite{naris04}, 
$500$ \cite{su19} or $5000$ \cite{tantos18} grid points employed 
in previous studies.

For simplicity, in this paper we only consider the Maxwell 
diffuse reflection model with complete accommodation at the 
bounding walls. The methodology can easily be extended to 
the case of more complex boundary conditions, such as the 
diffuse-specular \cite{sharipov16} and the Cercignani-Lampis 
\cite{cercignani71} boundary models.

This paper is organised as follows.
The kinetic models and the connection to the DSMC simulations 
via the transport coefficients is presented in Sec.~\ref{sec:boltz}.
The FDLB algorithm is summarized in Sec.~\ref{sec:FDLB}
and the simulation methodology employed in the frame
of the FDLB and DSMC approaches is summarized in 
Sec.~\ref{sec:meth}. \ref{app:hydro} discusses the application of 
the FDLB method to the hydrodynamic regime.
Sections~\ref{sec:ht}, 
\ref{sec:couette} and \ref{sec:htsh} present the numerical 
results for the heat transfer between stationary plates, the
Couette flow and the heat transfer between moving plates problems, 
respectively. Section~\ref{sec:conc} concludes this paper.

\section{Kinetic models and connection to DSMC}\label{sec:boltz}

Subsection~\ref{sec:boltz:RTA} introduces briefly the 
Shakhov and ellipsoidal-BGK models. Subsection~\ref{sec:boltz:fred}
introduces the reduced distribution functions
employed in the context of the channel flows discussed in this 
paper.
The implementation of the transport coefficients using the numerical 
data obtained from {\it ab initio} potentials at the level of the 
model equations is discussed in Subsec.~\ref{sec:boltz:tcoeff}.
Finally, our non-dimensionalization conventions are 
summarized in Subsec.~\ref{sec:boltz:adim}.

\subsection{Model equations in the relaxation time approximation} \label{sec:boltz:RTA}
In this paper, we focus on the study of channel flows between parallel plates.
The coordinate system is chosen such that the $\widetilde{x}$ axis is perpendicular 
to the walls. The discussion in this section is presented at 
the level of dimensional
quantities, which are denoted explicitly via an overhead tilde.
The origin
of the coordinate system is taken to be on the channel centerline, such that the
left and right walls are located at $\widetilde{x} = -\widetilde{L}/2$ and 
$\widetilde{x} = \widetilde{L}/2$, respectively.
The flow is studied in the Galilean frame where the left and right plates move with
velocities $-\widetilde{u}_w$ and $\widetilde{u}_w$, respectively 
($\widetilde{u}_w = 0$ for the heat transfer
problem between stationary plates discussed in Sec.~\ref{sec:ht}).
The temperatures of the left and right plates are set to
$\widetilde{T}_{\rm left} = \widetilde{T}_{\rm ref} - \widetilde{\Delta T} / 2$ and 
$\widetilde{T}_{\rm right} = \widetilde{T}_{\rm ref} + \widetilde{\Delta T} / 2$, respectively
($\widetilde{\Delta T} = 0$ for the Couette flow problem discussed in Sec.~\ref{sec:couette}).
In this case, the Boltzmann equation in the relaxation time approximation for the 
collision term can be written as follows:
\begin{equation}
 \frac{\partial \widetilde{f}}{\partial \widetilde{t}} + 
 \frac{\widetilde{p}_x}{\widetilde{m}} \frac{\partial \widetilde{f}}
 {\partial \widetilde{x}} = -\frac{1}{\widetilde{\tau}_*}(
 \widetilde{f} - \widetilde{f}_*),\label{eq:boltz}
\end{equation}
where $\widetilde{f}$ is the particle distribution function, 
$\widetilde{p}_x$ is the particle momentum along the direction 
perpendicular to the walls, $\widetilde{m}$ is the particle 
mass and $\widetilde{\tau}_*$ is the relaxation time. The collision 
term governs the relaxation of $\widetilde{f}$ towards
the local equilibrium distribution function $\widetilde{f}_*$. 
The star subscript in Eq.~\eqref{eq:boltz} distinguishes between 
the two models that we consider in this paper, namely the Shakhov 
model ($* = {\rm S}$) and the ellipsoidal-BGK ($* = {\rm ES}$) model.
We consider in this paper only monatomic 
ideal gases, for which $\widetilde{f}_*$ reduces at global thermodynamic 
equilibrium to the Maxwell-Boltzmann distribution function 
$\widetilde{f}_{\rm MB}$:
\begin{align}
 \widetilde{f}_{\rm MB}(\widetilde{n}, \widetilde{\bm{u}}, \widetilde{T}) =& 
 \widetilde{n} \widetilde{g}(\widetilde{p}_x, \widetilde{u}_x, \widetilde{T}) 
 \widetilde{g}(\widetilde{p}_y, \widetilde{u}_y, \widetilde{T}) 
 \widetilde{g}(\widetilde{p}_z, \widetilde{u}_z, \widetilde{T}), \nonumber\\
 \widetilde{g}(\widetilde{p}, \widetilde{u}, \widetilde{T}) =& 
 \frac{1}{\sqrt{2\pi \widetilde{m} \widetilde{K}_B \widetilde{T}}} 
 \exp\left[-\frac{(\widetilde{p} - \widetilde{m}\widetilde{u})^2}
 {2\widetilde{m}\widetilde{K}_B \widetilde{T}}\right].
 \label{eq:feq}
\end{align}
Here $\widetilde{n}$ is the particle number density, 
$\widetilde{T}$ is the temperature and $\widetilde{u}_\alpha$ 
($\alpha \in \{x, y, z\}$) are the components of 
the macroscopic velocity. These quantities are obtained as moments 
of $\widetilde{f}$ and $\widetilde{f}_*$ via the following relations:
\begin{equation}
 \begin{pmatrix}
  \widetilde{n}\\ \widetilde{\rho} \widetilde{\bm{u}} \\ 
  \frac{3}{2} \widetilde{n} \widetilde{K}_B \widetilde{T}
 \end{pmatrix} = 
 \int d^3 \widetilde{p}
 \begin{pmatrix}
  1\\ \widetilde{\bm{p}} \\ 
  \widetilde{\bm{\xi}}^2 / 2 \widetilde{m}
 \end{pmatrix} \widetilde{f} = \int d^3 \widetilde{p}
 \begin{pmatrix}
  1\\ \widetilde{\bm{p}} \\ 
  \widetilde{\bm{\xi}}^2 / 2\widetilde{m}
 \end{pmatrix} \widetilde{f}_*,
 \label{eq:macro_inv}
\end{equation}
where $\widetilde{\bm{\xi}} = \widetilde{\bm{p}} -\widetilde{m}\widetilde{\bm{u}}$ 
is the peculiar momentum. The last equality above is a statement that 
the model equations preserve the collision invariants, 
$\psi \in \{1, \bm{p}, \bm{p}^2 / 2m\}$, of the original Boltzmann 
collision operator.

In the case of the Shakhov (S) model, the local equilibrium can be written as
\cite{shakhov68a,shakhov68b,ambrus12pre,ambrus18pre,ambrus19rgd}:
\begin{equation}
 \widetilde{f}_{\rm S} = \widetilde{f}_{\rm MB}(1 + \mathbb{S}), \qquad
 \mathbb{S} = \frac{1-{\rm Pr}}
 {\widetilde{n} \widetilde{K}_B^2 \widetilde{T}^2}
 \left(\frac{\widetilde{\bm{\xi}}^{2}}
 {5\widetilde{m}\widetilde{K}_B \widetilde{T}}-1\right)
 \widetilde{\bm{q}} \cdot \widetilde{\bm{\xi}},\label{eq:sdef}
\end{equation}
where the heat flux $\widetilde{\bm{q}}$ is obtained via
\begin{equation}
 \widetilde{\bm{q}} = \int d^3 \widetilde{p}\,
 \widetilde{f} \frac{\widetilde{\bm{\xi}}^2}{2\widetilde{m}}
 \frac{\widetilde{\bm{\xi}}}{\widetilde{m}}.
 \label{eq:macro_q}
\end{equation}
In the S model, the dynamic viscosity and the heat conductivity 
are controlled by the relaxation time $\widetilde{\tau}_{\rm S}$ and 
the Prandtl number ${\rm Pr}$ through
\begin{equation}
 \widetilde{\mu}_{\rm S} = \widetilde{\tau}_{\rm S} \widetilde{P}, \qquad 
 \widetilde{\kappa}_{\rm S} = \frac{\widetilde{c}_p \widetilde{\mu}_{\rm S}}{\rm Pr} = 
 \frac{5 \widetilde{K}_B \widetilde{\tau}_{\rm S} \widetilde{P}}{2 \widetilde{m} {\rm Pr}},
 \label{eq:Stcoeff}
\end{equation}
where $\widetilde{c}_p= 5 \widetilde{K}_B / 2\widetilde{m}$ 
is the specific heat at constant pressure for an ideal monatomic gas.

In the ellipsoidal-BGK (ES) model, the equilibrium 
distribution $\widetilde{f}_{\rm ES}$ can 
be written as \cite{holway66,meng13,sharipov16,zhang18}:
\begin{equation}
 \widetilde{f}_{\rm ES} = \frac{\widetilde{n}}
 {(2 \pi \widetilde{m} \widetilde{K}_B \widetilde{T})^{3/2} 
 \sqrt{{\rm det} \mathbb{B}}}
 \exp\left(-\frac{\mathbb{B}_{\alpha\beta}^{-1} \widetilde{\xi}_\alpha \widetilde{\xi}_\beta}
 {2 \widetilde{m} \widetilde{K}_B \widetilde{T}}\right),
 \label{eq:fES}
\end{equation}
where $\mathbb{B}$ is an invertible $3\times 3$ matrix 
($1 \le \alpha, \beta \le D = 3$) having 
the following components:
\begin{equation}
 \mathbb{B}_{\alpha\beta} = \frac{1}{\rm Pr} 
 \left[\delta_{\alpha\beta} - (1 - {\rm Pr}) 
 \frac{\widetilde{T}_{\alpha\beta}}{\widetilde{P}}\right].
\end{equation}
In the above, $\widetilde{P} = \widetilde{n} \widetilde{K}_B \widetilde{T}$
is the ideal gas pressure, while
the Cartesian components $\widetilde{T}_{\alpha\beta}$
of the pressure tensor are obtained as 
second order moments of $\widetilde{f}$:
\begin{equation}
 \widetilde{T}_{\alpha\beta} = \int d^3 \widetilde{p}\,
 \widetilde{f} \frac{\widetilde{\xi}_\alpha \widetilde{\xi}_\beta}
 {\widetilde{m}}.
 \label{eq:macro_Tij}
\end{equation}
In the ES model, the transport coefficients are retrieved through:
\begin{equation}
 \widetilde{\mu}_{\rm ES} = \widetilde{\tau}_{\rm ES} {\rm Pr}\, \widetilde{P}, \qquad 
 \widetilde{\kappa}_{\rm ES} = 
 \frac{\widetilde{c}_p \widetilde{\mu}_{\rm ES}}{\rm Pr} = 
 \frac{5 \widetilde{K}_B \widetilde{\tau}_{\rm ES} \widetilde{P}}{2\widetilde{m}}.
 \label{eq:EStcoeff}
\end{equation}

Eq.~\eqref{eq:boltz} is supplemented by boundary conditions. 
In this paper, we restrict the analysis to the case of diffuse reflection
with complete accommodation, such that the distribution of the particles 
emerging from the wall back into the fluid is described by the 
Maxwell-Boltzmann distribution \cite{sharipov16}:
\begin{align}
 \widetilde{f}(-\widetilde{L}/2, \widetilde{p}_x > 0, \widetilde{t}) =& 
 \widetilde{f}_{\rm MB}(\widetilde{n}_{\rm left}, -\widetilde{\bm{u}}_w, 
 \widetilde{T}_{\rm left}), \nonumber\\
 \widetilde{f}(\widetilde{L}/2, \widetilde{p}_x < 0, \widetilde{t}) =& 
 \widetilde{f}_{\rm MB}(\widetilde{n}_{\rm right}, \widetilde{\bm{u}}_w, 
 \widetilde{T}_{\rm right}),
 \label{eq:diffuse_f}
\end{align}
where $\widetilde{n}_{\rm left}$ and $\widetilde{n}_{\rm right}$ 
are determined by imposing zero mass flux through the walls:
\begin{equation}
 \int d^3 \widetilde{p}\, \widetilde{f}(\pm \widetilde{L}/2, \widetilde{\bm{p}}, t) 
 \widetilde{p}_x = 0.
 \label{eq:diffuse_mflux}
\end{equation}
Substituting Eq.~\eqref{eq:diffuse_f} into Eq.~\eqref{eq:diffuse_mflux}
gives \cite{graur09}:
\begin{align}
 \widetilde{n}_{\rm left} =& -\sqrt{\frac{2 \pi}{\widetilde{m} 
 \widetilde{K}_B \widetilde{T}_{\rm left}}}
 \int_{\widetilde{p}_x < 0} d^3\widetilde{p} \, 
 \widetilde{f}(-\widetilde{L}/2,\widetilde{\bm{p}}, \widetilde{t}) \widetilde{p}_x,\nonumber\\
 \widetilde{n}_{\rm right} =& \sqrt{\frac{2 \pi}{\widetilde{m} 
 \widetilde{K}_B \widetilde{T}_{\rm right}}}
 \int_{\widetilde{p}_x > 0} d^3\widetilde{p} \, 
 \widetilde{f}(\widetilde{L}/2,\widetilde{\bm{p}}, \widetilde{t}) \widetilde{p}_x. 
 \label{diffuse_nw}
\end{align}

\subsection{Reduced distributions}\label{sec:boltz:fred}

In the context of the channel flows considered in this paper, 
the dynamics along the $z$ direction is trivial. Moreover, 
in the heat transfer problem without shear,
the dynamics along the $y$ axis also
becomes trivial. In this context, it is convenient to 
integrate out the trivial momentum space degrees of freedom
at the level of the model equation.

For notational convenience, let $D = 3$ represent the total number 
of degrees of freedom of the momentum space. 
Denoting by $d$ the number of non-trivial momentum space degrees of 
freedom, the $D - d$ degrees of freedom can be integrated out and two 
reduced distribution functions, $\widetilde{\phi}$ and 
$\widetilde{\chi}$, can be introduced 
as follows \cite{graur09,ambrus18pre,li04,ambrus19COST,busuioc19pre,meng13jcp}:
\begin{equation}
 \widetilde{\phi} = \int d^{D-d} \widetilde{p} \widetilde{f},\qquad
 \widetilde{\chi} = \int d^{D-d} \widetilde{p}\, 
 \frac{\widetilde{p}_{d+1}^2 + \cdots \widetilde{p}_{D}^2}{\widetilde{m}}
 \widetilde{f}.
 \label{eq:fred_def}
\end{equation}
The evolution equations for $\widetilde{\phi}$ and $\widetilde{\chi}$ 
can be obtained by multiplying Eq.~\eqref{eq:boltz} with the appropriate 
factors and integrating with respect to the $D - d$ trivial momentum 
space degrees of freedom:
\begin{equation}
 \frac{\partial}{\partial \widetilde{t}} 
 \begin{pmatrix}
  \widetilde{\phi}\smallskip \\ \widetilde{\chi}
 \end{pmatrix} + \frac{\widetilde{p}_x}{\widetilde{m}} 
 \frac{\partial}{\partial \widetilde{x}}
 \begin{pmatrix}
  \widetilde{\phi}\smallskip \\ \widetilde{\chi}
 \end{pmatrix} = -\frac{1}{\widetilde{\tau}_*}
 \begin{pmatrix}
  \widetilde{\phi} - \widetilde{\phi}_* \smallskip \\
  \widetilde{\chi} - \widetilde{\chi}_*
 \end{pmatrix}.
 \label{eq:boltz_red}
\end{equation}

Denoting using latin indices $i$ and $j$ the components 
corresponding to the non-trivial directions ($1 \le i, j \le d$),
the macroscopic moments given in Eqs.~\eqref{eq:macro_inv},
\eqref{eq:macro_Tij} and \eqref{eq:macro_q} can be obtained 
through:
\begin{align}
 \begin{pmatrix}
  \widetilde{n} \smallskip \\ \widetilde{\rho} \widetilde{u}_i
  \smallskip \\ \widetilde{T}_{ij} 
 \end{pmatrix} =& \int d^d \widetilde{p} 
 \begin{pmatrix}
  1 \smallskip \\ \widetilde{p}_i \smallskip \\ \widetilde{\xi}_i \widetilde{\xi}_j / \widetilde{m}
 \end{pmatrix} \widetilde{\phi}, \nonumber\\
 \begin{pmatrix}
  \frac{3}{2} \widetilde{n} \widetilde{K}_B \widetilde{T} \smallskip \\ 
  \widetilde{q}_i
 \end{pmatrix} =& \int d^d\widetilde{p} 
 \begin{pmatrix}
  1 \smallskip \\ \widetilde{\xi}_i / \widetilde{m}
 \end{pmatrix} \left(
 \frac{\widetilde{\xi}_j \widetilde{\xi}_j}{2\widetilde{m}} 
 \widetilde{\phi} + \frac{1}{2} \widetilde{\chi}\right),
 \label{eq:macro_red}
\end{align}
where the summation over the repeated index $j$ is implied.

For the Shakhov model, $\widetilde{\phi}_{\rm S}$ and 
$\widetilde{\chi}_{\rm S}$ are given by \cite{ambrus18pre}:
\begin{align}
 \widetilde{\phi}_{\rm S} =& \widetilde{\phi}_{\rm MB}(1 + \mathbb{S}_\phi), &
 \mathbb{S}_\phi =& \frac{1-{\rm Pr}}
 {(D + 2) \widetilde{n} \widetilde{K}_B^2 \widetilde{T}^2}
 \left(\frac{\widetilde{\xi}_j \widetilde{\xi}_j}
 {\widetilde{m} \widetilde{K}_B \widetilde{T}}-d-2 \right)
 \widetilde{q}_i \widetilde{\xi}_i,\nonumber\\
 \widetilde{\chi}_{\rm S} =& (D-  d) \widetilde{K}_B \widetilde{T} 
 \widetilde{\phi}_{\rm MB}(1 + \mathbb{S}_\chi), &
 \mathbb{S}_\chi =& \frac{1-{\rm Pr}}
 {(D + 2) \widetilde{n} \widetilde{K}_B^2 \widetilde{T}^2}
 \left(\frac{\widetilde{\xi}_j \widetilde{\xi}_j}
 {\widetilde{m} \widetilde{K}_B \widetilde{T}}-d \right)
 \widetilde{q}_i \widetilde{\xi}_i, \label{eq:phiS}
\end{align}
where again the summation over the repeated index $j$ is implied.
The reduced Maxwell-Boltzmann distribution 
$\widetilde{\phi}_{\rm MB}$ is:
\begin{equation}
 \widetilde{\phi}_{\rm MB} = \widetilde{n} 
 \widetilde{g}_x(\widetilde{p}_x, \widetilde{u}_x, \widetilde{T}) \cdots 
 \widetilde{g}_d(\widetilde{p}_d, \widetilde{u}_d, \widetilde{T}).
 \label{eq:phiMB}
\end{equation}

Before discussing the ES model, we first mention 
that the representation as a $D \times D$ matrix 
of the pressure tensor $\widetilde{T}_{\alpha\beta}$ admits 
the following block decomposition:
\begin{equation}
 \widetilde{T}_{\alpha\beta} = 
 \begin{pmatrix}
  \widetilde{T}_{ij} & 0_{ib}\\
  0_{aj} & \widetilde{P}_{\rm red} \delta_{ab}
 \end{pmatrix},\label{eq:Tblock}
\end{equation}
where the latin indices at the beginning of the alphabet run over the trivial 
degrees of freedom, i.e. $d < a, b \le D$. With this convention, 
the top left and bottom right blocks are $d \times d$ and 
$(D - d) \times (D -d)$ matrices with components $\widetilde{T}_{ij}$ and 
$\widetilde{P}_{\rm red} \delta_{ab}$, respectively, while 
the top right and bottom left blocks are $d\times (D-d)$ and 
$(D - d) \times d$ null matrices, respectively.
The Kronecker delta $\delta_{ab}$ takes the value $1$ when $a = b$ and 
$0$ otherwise. The scalar quantity $\widetilde{P}_{\rm red}$ 
is obtained from Eq.~\eqref{eq:macro_red}:
\begin{equation}
 \widetilde{P}_{\rm red} = 
 \frac{D \widetilde{P} - \sum_{j = 1}^d \widetilde{T}_{jj}}{D - d}.
\end{equation}
Using the same decomposition as in Eq.~\eqref{eq:Tblock}, 
the matrix $\mathbb{B}_{\alpha\beta}$ can be written as:
\begin{equation}
 \mathbb{B}_{\alpha\beta} = 
 \begin{pmatrix}
  \mathcal{B}_{ij} & 0_{ib} \smallskip \\
  0_{aj} & \mathbb{B}_{\rm red} \delta_{ab},
 \end{pmatrix},
\end{equation}
where 
\begin{equation}
 \mathcal{B}_{ij} = \frac{1}{\rm Pr} \delta_{ij} - 
 \frac{1 - {\rm Pr}}{\rm Pr} \frac{\widetilde{T}_{ij}}{\widetilde{P}}.
\end{equation}
The scalar quantity $\mathbb{B}_{\rm red}$ is given by:
\begin{equation}
 \mathbb{B}_{\rm red} = 
 \frac{1}{\rm Pr} - \frac{1 - {\rm Pr}}{\rm Pr} 
 \frac{\widetilde{P}_{\rm red}}{\widetilde{P}}.
\end{equation}
It can be seen that the determinant of $\mathbb{B}$ can be written as:
\begin{equation}
 {\rm det}\,\mathbb{B} = \mathbb{B}_{\rm red}^{D - d} 
 {\rm det}\, \mathcal{B}.
\end{equation}
This allows the integral of $\widetilde{f}_{\rm ES}$ over the $D - d$ 
trivial degrees of freedom to be performed analytically, giving:
\begin{equation}
 \widetilde{\phi}_{\rm ES} = 
 \frac{\widetilde{n}}{(2 \pi \widetilde{m} \widetilde{K}_B\widetilde{T})^{d/2} 
 \sqrt{{\rm det}\, \mathcal{B}}} 
 \exp\left(-\frac{\mathcal{B}_{ij}^{-1} \widetilde{\xi}_i \widetilde{\xi}_j}
 {2\widetilde{m} \widetilde{K}_B \widetilde{T}}\right),
 \label{eq:phiES}
\end{equation}
while $\widetilde{\chi}_{\rm ES} = (D - d) \widetilde{K}_B \widetilde{T}_{\rm red} 
\widetilde{\phi}_{\rm ES}$ and
$\widetilde{K}_B \widetilde{T}_{\rm red} = \widetilde{P}_{\rm red} / \widetilde{n}$.

\subsection{{\it Ab initio} transport coefficients}\label{sec:boltz:tcoeff}

In this paper, we consider a series of comparisons between the 
results obtained in the frame of the model equations introduced in 
the previous subsections and the
results obtained using the DSMC method 
with {\it ab initio} particle interactions. The connection between 
these two formulations can be made at the level of the transport 
coefficients. The basis for the approach that we take in this paper 
is to note that in the variable hard spheres model, the viscosity
has a temperature dependence of the form \cite{bird94}
\begin{equation}
 \widetilde{\mu} = \widetilde{\mu}_{\rm ref} 
 (\widetilde{T} / \widetilde{T}_{\rm ref})^{\omega},
 \label{eq:mu}
\end{equation}
where the tilde denotes dimensionful quantities, as discussed in 
the previous subsection. The viscosity index $\omega$ introduced 
above takes the values $1/2$ and $1$ for hard sphere and Maxwell 
molecules, respectively. For real gases, $\omega$
is in general temperature-dependent. This temperature dependence 
is not known analytically, however the values of 
$\widetilde{\mu}$ and $\widetilde{\kappa}$ corresponding to a gas 
comprised of molecules interacting via {\it ab initio} potentials
can be computed numerically. 
The supplementary material in Ref.~\cite{cencek12}
contains the data corresponding to ${}^3{\rm He}$ and 
${}^4{\rm He}$ consistuents in the temperature ranges 
$1\ {\rm K}\le \widetilde{T} \le 10000\ {\rm K}$,
while the data for ${\rm Ne}$ covering the range 
$20\ {\rm K} \le \widetilde{T} \le 10000\ {\rm K}$ 
can be found in the tables reported in Ref.~\cite{sharipov17}.
In order to perform simulations of the heat transfer problem 
(discussed in Sec.~\ref{sec:htsh})
at $\widetilde{T}_{\rm ref} = 1\ {\rm K}$ (for ${\rm He}$ constituents) 
and $20\ {\rm K}$ (for ${\rm Ne}$ constituents), we require data 
for the transport coefficients also in the temperature 
range $0.25\ {\rm K} \le \widetilde{T} \le 1\ {\rm K}$ and 
$5\ {\rm K} \le \widetilde{T} \le 20\ {\rm K}$, respectively.
These data were obtained by the method 
described in Ref.~\cite{sharipov17}.

The temperature dependence of the viscosity index $\omega$ is accounted 
for by employing Eq.~\eqref{eq:mu} in a piecewise fashion.
Let $n$ ($1 \le n \le N$) represent the index of the tabulated values
$\widetilde{T}_1 < \widetilde{T}_2 < \dots \widetilde{T}_N$ of the 
temperature, where $N$ is the total number of available entries.
Considering a temperature interval 
$\widetilde{T}_n \le \widetilde{T} \le \widetilde{T}_{n+1}$,
we define
\begin{equation}
 \widetilde{\mu}^{(n)}(\widetilde{T}) = \widetilde{\mu}_n 
 \left(\frac{\widetilde{T}}{\widetilde{T}_n}\right)^{\omega_n}, \qquad
 \omega_n = \frac{\ln(\widetilde{\mu}_{n+1} / \widetilde{\mu}_n)}
 {\ln(\widetilde{T}_{n+1} / \widetilde{T}_n)},
 \label{eq:mun}
\end{equation}
where $\widetilde{\mu}_n$ and $\widetilde{\mu}_{n+1}$ are the 
tabulated values of the viscosity corresponding to the temperatures
$\widetilde{T}_n$ and $\widetilde{T}_{n+1}$, respectively.
The above formula ensures that the function $\widetilde{\mu}^{(n)}$ satisfies
$\widetilde{\mu}^{(n)}(\widetilde{T}_n) = \widetilde{\mu}_n$ and 
$\widetilde{\mu}^{(n)}(\widetilde{T}_{n+1}) = \widetilde{\mu}_{n+1}$.

The Prandtl number ${\rm Pr}$ is also defined in a 
piecewise fashion. For the temperature range 
$\widetilde{T}_n \le \widetilde{T} < \widetilde{T}_{n+1}$, 
we define ${\rm Pr}_{n}$ as
\begin{equation}
 {\rm Pr}_n = \frac{\widetilde{c}_p \widetilde{\mu}_n}{\widetilde{\kappa}_n},
\end{equation}
where $\widetilde{\kappa}_n$ is the heat conductivity corresponding to
the temperature $\widetilde{T} = \widetilde{T}_n$, retrieved from the 
tabulated data mentioned above.

In general, the temperatures encountered in our simulations are within the bounds of 
the temperature range for which data is available for interpolation. For completeness, 
we present a possible extension of the above procedure for values of the
temperature which are outside the range spanned by the tabulated data.
In the case when $\widetilde{T} < \widetilde{T}_2$, we propose to use
$\widetilde{\mu}(\widetilde{T}) = \widetilde{\mu}^{(1)}(\widetilde{T})$
and ${\rm Pr}(T) = {\rm Pr}_1$. For $\widetilde{T} > \widetilde{T}_N$, 
where $\widetilde{T}_N$ is the highest available 
temperature in the tabulated data, we propose to use 
$\widetilde{\mu}(\widetilde{T}) = \widetilde{\mu}^{(N - 1)}(\widetilde{T})$
and ${\rm Pr}(\widetilde{T}) = {\rm Pr}_N$. 

The algorithm described in this section can be summarized through \cite{ambrus19rgd}:
\begin{align}
 \widetilde{\mu}(\widetilde{T}) =&
 \begin{cases}
  \widetilde{\mu}^{(1)}(\widetilde{T}), & \widetilde{T} < \widetilde{T}_2, \\
  \widetilde{\mu}^{(n)}(\widetilde{T}), & \widetilde{T}_n < \widetilde{T} < \widetilde{T}_{n+1}, \\
  \widetilde{\mu}^{(N-1)}(\widetilde{T}), & \widetilde{T}_{N} < \widetilde{T},
 \end{cases}, \nonumber\\
 {\rm Pr}(\widetilde{T}) =&
 \begin{cases}
  {\rm Pr}_1, & \widetilde{T} < \widetilde{T}_2, \\
  {\rm Pr}_n, & \widetilde{T}_n < \widetilde{T} < \widetilde{T}_{n+1}, \\
  {\rm Pr}_{N}, & \widetilde{T}_{N} < \widetilde{T},
 \end{cases},
 \label{eq:interp}
\end{align}
where $n = 2, 3, \dots N-1$ refers to the index of the tabulated data.


\subsection{Non-dimensionalization convention}\label{sec:boltz:adim}

All simulation results reported in this paper are based on the 
nondimensionalization conventions employed in Ref.~\cite{ambrus18pre},
which are summarized here for completeness. 
In general, the dimensionless form $A$ of a
dimensional quantity 
$\widetilde{A}$ is obtained by dividing the latter with respect 
to its reference value, $\widetilde{A}_{\rm ref}$:
\begin{equation}
 A = \frac{\widetilde{A}}{\widetilde{A}_{\rm ref}}.
\end{equation}
We employ the convention that dimensionless quantities are denoted without 
the overhead tilde encountered for their dimensionful counterparts. The 
reference temperature is taken as the average of the wall temperatures:
\begin{equation}
 \widetilde{T}_{\rm ref} = \frac{\widetilde{T}_{\rm left} + 
 \widetilde{T}_{\rm right}}{2}.
 \label{eq:Tref}
\end{equation}
The reference speed is defined through:
\begin{equation}
 \widetilde{c}_{\rm ref} = \sqrt{\frac{\widetilde{K}_B \widetilde{T}_{\rm ref}}{\widetilde{m}}},
 \label{eq:cref}
\end{equation}
where the particle mass $\widetilde{m}$ takes the values 
$5.0082373 \times 10^{-27}\,{\rm kg}$, 
$6.6464764 \times 10^{-27}\,{\rm kg}$, and 
$3.3509177 \times 10^{-26}\,{\rm kg}$ for 
${}^3{\rm He}$, ${}^4{\rm He}$, and 
${\rm Ne}$, respectively.

The reference particle number density is taken as the average particle number 
density over the channel:
\begin{equation}
 \widetilde{n}_{\rm ref} = \frac{1}{\widetilde{L}}
 \int_{-\widetilde{L}/2}^{\widetilde{L}/2} d\widetilde{x} \, \widetilde{n}.
\end{equation}
The reference length is taken to be the channel width:
\begin{equation}
 \widetilde{L}_{\rm ref} = \widetilde{L}.
\end{equation}
Finally, the reference time is
\begin{equation}
 \widetilde{t}_{\rm ref} = \frac{\widetilde{L}_{\rm ref}}{\widetilde{c}_{\rm ref}} = 
 \widetilde{L} \sqrt{\frac{\widetilde{m}}{\widetilde{K}_B \widetilde{T}_{\rm ref}}}.
\end{equation}

The dimensionless relaxation time 
$\tau_* = \widetilde{\tau}_* / \widetilde{t}_{\rm ref}$
in the S and ES models becomes:
\begin{equation}
 \tau_{\rm S} = \frac{\mu(T)}{P \delta \sqrt{2}}, \qquad 
 \tau_{\rm ES} = \frac{\mu(T)}{{\rm Pr} \, P \delta \sqrt{2}},
 \label{eq:tau}
\end{equation}
where the rarefaction parameter $\delta$ is
defined through \cite{sharipov18}:
\begin{equation}
 \delta = \frac{\widetilde{L} \widetilde{P}_{\rm ref}}
 {\widetilde{\mu}_{\rm ref} \widetilde{c}_{\rm ref} \sqrt{2}},
 \label{eq:delta}
\end{equation}
In the above, $\widetilde{\mu}_{\rm ref} = 
\widetilde{\mu}(\widetilde{T}_{\rm ref})$ and
$\widetilde{P}_{\rm ref} = \widetilde{n}_{\rm ref} 
\widetilde{K}_B \widetilde{T}_{\rm ref}$.

The distribution function $\widetilde{f}$ is nondimensionalized via
\begin{equation}
 f = \frac{\widetilde{f} \widetilde{p}_{\rm ref}^D}{\widetilde{n}_{\rm ref}},
\end{equation}
where $\widetilde{p}_{\rm ref} = \sqrt{\widetilde{m} \widetilde{K}_B 
\widetilde{T}_{\rm ref}}$.
The reduced distributions can be nondimensionalized in a similar fashion:
\begin{equation}
 \phi = \frac{\widetilde{\phi} \widetilde{p}_{\rm ref}^d}{\widetilde{n}_{\rm ref}}, \qquad 
 \chi = \frac{\widetilde{\chi} \widetilde{p}_{\rm ref}^d}{\widetilde{P}_{\rm ref}}.
\end{equation}
This allows Eq.~\eqref{eq:boltz_red} to be written as:
\begin{equation}
 \frac{\partial}{\partial t} 
 \begin{pmatrix}
  \phi \smallskip \\ \chi
 \end{pmatrix} + \frac{p_x}{m}
 \frac{\partial}{\partial x}
 \begin{pmatrix}
  \phi \smallskip \\ \chi
 \end{pmatrix} = -\frac{1}{\tau_*}
 \begin{pmatrix}
  \phi - \phi_* \smallskip \\
  \chi - \chi_*
 \end{pmatrix}.
 \label{eq:boltz_adim}
\end{equation}

\section{Finite difference lattice Boltzmann models with mixed quadratures}\label{sec:FDLB}

In this section, the FDLB algorithm employed to solve Eq.~\eqref{eq:boltz_adim}
is briefly described. There are three pieces to the algorithm, which will be 
outlined in the following subsections. 
The first piece concerns the implementation of both the
time stepping and the advection, which
will be addressed in Subsec.~\ref{sec:FDLB:num}.
The second concerns the discretization of the momentum space using 
the full-range and half-range Gauss-Hermite quadratures. This 
will be discussed in Subsec.~\ref{sec:FDLB:quad}.
The third and final piece is the projection of the collision term 
in the model equation on the space generated by the full-range 
Hermite polynomials for the direction parallel to the wall.
Details will be given in Subsec.~\ref{sec:FDLB:feq}.

\subsection{Time stepping and advection}\label{sec:FDLB:num}

In order to describe the time stepping algorithm, 
Eq.~\eqref{eq:boltz_adim} is written as:
\begin{equation}
 \partial_t F = L[F],
\end{equation}
where $F \in \{\phi, \chi\}$ represents the 
reduced distributions. Considering the equidistant discretization 
of the time variable using intervals $\delta t$ and using 
$t_n = n \delta t$ to denote the time coordinate after $n$ 
iterations, we employ the third order total variation diminishing 
Runge-Kutta scheme to obtain $F_{n+1}$ at time $t_{n+1}$ through 
two intermediate steps \cite{shu88,gottlieb98,trangenstein07}:
\begin{align}
 F^{(1)}_n =& F_n + \delta t L[F_n], \nonumber\\
 F^{(2)}_n =& \frac{3}{4} F_n + \frac{1}{4} F^{(1)}_n +  
 \frac{1}{4} \delta t L[F^{(1)}_n],\nonumber\\
 F_{n+1} =& \frac{1}{3} F_n + \frac{2}{3} F^{(2)}_n +  
 \frac{2}{3} \delta t L[F^{(2)}_n].
\end{align}

As pointed out by various authors \cite{mei98jcph,guo03,busuioc19pre},
an accurate account for the Knudsen layer phenomena requires 
a sufficiently fine grid close to the wall. 
This can be achieved by performing 
a coordinate change from $x = \widetilde{x} / \widetilde{L}$ 
to the coordinate $\eta$, defined through 
\cite{ambrus18pre,ambrus19COST,busuioc19pre,ambrus19rgd}:
\begin{equation}
 x = \frac{\tanh \eta}{2 A},
 \label{eq:stretch}
\end{equation}
where the stretching parameter $A$ controls the grid refinement.
When $A \rightarrow 0$, the grid becomes equidistant, while 
as $A \rightarrow 1$, the grid points accumulate towards the boundaries 
at $x = \pm 1/2$. The channel walls are located 
at $\eta = \pm {\rm arctanh}\, A$. In this paper, we always use the 
stretching corresponding to $A = 0.98$. 

The $\eta$ coordinate is discretized symmetrically with respect 
to the channel centerline (where $x = 0$ and $\eta = 0$).
On the right half of the channel, $S$ equidistant intervals 
of size $\delta \eta = {\rm arctanh}\, A / S$ are employed. 
In the case of the Couette flow, which is symmetric 
with respect to the channel centerline, the simulation
setup contains only the domain 
$0 \le \eta \le A$ and the total number of grid points 
is equal to $S$ \cite{ambrus18pre,ambrus19COST}. The 
center of cell $s$ ($1 \le s \le S$ for the right half of the 
channel and $-S < s \le 0$ for its left half)
is located at $\eta_s = (s - \frac{1}{2}) \delta \eta$. 
At each node $s$, the advection term is 
computed using the third order upwind method, implemented 
using a flux-based approach:
\begin{equation}
 \left(\frac{p_x}{m} \frac{\partial F}{\partial x}\right)_s 
 = \frac{p_x}{m} \left(\frac{\partial \eta}{\partial x}\right)_s
 \left(\frac{\partial F}{\partial \eta}\right)_s
 = 2 A \cosh^2 \eta_s 
 \frac{\mathcal{F}_{s+1/2} - \mathcal{F}_{s-1/2}}{\delta \eta}
  + O(\delta \eta^3).\label{eq:fluxes}
\end{equation}
The stencil employed for the flux $\mathcal{F}_{s+1/2}$ is chosen 
depending on the sign of the advection velocity $p_x / m$:
\begin{equation}
 \mathcal{F}_{s+1/2} = \frac{p_x}{m}
 \begin{cases}
  {\displaystyle \frac{1}{3} F_{s+1} + \frac{5}{6} F_s - \frac{1}{6} F_{s-1}}, & 
  p_x > 0,\smallskip\\
  {\displaystyle \frac{1}{3} F_{s} + \frac{5}{6} F_{s+1} - \frac{1}{6} F_{s+2}}, & 
  p_x < 0.
 \end{cases}
 \label{eq:u3flux}
\end{equation}

The diffuse reflection boundary conditions in Eq.~\eqref{eq:diffuse_f}
specify the distributions $\phi$ and $\chi$ on the channel walls.
For definiteness, we will refer to the right boundary, 
which is located at $\eta_{S + 1/2} = {\rm arctanh} A$. 
In order to perform the advection at node $S$ for the particles 
traveling towards the wall (having $p_x > 0$),
the value of the distribution function in the node
$s = S + 1$ is required. This value
can be obtained using a third order 
extrapolation from the fluid nodes:
\begin{equation}
 F_{S + 1}^{p_x > 0} = 4 F_S - 6 F_{S-1} + 4 F_{S-2} - F_{S-3}.
 \label{eq:ghostplus}
\end{equation}
It can be shown that the third order accuracy in the sense of Eq.~\eqref{eq:fluxes}
is preserved when the fluxes $F_{S+1/2}$ and $F_{S-1/2}$ are computed 
using Eq.~\eqref{eq:u3flux}.
For the particles traveling towards the fluid ($p_x < 0$), the 
nodes at $S + 1$ and $S+ 2$ must be populated. 
According to the diffuse reflection concept, summarized in 
Eq.~\eqref{eq:diffuse_f}, the reduced distributions at 
$s = S + 1/2$ are set to:
\begin{align}
 \phi_{S+1/2} =& \phi_{\rm MB}(n_{\rm right}, \bm{u}_w, T_{\rm right}), \nonumber\\
 \chi_{S+1/2} =& (D - d) T_{\rm right} \phi_{S+1/2},
 \label{eq:bcs}
\end{align}
where $T_{\rm right} = 1 + \Delta T / 2$ is the temperature on the right 
wall ($\Delta T = 0$ in the case of Couette flow).
The distributions in the ghost nodes at $S + 1$ and $S + 2$ 
can be set to \cite{ambrus19luo}:
\begin{align}
 F_{S+1}^{p_x < 0} =& \frac{16}{5} F_{S+1/2} - 3 F_S + F_{S-1} - \frac{1}{5} F_{S-2}, 
 \nonumber\\
 F_{S+2}^{p_x < 0} =& 4 F_{S+1} - 6 F_{S} + 4 F_{S-1} - F_{S-2}.\label{eq:ghostminus}
\end{align}
The expression for $F_{S+2}$ can be seen to represent a third 
order extrapolation from the nodes with $S- 2 \le s \le S + 1$. 
In the expression for $F_{S+1}$, the distribution at the wall, 
$F_{S+1/2}$ is employed. It can be checked by direct substitution 
in Eq.~\eqref{eq:fluxes} that the third order accuracy is preserved 
when the ghost nodes are populated as indicated above.

The density $n_{\rm right}$ in Eq.~\eqref{eq:bcs}
can be computed using the discrete equivalent of Eq.~\eqref{eq:diffuse_mflux}:
\begin{equation}
 \int d^d p\, \Phi_{S+1/2} = 0,\label{eq:diffuse_mflux_flux}
\end{equation}
where $\Phi_{S+1/2}$ is the flux corresponding to the reduced 
distribution $\phi$, computed using Eq.~\eqref{eq:u3flux}.
Using Eq.~\eqref{eq:ghostminus}, the flux for outgoing particles is:
\begin{align}
 \Phi_{S+1/2}^{p_x<0} =& \frac{p_x}{m} 
 \left(\frac{8}{15} \phi_{S+1/2} + \frac{5}{6} \phi_S - 
 \frac{1}{2} \phi_{S-1} + \frac{2}{15} \phi_{S-2}\right),\nonumber\\
 \Phi_{S-1/2}^{p_x<0} =& \frac{p_x}{m} \left(-\frac{8}{15} \phi_{S+1/2} + 
 \frac{4}{3} \phi_S + \frac{1}{6} \phi_{S-1} + \frac{1}{30} \phi_{S-2}\right),
 \label{eq:bcs_Phiw_minus}
\end{align}
where the flux $\Phi_{S-1/2}$ is given above for completeness.
Due to the above expression for $\Phi_{S+1/2}^{p_x < 0}$, it can be 
seen that the unknown density, $n_{\rm right}$, enters 
Eq.~\eqref{eq:diffuse_mflux_flux} through the distribution 
$\phi_{S+1/2}$, which is fixed by boundary conditions for 
momenta pointing towards the fluid ($p_x < 0$), according 
to Eq.~\eqref{eq:bcs}. Splitting the integration domain in
Eq.~\eqref{eq:diffuse_mflux_flux} in two domains, corresponding 
to positive and negative values of $p_x$, the integral
for $p_x < 0$ of $\phi_{S+1/2}$ can be computed as follows:
\begin{equation}
 \int_{p_x < 0} d^dp\, \frac{p_x}{m} \phi_{S+1/2} = 
 -n_{\rm right} \sqrt{\frac{T_{\rm right}}{2\pi m}}.
 \label{eq:bcs_nw_aux}
\end{equation}
Taking into account Eqs.~\eqref{eq:bcs_Phiw_minus} and \eqref{eq:bcs_nw_aux},
the following expression is obtained for $n_{\rm right}$:
\begin{multline}
 n_{\rm right} = \frac{15}{8}\sqrt{\frac{2\pi m}{T_{\rm right}}} 
 \left[ \int_{p_x > 0} d^dp \, \Phi_{S+1/2} \right.\\
 \left.- \int_{p_x < 0} d^dp \frac{p_x}{m} \left(\frac{5}{6} \phi_S - 
 \frac{1}{2} \phi_{S-1} + \frac{2}{15} \phi_{S-2}\right)\right].
\end{multline}
For completeness, we also give below the expressions for 
$\Phi_{S+1/2}$ when $p_x > 0$:
\begin{equation}
 \Phi_{S+1/2}^{p_x>0} = \frac{p_x}{m} \left(\frac{13}{6} \phi_S - 
 \frac{13}{6} \phi_{S-1} + \frac{4}{3} \phi_{S-2} - \frac{1}{3} \phi_{S-3}\right).
\end{equation}

In the case of the Couette flow, only the nodes with 
$1 \le s \le S$ comprise the fluid domain, while 
bounce-back boundary conditions are imposed
on the channel centerline ($s = 1/2$). The nodes 
with $s < 1$ become ghost nodes, which are populated 
according to:
\begin{align}
 p_x <& 0:
 F_0(\bm{p}) = -F_1(-\bm{p}), &
 p_x >& 0:
 \begin{cases}
  F_0(\bm{p}) = -F_1(-\bm{p}), \smallskip \\
  F_{-1}(\bm{p}) = -F_2(-\bm{p}).
 \end{cases} 
\end{align}

\subsection{Momentum space discretization}\label{sec:FDLB:quad}

Through the discretization of the momentum space, the integrals
defining the macroscopic moments in Eq.~\eqref{eq:macro_red} 
are replaced by quadrature sums, i.e.:
\begin{align}
 \begin{pmatrix}
  n \smallskip \\ \rho u_i \smallskip \\ T_{ij}
 \end{pmatrix} \simeq& \sum_{\bm{\kappa}}
 \begin{pmatrix}
  1 \smallskip \\
   p_{\bm{\kappa};i} \smallskip \\ \xi_{\bm{\kappa};i} \xi_{\bm{\kappa};j} / m
 \end{pmatrix} \phi_{\bm{\kappa}},\nonumber\\
 \begin{pmatrix}
  \frac{3}{2} n T \smallskip \\ q_i
 \end{pmatrix} \simeq& \sum_{\bm{\kappa}}
 \begin{pmatrix}
  1 \smallskip \\ \xi_{\bm{\kappa};i} / m
 \end{pmatrix} \frac{\xi_{\bm{\kappa};j} \xi_{\bm{\kappa};j}}{2m} 
 \phi_{\bm{\kappa}} + \frac{1}{2} \sum_{\bm{\sigma}}
 \begin{pmatrix}
  1 \smallskip \\ \xi_{\bm{\sigma};i} / m
 \end{pmatrix} \chi_{\bm{\sigma}},
 \label{eq:macro_quad}
\end{align}
where $\bm{\kappa}$ and $\bm{\sigma}$ collectively denote the indices 
labeling the momenta corresponding to the discrete populations 
$\phi_{\bm{\kappa}}$ and $\chi_{\bm{\sigma}}$.

The discretization on the axis perpendicular to the walls 
(the $x$ axis), is performed using the half-range Gauss-Hermite 
quadrature prescription \cite{ambrus16jcp}. 
In principle, $p_x$ can be discretized separately 
for the $\phi$ and $\chi$ distributions. We take advantage 
of this freedom when deriving the FDLB models for the hydrodynamic 
regime, which is discussed in \ref{app:hydro}. For the flows considered 
in Sections~\ref{sec:ht}, \ref{sec:couette} and \ref{sec:htsh}, we 
consider the same quadrature order $Q_x^\phi = Q_x^\chi \equiv Q_x$ on 
each $x$ semiaxis. For definiteness, we assume $Q_x^\phi = Q_x^\chi$ 
henceforth, unless otherwise stated. Focusing on the distribution $\phi$, 
the discrete momentum components 
$p_{x,k_x}$ ($1 \le k_x \le 2Q_x$) are linked to the roots of
the half-range Hermite polynomial $\hh_{Q_x}(z)$
of order $Q_x$ via:
\begin{equation}
 p_{x,k_x} = 
 \begin{cases}
  p_{0,x} z_{k_x}, & 1 \le k_x \le Q_x, \smallskip \\
  -p_{0,x} z_{k_x - Q_x}, & Q_x < k_x \le 2Q_x,
 \end{cases}
\end{equation}
where $\hh_{Q_x}(z_{k_x}) = 0$ for $1 \le k_x \le Q_x$, 
while $p_{0,x}$ represents a constant momentum scale 
(we set $p_{0,x}=1$ for the rest of this paper).
The same considerations apply for the distribution $\chi$,
after replacing $k_x$ with the index $s_x$ ($1 \le s_x \le 2Q_x$).

When $d = 1$, the populations $\phi_{\bm{\kappa}} \equiv \phi_{k_x}$ 
and $\chi_{\bm{\sigma}} \equiv \chi_{s_x}$ 
are linked to the continuum distributions $\phi$ and $\chi$ through:
\begin{align}
 \phi_{k_x} =& \frac{p_{0,x} w_{k_x}^\hh(Q_x)}{\omega(\overline{p}_{x,k_x})} 
 \phi(\overline{p}_{x,k_x}), &
 \chi_{s_x} =& \frac{p_{0,x} w_{s_x}^\hh(Q_x)}{\omega(\overline{p}_{x,s_x})} 
 \chi(\overline{p}_{x,s_x}),
 \label{eq:Fkx_from_F}
\end{align}
where $\overline{p}_x \equiv p_x / p_{0,x}$, while 
the weight function $\omega(z)$ is defined through:
\begin{equation}
 \omega(z)= \frac{1}{\sqrt{2\pi}} e^{-z^2/2}.
 \label{eq:omega}
\end{equation}
The quadrature weights $w_k^{\hh}(Q)$ can be computed 
using \cite{ambrus16jcp,ambrus16jocs}
\begin{align}
 w_k^{\hh}(Q) =& \frac{\overline{p}_{x,k} a_Q^2}
 {\hh_{Q+1}^2(\overline{p}_{x,k}) [\overline{p}_{x,k} + 
 \hh_Q^2(0) / \sqrt{2\pi}]}\nonumber\\
 =& \frac{\overline{p}_{x,k} a_{Q-1}^2}
 {\hh_{Q-1}^2(\overline{p}_{x,k}) [\overline{p}_{x,k} + 
 \hh_Q^2(0) / \sqrt{2\pi}]}.
\end{align}
In the above, $a_Q = \hh_{Q+1,Q+1} / \hh_{Q,Q}$ represents the 
ratio of the coefficients of the leading power of 
$\overline{p}_{x}$ in $\hh_{Q+1}(\overline{p}_{x})$ and 
$\hh_Q(\overline{p}_{x})$. Specifically, the 
notation $\hh_{\ell,s}$ refers to the coefficient 
of $x^s$ appearing in $\hh_\ell(\overline{p}_{x})$, namely:
\begin{equation}
 \hh_\ell(\overline{p}_{x}) = \sum_{s = 0}^\ell \hh_{\ell,s} 
 \overline{p}_x^s.\label{eq:hh_exp}
\end{equation}

In the case when the boundaries are moving, $d = 2$ and 
the momentum component $p_y$ is discretized using the 
full-range Gauss-Hermite quadrature prescription.
As remarked in Refs.~\cite{ambrus18pre,ambrus16jcp}, 
a small quadrature order is sufficient to ensure the 
exact recovery of the dynamics along this axis. 
To assess the quadrature orders for the $\phi$ and 
$\chi$ distributions, we consider the expansions of 
$\phi$ and $\chi$ with respect to the full-range Hermite polynomials 
for the $p_y$ degree of freedom:
\begin{align}
 \begin{pmatrix}
  \phi \\ \chi 
 \end{pmatrix} =& \frac{\omega(\overline{p}_y) }{p_{0,y}}
 \sum_{\ell = 0}^\infty \frac{1}{\ell!} H_\ell(\overline{p}_y) 
 \begin{pmatrix}
  \Phi_{\ell} \\ X_\ell
 \end{pmatrix},&
 \begin{pmatrix}
  \Phi_\ell \\ X_\ell
 \end{pmatrix} =& \int_{-\infty}^\infty dp_y H_\ell(\overline{p}_y)
 \begin{pmatrix}
  \phi \\ \chi
 \end{pmatrix},\label{eq:F_projy}
\end{align}
where $\Phi_\ell$ and $X_\ell$ are expansion coefficients [not to be confused 
with the fluxes in Eq.~\eqref{eq:diffuse_mflux_flux}].
Substituting the above expansions in Eq.~\eqref{eq:boltz_adim} gives:
\begin{equation}
 \left(\frac{\partial}{\partial t} + 
 \frac{p_x}{m} \frac{\partial}{\partial x} \right)
 \begin{pmatrix}
  \Phi_\ell \\ X_\ell
 \end{pmatrix} = -\frac{1}{\tau_*}
 \begin{pmatrix}
  \Phi_\ell - \Phi^*_\ell \\
  X_\ell - X^*_\ell
 \end{pmatrix}.\label{eq:boltz_projy}
\end{equation}
It can be seen that the moments $\Phi_\ell$ and $X_\ell$ of order
$\ell$ are coupled with those of order $\ell' \neq \ell$ only through 
the collision term. However, the equilibrium populations 
$\phi_*$ and $\chi_*$ are determined exclusively by the macroscopic
quantities corresponding to the collision invariants, $n$, $\bm{u}$ and $T$, 
as well as $T_{ij}$ (for the ES model) and $q_i$ (for the S model). 
These quantities can be written in terms of the coefficients $\Phi_{\ell'}$ 
with $0 \le \ell' \le 3$ and $X_{\ell'}$ with $0 \le \ell' \le 1$, as follows:
\begin{align}
 \begin{pmatrix}
  n \\ \rho u_x \\ \rho u_y
 \end{pmatrix} =& \int_{-\infty}^\infty dp_x\, 
 \begin{pmatrix}
  \Phi_0 \\ 
  \Phi_0 p_x \\ 
  \Phi_1 p_{0,y}
 \end{pmatrix},\nonumber\\
 \begin{pmatrix}
 T_{xx}  \\ T_{xy} \\ T_{yy}
 \end{pmatrix} =& \int_{-\infty}^\infty \frac{dp_x}{m}
 \begin{pmatrix}
  \xi_x^2 \Phi_0 \\
  \xi_x (\Phi_1 p_{0,y} - m u_y \Phi_0),\\
  \Phi_2 p_{0,y}^2 - 2m p_{0,y} u_y \Phi_1 
 + (p_{0,y}^2 + m^2 u_y^2) \Phi_0
 \end{pmatrix},\nonumber\\
 \begin{pmatrix}
  \frac{3}{2} n T \\ q_x 
 \end{pmatrix} =& \int_{-\infty}^\infty dp_x 
 \begin{pmatrix}
  1 \\ \xi_x / m
 \end{pmatrix}
 \left(\frac{\xi_x^2 + p_{0,y}^2 + m^2 u_y^2}{2m} \Phi_0 
 - p_{0,y} u_y \Phi_1 + \frac{p_{0,y}^2}{2m} \Phi_2 + 
 \frac{1}{2} X_0\right),\nonumber\\
 q_y =& \int_{-\infty}^\infty dp_x 
 \left[\frac{p_{0,y}^3}{2m^2} \Phi_3
 -\frac{3p_{0,y}^2 u_y}{2m} \Phi_2 
 + \frac{p_{0,y}}{2m^2}(\xi_x^2 + 3p_{0,y}^2 + 3m^2 u_y^2) \Phi_1 \right.\nonumber\\
 &\left.\hspace{20pt}- \frac{u_y}{2m} (\xi_x^2 + 3p_{0,y}^2 + m^2 u_y^2) \Phi_0 
 + \frac{p_{0,y}}{2m} X_1 - \frac{u_y}{2} X_0\right].
 \label{eq:moms_projy}
\end{align}
It can be seen that for a given value of $\ell$, Eq.~\eqref{eq:boltz_projy} 
involves only terms with $\ell'$ such that $0 \le \ell' \le {\rm max}(\ell,3)$ 
for $\Phi_\ell$ and $0 \le \ell' \le {\rm max}(\ell, 1)$ for $X_\ell$. Thus,
it can be concluded that the moment system with respect to the $p_y$ degree 
of freedom is closed when the terms up to $\ell = 3$ and $1$ in the series 
expansions of $\phi$ and $\chi$, respectively, are included. Moreover,
the dynamics (and therefore stationary state properties) of the 
moments in Eq.~\eqref{eq:moms_projy} is recovered exactly when 
the series for $\phi$ and $\chi$ in Eq.~\eqref{eq:F_projy} are 
truncated at $\ell = 3$ and $1$, respectively.
This truncation is equivalent to considering the quadrature orders 
$Q_y^\phi = 4$ and $Q_y^\chi = 2$, in the sense that employing higher 
order quadratures yields results which are exactly equivalent (up to 
numerical errors due to finite machine precision) to those obtained using 
$Q_y^\phi= 4$ and $Q_y^\chi = 2$. We discuss below the discretization 
corresponding to these quadrature orders.

The roots of the Hermite polynomial $H_4(z) = z^4 - 6z^2 + 3$ of order $4$ 
are known analytically \cite{shan06,ansumali03,bardow06,bardow08,biciusca15,sofonea18pre}:
\begin{align}
 \overline{p}^\phi_{y,1} =& -\sqrt{3 + \sqrt{6}}, &
 \overline{p}^\phi_{y,2} =& -\sqrt{3 - \sqrt{6}}, \nonumber\\
 \overline{p}^\phi_{y,3} =& \sqrt{3 - \sqrt{6}}, &
 \overline{p}^\phi_{y,4} =& \sqrt{3 + \sqrt{6}}, 
\end{align}
where $\overline{p}_{y,k_y}^\phi \equiv 
p_{y,k_y}^\phi / p_{0,y}^\phi$ is normalized with respect to an 
arbitrary scaling factor $p_{0,y}^\phi$, which we set to $1$ in 
this paper.
For the $\chi$ populations, the discrete momentum components
along the $y$ axis can be found via the roots of 
$H_2(z) = z^2 - 1$:
\begin{equation}
 \overline{p}^\chi_{y,1} = -1, \qquad 
 \overline{p}^\chi_{y,2} = 1,
\end{equation}
where $\overline{p}_{y,s_y}^\chi \equiv 
p_{y,s_y}^\chi / p_{0,y}^\chi$ and $p_{0,y}^\chi = 1$.

The connection between the discrete populations 
$\phi_{\bm{\kappa}}$ and $\chi_{\bm{\sigma}}$ and their 
continuous counterparts is given by the $2D$ extension of 
Eq.~\eqref{eq:Fkx_from_F}:
\begin{align}
 \phi_{\bm{\kappa}} =& \frac{p_{0,x} w_{k_x}^\hh(Q_x)}{\omega(\overline{p}_{x,k_x})} 
 \frac{p_{0,y}^\phi w_{k_y}^H(Q_y^\phi)}{\omega(\overline{p}^\phi_{y,k_y})}
 \phi(\overline{p}_{x,k_x}, \overline{p}^\phi_{y,k_y}),\nonumber\\ 
 \chi_{\bm{\sigma}} =& \frac{p_{0,x} w_{s_x}^\hh(Q_x)}{\omega(\overline{p}_{x,s_x})} 
 \frac{p_{0,y}^\chi w_{s_y}^H(Q_y^\chi)}{\omega(\overline{p}^\chi_{y,s_y})}
 \chi(\overline{p}_{x,s_x}, \overline{p}^\chi_{y,s_y}),
 \label{eq:phi_kappa}
\end{align}
where $\omega(z)$ is defined in Eq.~\eqref{eq:omega}.
The quadrature weights for the full-range Gauss-Hermite quadrature
can be computed via \cite{ambrus16jcp,hildebrand87,shizgal15}:
\begin{equation}
 w_{k}^H(Q_y^*) = \frac{Q_y^*!}{[H_{Q_y^*+1}(z_k)]^2},
\end{equation}
where $z_k$ ($1 \le k \le Q_y^*$) is the $k$'th root of $H_{Q_y^*}(z)$.
In particular, the weights for $Q_y^\phi = 4$ and $Q^y_\chi = 2$ are 
given by:
\begin{align}
 w_{y,1}^H(4) =& w_{y,4}^H(4) = \frac{3 - \sqrt{6}}{12}, \nonumber\\
 w_{y,2}^H(4) =& w_{y,3}^H(4) = \frac{3 + \sqrt{6}}{12}, \nonumber\\
 w_{y,1}^H(2) =& w_{y,2}^H(2) = \frac{1}{2}.
\end{align}

We now summarize the procedure described above. 
In the case of the heat transfer problem, the 
one-dimensional momentum space is discretized following 
the half-range Gauss-Hermite quadrature prescription using 
$Q_x^\phi = Q_x^\chi = Q_x$ quadrature points on each semiaxis 
for both $\phi$ and $\chi$.

For the shear flow problems, the $y$ axis of the 
momentum space is discretized separately for $\phi$ 
and $\chi$. The total number of quadrature points 
used to discretize the momentum space for $\phi$ 
is $2Q_x Q_y^\phi = 8Q_x$, 
while for $\chi$, $2Q_x Q_y^\chi = 4 Q_x$ 
quadrature points are required, resulting in a total number 
of $12 Q_x$ discrete populations.

\subsection{Projection of the collision term}\label{sec:FDLB:feq}

Part of the lattice Boltzmann paradigm is to replace 
the local equilibrium distribution by a polynomial expansion,
such that the collision invariants $\psi \in \{1, p_i, \bm{p}^2 / 2m\}$ 
are exactly preserved. This requires that, after the discretization
of the momentum space, 
the folllowing quadrature sums are exact:
\begin{align}
 \sum_{\bm{\kappa}} 
 \begin{pmatrix}
 1 \\ p_{\bm{\kappa}; i} 
 \end{pmatrix} \phi_{*;\bm{\kappa}} =& 
 \begin{pmatrix}
  n \\ \rho u_i
 \end{pmatrix}, &
 \sum_{\bm{\kappa}} \frac{\xi_{\bm{\kappa}; i} \xi_{\bm{\kappa}; i}}{2m}
 \phi_{*; \bm{\kappa}} + 
 \frac{1}{2} \sum_{\bm{\sigma}} \chi_{*;\bm{\sigma}} =& 
 \frac{3}{2} n T.
 \label{eq:moms_eq}
\end{align}
The above relations can be exactly ensured by 
first expanding $\phi_*$ and $\chi_*$ with respect to the 
Hermite polynomials (half-range on the $x$ 
and full-range on the $y$ axes, if required), followed by 
a truncation of the sums at orders 
$0 \le N_i^{\phi/\chi} < Q_i^{\phi/\chi}$. 
This approach is followed in \ref{app:hydro} in order to 
tackle flows in the hydrodynamic regime.

For the flows with $0.1 \le \delta \le 10$ considered in 
Sections~\ref{sec:ht}, \ref{sec:couette} and \ref{sec:htsh},
we follow a hybrid approach. Namely, the equilibrium 
distributions $\phi_*$ and $\chi_*$ are projected onto the 
set of full-range Hermite polynomials with respect to the axis parallel 
to the walls (no projection is required in the case of the heat 
transfer between stationary plates problem). Then, the 
expansion coefficients are evaluated directly, following 
the standard DVM approach. This hybrid approach is motivated 
as follows.

On the $x$ axis, the quadrature order $Q_x$ is 
considered to be equal for both $\phi$ and $\chi$. 
Since we are interested in performing simulations in 
the slip flow and transition regime, we need in general 
high values of $Q_x$ (i.e., $Q_x \ge 7$ will be required \cite{ambrus18pre}). 
Let us now assume the equilibrium distributions are 
expanded with respect to the half-range Hermite polynomials
up to order $N_x = Q_x - 1$. It is expected that the coefficients of the 
expansion grow with $N_x$ as $\sim N_x! {\rm Ma}^{N_x}$. Since the simulations 
that we are considering are performed in the non-linear regime,
where ${\rm Ma} > 1$, high expansion orders may be required
(we use $Q_x  = 50$ at ${\delta = 0.1}$), such that the individual 
terms in the series expansion 
can be large. The addition and subtraction of these terms typically
leads to a significantly smaller remainder, which can easily be 
poluted by numerical errors due to finite numerical precision.
It is a well-known limitation of the (FD)LB algorithm that 
the polynomial expansion of the equilibrium distribution 
is not well suited for high-Mach number flows. On the other hand,
directly evaluating the equilibrium distributions discussed 
in Sec.~\ref{sec:boltz:RTA} when computing the equilibrium 
moments in Eq.~\eqref{eq:moms_eq} at $Q_x \ge 7$ is already 
quite accurate for $\delta \lesssim 10$ when the half-range 
Gauss-Hermite quadrature 
is employed (in this case, $2Q_x \ge 14$ quadrature points 
are employed on the $p_x$ axis). Thus, we find the loss in precision 
due to the integration using Gauss quadratures of non-polynomial
functions via Eq.~\eqref{eq:moms_eq} to be irrelevant. 

At larger values of $\delta$, the physical time to 
reach the steady state increases. Over a longer time interval, the 
errors in the recovery of the conservation laws due to the 
inaccurate integration of the equilibrium distribution accumulate,
affecting the accuracy of the properties of the stationary state. 
This problem can be alleviated by projecting the equilibrium 
distribution onto the space of half-range Hermite polynomials 
also on the $p_x$ direction, as discussed in \ref{app:hydro}.

We further discuss in detail the implementation of the S and 
ES collision terms for the $d = 1$ case encountered in the heat transfer 
between stationary plates problem (Subsec.~\ref{sec:FDLB:feq:1D}).
In the $d = 2$ case, encountered for the Couette flow and heat 
transfer between moving plates problem, the implementation 
of the ES and S models is discussed separately 
in Subsecs.~\ref{sec:FDLB:feq:2DES} and \ref{sec:FDLB:feq:2DS},
respectively.

\subsubsection{$d = 1$ case} \label{sec:FDLB:feq:1D}

In the case of the ES model, the equilibrium distribution
functions 
can be found from Eq.~\eqref{eq:phiES}.
When  $d = 1$, the equilibrium distribution
function  is
\begin{equation}
 \phi_{\rm ES} = 
 \frac{n}{\sqrt{2 \pi m T \mathcal{B}_{xx}}} 
 \exp\left[-\frac{(p_x - mu_x)^2}
 {2m T \mathcal{B}_{xx}}\right],
 \label{eq:phiES1D}
\end{equation}
while $\chi_{\rm ES} = 2 T_{\rm red} \phi_{\rm ES}$,
where $T_{\rm red} = P_{\rm red} / n$. In the above,
$\mathcal{B}_{xx}$ and $P_{\rm red}$ are given by:
\begin{equation}
 \mathcal{B}_{xx} = \frac{1}{\rm Pr} - 
 \frac{1 - {\rm Pr}}{\rm Pr} \frac{T_{xx}}{P},\qquad 
 P_{\rm red} = \frac{3}{2} P - \frac{1}{2} T_{xx}.
 \label{eq:phiES1D_Bxx}
\end{equation}

The transition to the discrete system is made via 
Eq.~\eqref{eq:Fkx_from_F}:
\begin{align}
  \phi_{{\rm ES}; k_x} =& \frac{p_{0,x} w_{k_x}^\hh(Q_x)}{\omega(\overline{p}_{x;k_x})}
  \phi_{\rm ES}(p_{x;k_x}),&
  \chi_{{\rm ES}; s_x} =& \frac{p_{0,x} w_{s_x}^\hh(Q_x)}{\omega(\overline{p}_{x;s_x})}
  \chi_{\rm ES}(p_{x;s_x}),
 \label{eq:phiES1Dkx}
\end{align}
where $1 \le k_x, s_x \le 2Q_x$ and 
$\omega(z)$ is defined in Eq.~\eqref{eq:omega}.

For the S model, the equilibrium distributions 
$\phi_{\rm S}$ and $\chi_{\rm S}$ can be obtained 
from Eq.~\eqref{eq:phiS}:
\begin{align}
 \phi_{\rm S} =& n g_x (1 + \mathbb{S}_\phi), &
 \mathbb{S}_\phi =& \frac{1-{\rm Pr}}
 {5 nT^2}
 \left(\frac{\xi_x^2}
 {mT}-3 \right)
 q_x \xi_x,\nonumber\\
 \chi_{\rm S} =& 2 P g_x (1 + \mathbb{S}_\chi),&
 \mathbb{S}_\chi =& \frac{1-{\rm Pr}}
 {5nT^2}
 \left(\frac{\xi_x^2}
 {mT}-1 \right)
 q_x \xi_x,
 \label{eq:phiS1D}
\end{align}
where $P = n T$ and $g_x \equiv g(p_x, u_x, T)$ is given through Eq.~\eqref{eq:feq}, 
while $\xi_x = p_x - mu_x$. As in Eq.~\eqref{eq:phiES1Dkx}, the equilibrum distributions
after discretization are computed using:
\begin{align}
  \phi_{{\rm S}; k_x} =& \frac{p_{0,x} w_{k_x}^\hh(Q_x)}{\omega(\overline{p}_{x;k_x})}
  \phi_{\rm S}(p_{x;k_x}),&
  \chi_{{\rm S}; s_x} =& \frac{p_{0,x} w_{s_x}^\hh(Q_x)}{\omega(\overline{p}_{x;s_x})}
  \chi_{\rm S}(p_{x;s_x}).
 \label{eq:phiS1Dkx}
\end{align}

\subsubsection{$d = 2$ case: ES model}\label{sec:FDLB:feq:2DES}

In the $d = 2$ case, the exponent $\mathcal{B}_{ij}^{-1}\xi_i \xi_j$ in 
Eq.~\eqref{eq:phiES} can be written as:
\begin{equation}
 \mathcal{B}_{ij}^{-1} \xi_i \xi_j = \mathcal{B}_{yy}^{-1} 
 \left(\xi_y + \frac{\mathcal{B}_{xy}^{-1}}{\mathcal{B}_{yy}^{-1}} \xi_x\right)^2  
 + \frac{\xi_x^2}{\mathcal{B}_{yy}^{-1}} \left(
 \mathcal{B}_{xx}^{-1} \mathcal{B}_{yy}^{-1} - (\mathcal{B}_{xy}^{-1})^2\right).
\end{equation}
Noting that the inverse of $\mathcal{B}_{ij}$ is given by:
\begin{equation}
 \mathcal{B}^{-1}_{ij} = \frac{1}{{\rm det}\mathcal{B}}
 \begin{pmatrix}
  \mathcal{B}_{yy} & -\mathcal{B}_{xy} \\
  -\mathcal{B}_{xy} & \mathcal{B}_{xx} 
 \end{pmatrix},
\end{equation}
$\phi_{\rm ES}$ can be factorized as follows:
\begin{equation}
 \phi_{\rm ES} = n g(p_x, u_x, T \mathcal{B}_{xx}) 
 g\left(p_y, u_y + \frac{\xi_x \mathcal{B}_{xy}}{m \mathcal{B}_{xx}},
 \frac{T\, {\rm det} \mathcal{B}}{\mathcal{B}_{xx}}\right).
 \label{eq:phiES2D}
\end{equation}
A similar factorization holds for $\chi_{\rm ES} = T_{\rm red} \phi_{\rm ES}$, 
where $T_{\rm red} = P_{\rm red} / n$ and
\begin{equation}
 P_{\rm red} = 3 P - T_{xx} - T_{yy}.
 \label{eq:ES2DPred}
\end{equation}

We now seek to replace $\phi_{\rm ES}$ and $\chi_{\rm ES}$ with the expansions 
$\phi_{\rm ES}^{(N^\phi_y)}$ and $\chi_{\rm ES}^{(N^\chi_y)}$ 
with respect to the Hermite polynomials $H_{\ell}(\overline{p}_y)$ containing only 
terms up to orders $N_y^\phi = Q_y^\phi - 1 = 3$ and 
$N_y^\chi = Q_y^\chi - 1 = 1$, respectively. Defining:
\begin{equation}
 \zeta_y = u_y + \frac{\xi_x \mathcal{B}_{xy}}{m \mathcal{B}_{xx}}, \qquad 
 T_y = \frac{{\rm det} \mathcal{B}}{\mathcal{B}_{xx}} T,
 \label{eq:phiES2D_zetaT}
\end{equation}
Eq.~\eqref{eq:phiES2D} reduces to 
$\phi_{\rm ES} = n g(p_x, u_x, T \mathcal{B}_{xx}) g(p_y, \zeta_y, T_y)$.
The trailing function $g(p_y, \zeta_y, T_y)$ is expanded with respect to 
$H_\ell(\overline{p}_y)$ up to order $N_y^* \in \{N_y^\phi, N_y^\chi\}$,
as follows:
\begin{equation}
 g^{(N_y^*)}(p_y, \zeta_y, T_y) = 
 \frac{\omega(\overline{p}_y)}{p_{0,y}} 
 \sum_{\ell = 0}^{N_y^*} \frac{1}{\ell!} H_\ell(\overline{p}_y) 
 \mathcal{G}^H_\ell(\zeta_y, T_y).\label{eq:gN}
\end{equation}
The expansion coefficients $\mathcal{G}^H_\ell(\zeta_y,T_y)$ were obtained analytically 
in Eq.~(C.13) in Ref.~\cite{ambrus16jcp}. Below, we reproduce the coefficients 
for $0 \le \ell \le 3$:
\begin{equation}
 \mathcal{G}_0^H = 1, \qquad 
 \mathcal{G}_1^H = \mathfrak{U}, \qquad 
 \mathcal{G}_2^H = \mathfrak{U}^2 + \mathfrak{I}, \qquad
 \mathcal{G}_3^H = \mathfrak{U}^3 + 3\mathfrak{U} \mathfrak{I}.
 \label{eq:Gell}
\end{equation}
Identifying 
$\mathfrak{U}$ and $\mathfrak{I}$ from Eq.~(C.16) of Ref.~\cite{ambrus16jcp} 
with the following expressions,
\begin{equation}
 \mathfrak{U}(\zeta_y) = \frac{m \zeta_y}{p_{0,y}}, \qquad 
 \mathfrak{I}(T_y) = \frac{m T_y}{p_{0,y}^2} - 1,\label{eq:Gell_UT}
\end{equation}
$g^{(1)}(p_y, \zeta_y, T_y)$ necessary for the construction of $\chi_{\rm ES}$
can be written as:
\begin{equation}
 g^{(1)}(p_y, \zeta_y, T_y) = \frac{\omega(\overline{p}_y)}{p_{0,y}}
 \left[H_0(\overline{p}_y) + H_1(\overline{p}_y) \mathfrak{U}(\zeta_y)\right]. 
\end{equation}
The function $g^{(3)}(p_y, \zeta_y, T_y)$ required for $\phi_{\rm ES}$, 
is given by:
\begin{multline}
 g^{(3)}(p_y, \zeta_y, T_y) = \frac{\omega(\overline{p}_y)}{p_{0,y}}
 \left\{H_0(\overline{p}_y) + H_1(\overline{p}_y) \mathfrak{U}(\zeta_y)
 + \frac{1}{2!} H_2(\overline{p}_y) \left[\mathfrak{U}^2(\zeta_y) 
 + \mathfrak{I}(T_y)\right]\right.\\
 \left.
 + \frac{1}{3!} H_3(\overline{p}_y) \left[\mathfrak{U}^3(\zeta_y) + 
 3 \mathfrak{U}(\zeta_y) \mathfrak{I}(T_y)\right]\right\}.
\end{multline}

With the above ingredients, after discretization, 
$\phi^{\rm ES}_{\bm{\kappa}}$ can be evaluated using:
\begin{equation}
 \phi^{\rm ES}_{\bm{\kappa}} = n \frac{p_{0,x} w_{k_x}^\hh(Q_x)}{\omega(\overline{p}_{x,k_x})}
 \frac{p_{0,y}^\phi w_{k_y}^H(Q_y^\phi)}{\omega(\overline{p}^\phi_{y,k_y})}
 g(p_{x,k_x}, u_x, T \mathcal{B}_{xx})
 g^{(3)}(p^\phi_{y,k_y}, \zeta_{y; k_x}, T_y),
 \label{eq:phiESd2k}
\end{equation}
where $\zeta_{y; k_x} = u_y + \frac{\mathcal{B}_{xy}}{m \mathcal{B}_{xx}} \xi_{x,k_x}$. 
Similarly, $\chi^{\rm ES}_{\bm{\sigma}}$ is:
\begin{equation}
 \chi^{\rm ES}_{\bm{\sigma}} = P_{\rm red}
 \frac{p_{0,x} w_{s_x}^\hh(Q_x)}{\omega(\overline{p}_{x,s_x})}
 \frac{p_{0,y}^\chi w_{s_y}^H(Q_y^\chi)}{\omega(\overline{p}^\chi_{y,s_y})}
 g(p_{x,s_x}, u_x, T \mathcal{B}_{xx}) 
 g^{(1)}(p^\chi_{y,s_y}, \zeta_{y; s_x}, T_y).
 \label{eq:chiESd2s}
\end{equation}

In Eqs.~\eqref{eq:phiESd2k} and \eqref{eq:chiESd2s},
the function $g(p_x, u_x, T \mathcal{B}_{xx})$ is evaluated directly.
Its expression is reproduced below for convenience:
\begin{equation}
 g(p_{x,k_x}, u_x, T \mathcal{B}_{xx}) = 
 \frac{1}
 {\displaystyle \sqrt{2\pi m T \mathcal{B}_{xx}}}
 \exp\left[-\frac{(p_{x,k_x} - mu_x)^2}{2m T \mathcal{B}_{xx}}\right].
\end{equation}

\subsubsection{$d = 2$ case: S model} \label{sec:FDLB:feq:2DS}

In the case of the Shakhov model, $\phi_{\rm S}$ and $\chi_{\rm S}$
can be written as:
\begin{align}
 \phi_{\rm S} =& n g_x g_y (1 + \mathbb{S}_\phi), &
 \mathbb{S}_\phi =& \frac{1-{\rm Pr}}{5 nT^2}
 \left(\frac{\xi_x^2 + \xi_y^2}{mT}-4 \right)
 (q_x \xi_x + q_y \xi_y),\nonumber\\
 \chi_{\rm S} =& n T g_x g_y(1 + \mathbb{S}_\chi), &
 \mathbb{S}_\chi =& \frac{1-{\rm Pr}}{5nT^2} 
 \left(\frac{\xi_x^2 + \xi_y^2} {mT}-2 \right)
 (q_x \xi_x + q_y \xi_y),
 \label{eq:phiS2D}
\end{align}
where $g_x \equiv g(p_x, u_x, T)$ and $g_y \equiv g(p_y, u_y, T)$ are 
one-dimensional Maxwell-Boltzmann distributions introduced in 
Eq.~\eqref{eq:feq}. The functions $\phi_{\rm S}$ and $\chi_{\rm S}$ 
can be expanded with respect to 
the full-range Hermite polynomials $H_\ell(\overline{p}_y)$, as follows:
\begin{equation}
 \begin{pmatrix}
  \phi_{\rm S} \\ \chi_{\rm S} 
 \end{pmatrix} = \frac{\omega(\overline{p}_y)}{p_{0,y}} 
 \sum_{\ell = 0}^\infty \frac{1}{\ell!} H_{\ell}(\overline{p}_y) 
 \begin{pmatrix}
  \mathcal{G}_{{\rm S};\ell}^{\phi;H} \\ \mathcal{G}_{{\rm S};\ell}^{\chi;H}
 \end{pmatrix}.
\end{equation}
The expansion coefficients $\mathcal{G}_{{\rm S};\ell}^{\phi/\chi;H}$ 
can be written as:
\begin{equation}
\begin{pmatrix}
  \mathcal{G}_{{\rm S};\ell}^{\phi;H} \\ \mathcal{G}_{{\rm S};\ell}^{\chi;H}
 \end{pmatrix} = g_x
 \begin{pmatrix}
  n \\ P
 \end{pmatrix}
 \left[\mathcal{G}^H_\ell(u_y, T) + \frac{1 - {\rm Pr}}{5 n T^2} 
 \begin{pmatrix}
  \mathfrak{G}^{\phi;H}_{{\rm S};\ell} \\
  \mathfrak{G}^{\chi;H}_{{\rm S};\ell}
 \end{pmatrix}\right].
 \label{eq:S2D_Gell}
\end{equation}
The coefficients $\mathcal{G}^H_\ell$ have the same form as in Eq.~\eqref{eq:Gell}, 
where the factors $\mathfrak{U} \equiv \mathfrak{U}(u_y)$ and 
$\mathfrak{I} \equiv \mathfrak{I}(T)$ are given by Eq.~(C.16) in 
Ref.~\cite{ambrus16jcp}:
\begin{equation}
 \mathfrak{U}(u_y) = \frac{m u_y}{p_{0,y}}, \qquad 
 \mathfrak{I}(T) = \frac{m T}{p_{0,y}^2} - 1.
 \label{eq:S2D_Gell_UT}
\end{equation}

Denoting:
\begin{equation}
 \begin{pmatrix}
  \mathcal{I}_s^\phi \\ 
  \mathcal{I}_s^\chi
 \end{pmatrix} = \int_{-\infty}^\infty dp_y\, g(p_y, u_y, T) 
 (q_x \xi_x + q_y \xi_y) 
 \left[\frac{\xi_x^2 + \xi_y^2}{m T} -
 \begin{pmatrix}
  4 \\ 2
 \end{pmatrix}\right] \xi_y^s,
 \label{eq:S2D_Is}
\end{equation}
the coefficients $\mathfrak{G}_{{\rm S};\ell}^{*;H}$ ($* \in \{\phi, \chi\}$) 
in Eq.~\eqref{eq:S2D_Gell} can be obtained as:
\begin{align}
 \mathfrak{G}_{{\rm S};0}^{*;H} =& \mathcal{I}_0^*, \qquad 
 \mathfrak{G}_{{\rm S};1}^{*;H} = \frac{1}{p_{0,y}} 
 (\mathcal{I}_1^* + mu_y \mathcal{I}_0^*), \nonumber\\
 \mathfrak{G}_{{\rm S};2}^{*;H} =& \frac{1}{p_{0,y}^2}[\mathcal{I}_2^* + 
 2mu_y \mathcal{I}_1^* + (m^2 u_y^2 - p_{0,y}^2) \mathcal{I}_0^*], \nonumber\\
 \mathfrak{G}_{{\rm S};3}^{*;H} =& \frac{1}{p_{0,y}^3} [\mathcal{I}_3^* + 
 3m u_y \mathcal{I}_2^* + 3(m^2 u_y^2 - p_{0,y}^2) \mathcal{I}_1^*\nonumber\\
 & + mu_y (m^2 u_y^2 - 3 p_{0,y}^2) \mathcal{I}_0^*].
 \label{eq:S2D_Gellfrak}
\end{align}
Finally, the terms $\mathcal{I}_s^*$ can be obtained by direct integration 
in Eq.~\eqref{eq:S2D_Is}:
\begin{align}
 \begin{pmatrix}
  I_0^\phi \\ I_0^\chi
 \end{pmatrix} =& \, q_x \xi_x \left[\frac{\xi_x^2}{mT} - 
 \begin{pmatrix}
  3 \\ 1
 \end{pmatrix}\right], &
 \begin{pmatrix}
  I_1^\phi \\ I_1^\chi
 \end{pmatrix} =& \, q_y m T \left[\frac{\xi_x^2}{mT} + 
 \begin{pmatrix}
  -1 \\ 1
 \end{pmatrix}\right],\nonumber\\
 \begin{pmatrix}
  I_2^\phi \\ I_2^\chi
 \end{pmatrix} =& \, q_x \xi_x m T \left[\frac{\xi_x^2}{mT} + 
 \begin{pmatrix}
  -1 \\ 1
 \end{pmatrix}\right],&
 \begin{pmatrix}
  I_3^\phi \\ I_3^\chi
 \end{pmatrix} =& \, 3 q_y (m T)^2 \left[\frac{\xi_x^2}{mT} + 
 \begin{pmatrix}
  1 \\ 3
 \end{pmatrix}\right].\label{eq:S2D_Is_explicit}
\end{align}

Putting the pieces together, the discrete populations $\phi_{{\rm S};\bm{\kappa}}$ and 
$\chi_{{\rm S};\bm{\sigma}}$ can be computed using:
\begin{align}
 \phi_{{\rm S};\bm{\kappa}} =& n\,\frac{p_{0,x} w_{k_x}^\hh(Q_x)}
 {\omega(\overline{p}_{x, k_x})} 
 g(p_{x,k_x}, u_x, T) w_{k_y}^H(4)
 \sum_{\ell = 0}^3 \frac{1}{\ell!} H_\ell(\overline{p}^\phi_{y,k_y}) 
 \left(\mathcal{G}^H_\ell + \frac{1 - {\rm Pr}}{5 nT^2} 
 \mathfrak{G}^{\phi;H}_{{\rm S};\ell}\right), \nonumber\\
 \chi_{{\rm S};\bm{\sigma}} =& n T \, \frac{p_{0,x} w_{s_x}^\hh(Q_x)}
 {\omega(\overline{p}_{x, s_x})} 
 g(p_{x,s_x}, u_x, T) w_{s_y}^H(2)
 \sum_{\ell = 0}^1 \frac{1}{\ell!} H_\ell(\overline{p}^\chi_{y,s_y}) 
 \left(\mathcal{G}^H_\ell + \frac{1 - {\rm Pr}}{5 nT^2} 
 \mathfrak{G}^{\chi;H}_{{\rm S};\ell}\right).
\end{align}

\section{Simulation methodology}\label{sec:meth}

This section briefly summarizes the methodology employed 
for obtaining the numerical results discussed in the
next sections. Three applications are considered 
in this paper, namely the heat transfer between stationary 
plates (Sec.~\ref{sec:ht}), the Couette flow between 
plates at the same temperature (Sec.~\ref{sec:couette}),
and the heat transfer between moving plates (Sec.~\ref{sec:htsh}). 

In all cases, the simulation results are presented for 
three values of the rarefaction parameter, namely 
$\delta = 10$, $1$ and $0.1$.
For all applications, we take the working gas to be 
comprised of ${}^3{\rm He}$ or ${}^4{\rm He}$ molecules.
Additionally, in the case of the heat transfer between moving 
plates problem, we also report results for ${\rm Ne}$.
The reference temperature $\widetilde{T}_{\rm ref}$, defined 
in Eq.~\eqref{eq:Tref}, varies between 
$1\ {\rm K}$ and $3000\ {\rm K}$ for the ${}^3{\rm He}$ 
and ${}^4{\rm He}$ constituents and between 
$20\ {\rm K}$ and $5000\ {\rm K}$ for the ${\rm Ne}$ 
constituents.

Quantitative comparisons are performed by considering a set
of dimensionless numbers. In the context of the flows between 
moving walls (discussed in Sections~\ref{sec:couette} and 
\ref{sec:htsh}), the shear stress is used to define the 
quantity \cite{sharipov18}
\begin{equation}
 \Pi = -\frac{\widetilde{T}_{xy} \widetilde{c}_{\rm ref}}
 {\widetilde{P}_{\rm ref} \widetilde{u}_w \sqrt{2}}.
 \label{eq:Pi}
\end{equation}
It can be shown that, in the stationary state, $\Pi$ is constant 
throughout the channel.
In order to access the non-linear regime, we set
the wall velocities to $\widetilde{u}_w = \widetilde{c}_{\rm ref} \sqrt{2} = 
\sqrt{2 \widetilde{K}_B \widetilde{T}_{\rm ref} / \widetilde{m}}$, such that 
the Mach number is
\begin{equation}
 {\rm Ma} = \frac{2 \widetilde{u}_w}{\widetilde{c}_s} \simeq 2.19,
 \label{eq:Ma}
\end{equation}
where $\widetilde{c}_s = \sqrt{\gamma \widetilde{K}_B \widetilde{T}_{\rm ref} 
/ \widetilde{m}}$ is the speed of sound and $\gamma = 5/3$ is the adiabatic 
index for a monatomic ideal gas.
After non-dimensionalization, $\Pi$ is computed through
\begin{equation}
 \Pi = -\frac{1}{2} T_{xy}.\label{eq:Pi_adim}
\end{equation}

In the heat transfer problems, discussed in Sections~\ref{sec:ht} 
and \ref{sec:htsh},
the longitudinal heat flux $\widetilde{q}_x$ (perpendicular to the 
$x$ axis) is used to introduce
\begin{equation}
 Q = -\frac{(\widetilde{q}_{x} + \widetilde{T}_{xy} \widetilde{u}_y) 
 \widetilde{T}_{\rm ref}}{\widetilde{P}_{\rm ref} 
 \widetilde{c}_{\rm ref} \widetilde{\Delta T} \sqrt{2}},
 \label{eq:Q}
\end{equation}
which is again constant throughout the channel.
The second term in the numerator vanishes when the 
walls are stationary (i.e., in Sec.~\ref{sec:ht}).
We consider the nonlinear regime, in which the ratio
between the temperature difference $\widetilde{\Delta T} = 
\widetilde{T}_{\rm right} - \widetilde{T}_{\rm left}$ and 
$\widetilde{T}_{\rm ref}$, defined in Eq.~\eqref{eq:Tref}, is
\begin{equation}
 \frac{\widetilde{\Delta T}}{\widetilde{T}_{\rm ref}} = 
 2\frac{\widetilde{T}_{\rm right} - \widetilde{T}_{\rm left}}
 {\widetilde{T}_{\rm right} + \widetilde{T}_{\rm left}}
 = 1.5.\label{eq:DT}
\end{equation}
After non-dimensionalization,  the wall temperatures are 
$T_{\rm left} = 0.25$ and $T_{\rm right} = 1.75$, while $Q$ 
is obtained via:
\begin{equation}
 Q = \frac{2\sqrt{2}}{3} \Pi u_y - \frac{\sqrt{2}}{3} q_x.
 \label{eq:Q_adim}
\end{equation}

In the context of the Couette flow, we further consider 
two more quantities. The first is the dimensionless half-channel 
heat flow rate, defined through
\begin{equation}
 Q_y = \frac{2}{\widetilde{L}} \int_0^{\widetilde{L}/2} d\widetilde{x}\, 
 \frac{\widetilde{q}_y}{\widetilde{P}_{\rm ref} \widetilde{u}_w}.
 \label{eq:Qy}
\end{equation}
The second is related to the heat transfer through the domain wall,
and is defined through:
\begin{equation}
 Q_w = \frac{\widetilde{q}_x(\widetilde{L}/2) \widetilde{c}_{\rm ref}}
 {\widetilde{P}_{\rm ref} \widetilde{u}_w^2 \sqrt{2}}
 = \frac{\widetilde{u}_y(1/2)}{\widetilde{u}_w} \Pi,
 \label{eq:Qw}
\end{equation}
where the second equality follows after noting that 
$\widetilde{q}_x + \widetilde{T}_{xy} \widetilde{u}_y = 0$ in the 
stationary state of the Couette flow.

In practice, the quantities $\Pi$ and $Q$ exhibit a mild coordinate 
dependence in the stationary state due to the errors of the numerical 
scheme. The values reported in the applications sections are obtained 
by averaging $\Pi$ and $Q$ over the simulation domain, as follows:
\begin{equation}
 \begin{pmatrix}
  \Pi \\ Q
  \end{pmatrix} = \frac{1}{\widetilde{L}} 
  \int_{-{\widetilde{L}}/2}^{\widetilde{L} / 2} d\widetilde{x} 
 \begin{pmatrix}
  \Pi(\widetilde{x}) \\ Q(\widetilde{x})
 \end{pmatrix}.
\end{equation}
In the case of the Couette flow, $\Pi(-\widetilde{x}) = \Pi(\widetilde{x})$ 
is used to reduce the integration domain to 
$0 \le \widetilde{x} \le \widetilde{L} /2$.

The FDLB methodology is discussed in 
Subsec.~\ref{sec:meth:FDLB} and the DSMC methodology is 
summarized in Subsec.~\ref{sec:meth:DSMC}.

\subsection{FDLB methodology}\label{sec:meth:FDLB}

For the heat transfer problems, the FDLB simulations 
are performed on grids comprised of $2S$ cells. For the Couette
flow simulations, we take advantage of the symmetry and use
only $S$ cells on the half-channel.
Each cell has the width $\delta \eta = {\rm arctanh A} / S$ 
with respect to the $\eta$ coordinate and the stretching 
parameter entering Eq.~\eqref{eq:stretch} is set to $A = 0.98$. 

\begin{table}
\begin{center}
\begin{tabular}{l|ccccc}
 $\delta$ & $1000$ & $100$ & $10$ & $1$ & $0.1$ \\\hline
 Nodes per half-channel $S$ & $32$ & $32$ & $32$ & $32$ & $16$ \\
 Time step $\delta t$ & $10^{-4}$ & $2.5 \times 10^{-4}$ & 
 $5 \times 10^{-4}$ & $5 \times 10^{-4}$ & $5 \times 10^{-4}$\\
 $Q_x$ & $11$ & $8$ & $7/8$ & $11$ & $50$ \\
 Discrete populations ($d = 1$) & $44$ & $32$ & $28/32$ & $44$ & $200$ \\ 
 Discrete populations ($d = 2$) & $132$ & $96$ & $84/96$ & $132$ & $600$
\end{tabular}
\end{center}
\caption{Discretisation details for the FDLB method for 
various values of the rarefaction parameter $\delta$.
For $\delta = 10$, both $Q_x = 7$ (for the heat transfer 
between stationary plates and the Couette flow) and 
$Q_x = 8$ (for the heat transfer between moving plates)
are employed.
\label{tab:discretisation}}
\end{table}

At $\delta = 10$, the quadrature order on the $x$ axis 
is set to $Q_x = 7$ for the Couette flow and heat transfer 
between stationary plates problems, while for the heat transfer 
between moving plates, $Q_x = 8$ is used. For $\delta = 1$ and 
$0.1$, the quadrature order is increased to $Q_x = 11$ and $50$, 
respectively, in order to capture the rarefaction effects.
The quadrature orders and total number of discrete 
populations are shown in Table~\ref{tab:discretisation}.
For completeness, Table~\ref{tab:discretisation} 
also includes information for the $\delta = 1000$ and $100$ 
cases. For these larger values of $\delta$, lower quadrature 
orders can be employed if the equilibrium distributions 
are projected onto the space of half-range Hermite polynomials, as discussed 
in \ref{app:hydro}.


The simulation is performed until the stationary state is 
achieved. The time steps $\delta t = 10^{-4}$, $2.5 \times 10^{-4}$ 
and $5 \times 10^{-4}$ were employed for 
$\delta = 1000$, $\delta = 100$ and $\delta \le 10$, respectively. 
The number of points on the half-channel is set to 
$S = 32$ and $16$ for $\delta \ge 1$ and $\delta = 0.1$, 
respectively. The number of iterations performed to 
reach the stationary state is 
$5 \times 10^6$, $2 \times 10^5$, $6 \times 10^4$, 
$4 \times 10^4$ and $2 \times 10^5$
for $\delta = 1000$, $100$, $10$, $1$ and $0.1$, respectively.

In order to assess the accuracy of the simulation results, 
another set of simulations is performed using 
$Q_x = 16$ for $\delta = 1000$ and $100$,  $Q_x = 40$ 
for $\delta = 10$ and $1$, while for $\delta = 0.1$, $Q_x = 200$
is employed. The spatial grid is refined by a factor of $4$, 
such that $S = 128$ is used for $\delta \ge 1$, while 
for $\delta = 0.1$, $S = 64$ points are used on the half-channel.
The time step in this case is set to $\delta t = 5 \times 10^{-5}$ 
for $\delta \ge 1$, and $\delta t = 4 \times 10^{-5}$ for 
$\delta = 0.1$. We compared 
the results obtained for $Q$, $\Pi$, $Q_w$ and $Q_y$ and 
found that the relative differences between the results 
obtained within the two sets of simulations were below $0.1\%$ 
for $\delta \le 10$.
At $\delta = 100$ and $1000$, the relative error has a 
larger magnitude, since it is computed at the level of quantities that 
tend to zero in the large $\delta$ limit. It can be seen in 
Figs.~\ref{fig:hydro:conv100} and \ref{fig:hydro:conv1000} that the 
absolute error is confortably under $0.1\%$ at $S = 32$. 

In order to compute the integrals over the discretized domain, 
a fourth order rectangle method is used, summarized below:
\begin{align}
 \frac{1}{\widetilde{L}} \int_{-\widetilde{L}/2}^{\widetilde{L}/2} d\widetilde{x} \,
 M(\widetilde{x}) =& \frac{1}{A} \int_{-{\rm arctanh}\,A}^{{\rm arctanh}\,A} 
 \frac{d\eta}{\cosh^2 \eta} M(\eta) \nonumber\\
 =& \frac{{\rm arctanh}\, A}{A S} \sum_{s = -S + 1}^S \mathfrak{f}_s 
 \frac{M_s}{\cosh^2 \eta_s},
 \label{eq:rect4}
\end{align}
where $M_s \equiv M(\eta_s)$ and 
\begin{equation}
 \mathfrak{f}_s = 
 \begin{cases}
  {\displaystyle \frac{13}{12}}, & s = 4i \text{ or } 4i + 1,\smallskip\\
  {\displaystyle \frac{11}{12}}, & s = 4i + 2 \text{ or } 4i + 3.
 \end{cases}
 \label{eq:rect4coef}
\end{equation}

A comparison with various approaches described in the literature confirms 
that our method is efficient. For example, in Ref.~\cite{naris04}, the heat
transfer between stationary plates problem is simulated in the linear regime 
using a full-range Gauss-Hermite discretisation of order $128$ (only one 
distribution is required in this case), which is above the number of 
velocities that we employ at $\delta = 0.1$, when $2Q_x = 100$ (we note that 
this discretisation allows us to access the non-linear regime).
Another example can be seen in Ref.~\cite{graur09}, where the $2D$ 
velocity space comprised of $p_x$ and $p_y$ is discretised using 
polar coordinates $(p,\theta)$. In the transition regime, 
a number of $16 \times 101 = 1616$ velocities are employed \cite{graur09},
compared to only $100$ with our approach (when $d = 2$, we employ $400$ and $200$ 
velocities for the $\phi$ and $\chi$ distributions, respectively). 

\subsection{DSMC methodology}\label{sec:meth:DSMC}

The DSMC calculations were carried out dividing the space 
$-\widetilde{L}/2 \le \widetilde{x} \le \widetilde{L}/2$ 
into $800$ cells, considering $200$ particles per cell in 
average, and using the time step $\delta \widetilde{t}$ equal 
to $0.002 \widetilde{L}/\sqrt{2}\widetilde{c}_{\rm ref}$,
where $\widetilde{c}_{\rm ref} = 
\sqrt{\widetilde{K}_B \widetilde{T}_{\rm ref} / \widetilde{m}}$
is defined in Eq.~\eqref{eq:cref}.
The shear stress $\Pi$ and heat flux $Q$, defined in Eqs.~\eqref{eq:Pi}
and \eqref{eq:Q}, were calculated by 
counting the momentum and energy brought and taken away by all 
particles on both surfaces.  
To reduce the statistical scattering, the macroscopic quantities 
were calculated by averaging over $5\times10^5$ samples. These 
parameters of the numerical scheme provide the total numerical error 
of $Q$ and $\Pi$ less than $0.1\%$, estimated by carrying out test 
calculations with the double number of cells, the double number of 
particles and reducing the time step by a factor of 2. The relative 
divergence of $\Pi$ and $Q$, calculated on the difference 
surface using an additional accuracy criterion, does not
exceed $0.01\%$. The details of the numerical scheme and the
method used to calculate the look-up tables can be found in
Ref. \cite{sharipov18}.

\subsection{Computational time analysis}

\begin{table}
\begin{center}
\begin{tabular}{c|cc|cc|cc}
& \multicolumn{2}{|c|}{HT} & \multicolumn{2}{|c|}{SH} & \multicolumn{2}{|c}{HT-SH} \\
 \hline\hline
 $\delta$ & $T_{\rm S} ({\rm s})$ & $T_{\rm ES} ({\rm s})$ & $T_{\rm S} ({\rm s})$ & $T_{\rm ES} ({\rm s})$ & $T_{\rm S} ({\rm s})$ & $T_{\rm ES} ({\rm s})$ \\\hline
 $1000$ & $2094$ & $1556$ & $4621$ & $3523$ & $8900$ & $6621$ \\
 $100$ & $62$ & $47$ & $133$ & $99$ & $272$ & $194$ \\
 $10$ & $16$ & $12$ & $35$ & $26$ & $78$ & $58$ \\
 $1$ & $16$ & $12$ & $36$ & $26$ & $72$ & $52$ \\
 $0.1$ & $170$ & $120$ & $400$ & $300$ & $790$ & $570$ \\\hline\hline
\end{tabular}
\end{center}
\caption{Computational times (in seconds) required 
to reach the steady state using the FDLB method
for the heat transfer between stationary plates (HT), Couette flow 
(SH) and heat transfer under shear (HT-SH) problems, considered in 
Sections~\ref{sec:ht}, \ref{sec:couette} and \ref{sec:htsh}.
The data for $\delta =100$ and $1000$ is added
for completeness.
\label{tab:comp}}
\end{table}

\begin{figure}
\begin{center}
\begin{tabular}{cc}
\includegraphics[angle=0,width=0.48\linewidth]{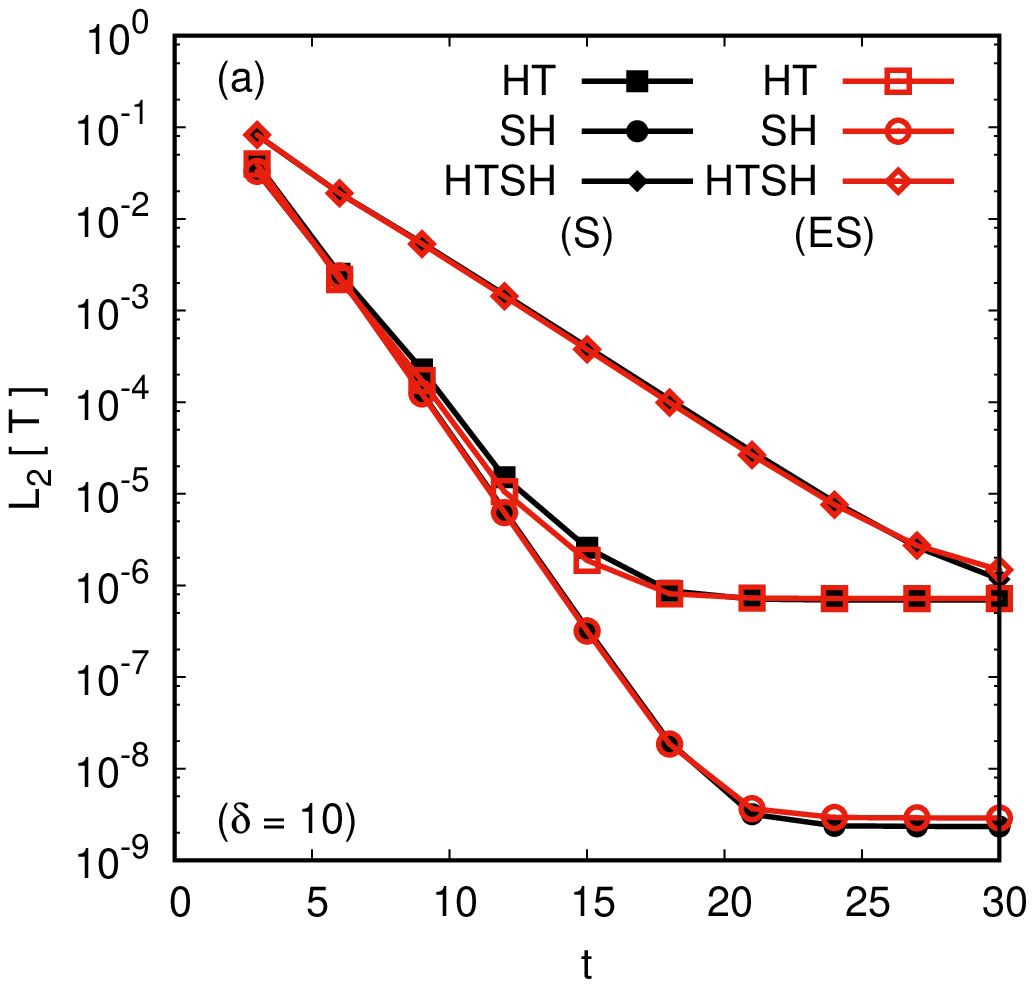} &
\includegraphics[angle=0,width=0.48\linewidth]{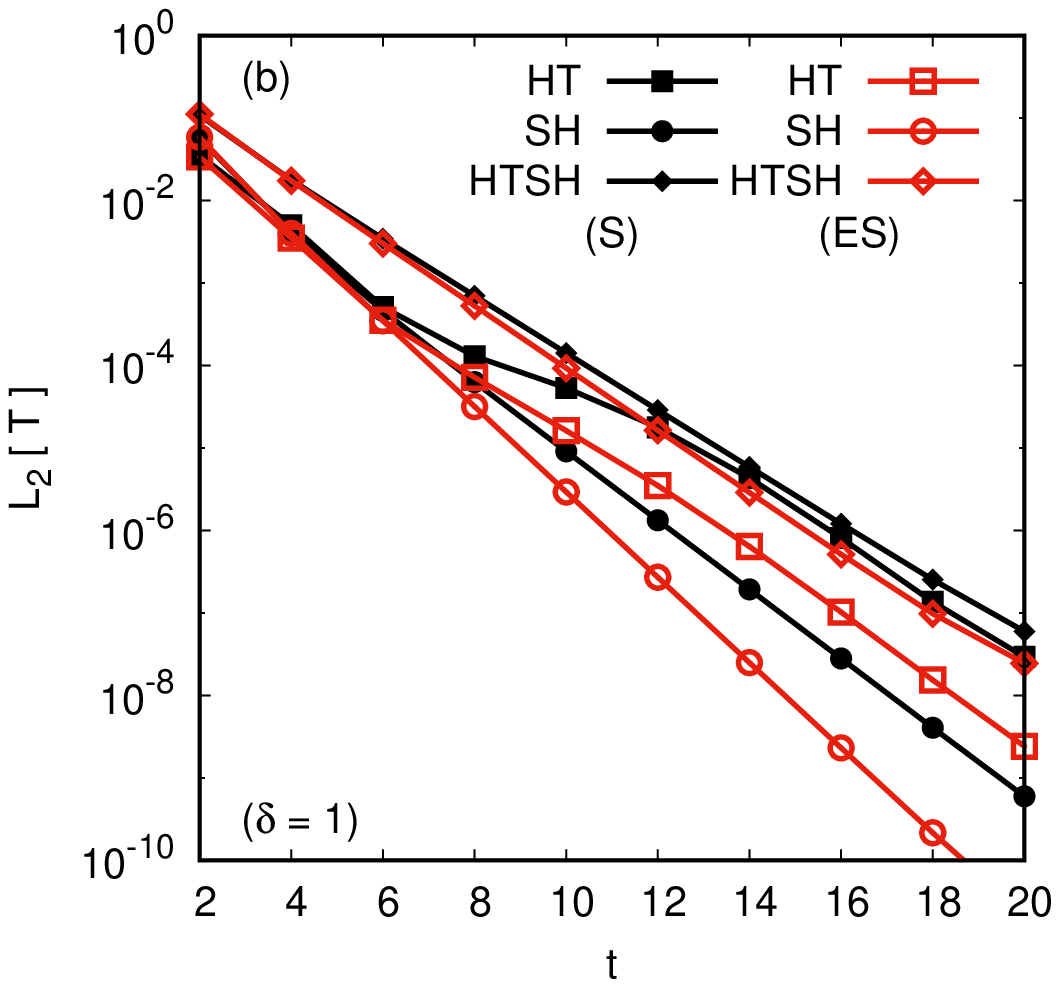}
\end{tabular}

\begin{tabular}{c}
\includegraphics[angle=0,width=0.48\linewidth]{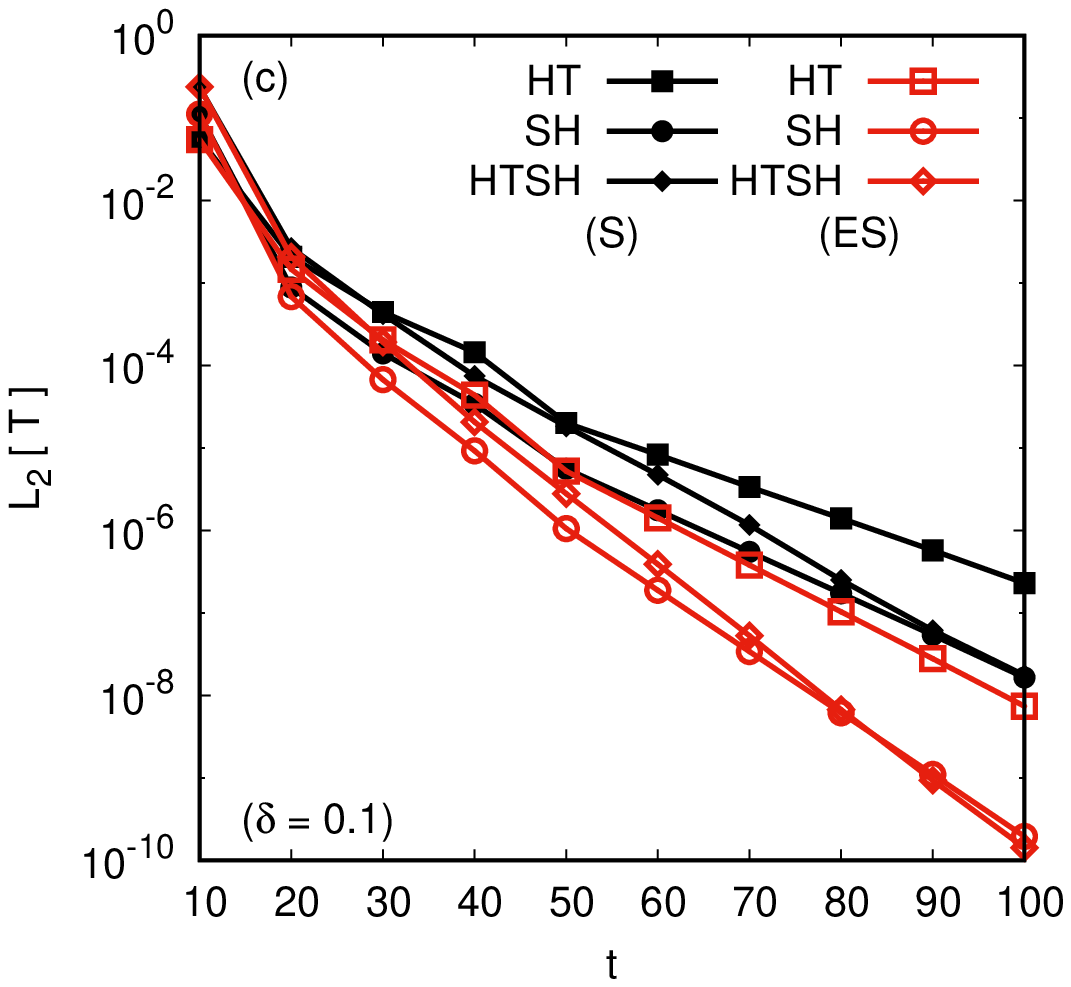}
\end{tabular}
\end{center}
\caption{Approach to steady state, assessed at the level of the 
temperature profile, at (a) $\delta = 10$, (b) $\delta = 1$ and (c) 
$\delta = 0.1$. The results for heat transfer under stationary walls 
(HT), Couette flow (SH) and heat transfer under shear (HTSH) 
are represented using squares, circles and rhombi, respectively.
The results for the S and ES models are represented using 
black lines with filled symbols and red lines with empty symbols, 
respectively.
\label{fig:steady}}
\end{figure}

It is known that the DSMC method suffers from stochastic noise, which persists 
after the steady state is reached. This noise can be eliminated
through averaging over a large number of time steps, which can be 
time consuming especially at large values of $\delta$. The time required 
to complete the DSMC simulations in this paper is about $20$ hours
using an MPI parallel code which runs on 32 processor cores.

In the case of the kinetic solver, we estimate the computational efficiency by 
considering simulations on a single core of an i7-4790K processor, running at a 
frequency of $4.0\ {\rm GHz}$. The simulation time is very short 
at $\delta = 10$ and $\delta =1$ -- of the order of one minute. This 
is because the quadrature order employed can be very small. At $\delta = 0.1$,
the quadrature must be increased, leading to computational times of 
the order of $5$--$10$ minutes on a single processor core. The exact figures are 
summarised in Table~\ref{tab:comp}. 
For completeness, in this table we included also the simulation times required 
to reach the stationary states when using the hybrid approach at 
$\delta = 100$ and $1000$, corresponding to the hydrodynamics regime. It can be seen that 
the hybrid approach described in this section becomes 
inefficient when $\delta$ increases.
This happens because, for $\delta \gtrsim 10$, the number of iterations
 required 
to reach the steady state increases dramatically with $\delta$, being around two orders 
of magnitude larger at $\delta = 1000$ than at
 $\delta = 10$.
It is noteworthy that 
the projection method discussed in \ref{app:hydro} performs better at 
larger values of $\delta$, as can be seen in Table~\ref{tab:comp2}.
In particular, the computing times required by the projection method at 
$\delta = 1000$ are shorter than those required by the hybrid method 
by factors of about $4$ and $5$ for the S and ES models, respectively. 
Further decreases in computational time at large 
values of $\delta$ can be expected when the explicit time-stepping 
method employed in this paper is replaced by, e.g., the implicit-explicit 
(IMEX) method that treats the collision term implicitly 
\cite{pareschi05,wang07,kis20}, or the iterative methods discussed 
in Refs.~\cite{valougeorgis03,wu17,su19,zhu20}.

Before ending this section, we briefly mention the procedure employed
to judge the approach to steady state. We consider 10 batches of $N_T$
iterations each, with a time step of $\delta t$. The time interval 
corresponding to one batch is $\Delta t = N_T \delta t$. Denoting via
$T^{n}(x)$ the temperature profile after the $n$th batch, we compute 
the $L_2$ norm of the relative difference between two successive batches, 
as follows:
\begin{equation}
 L_2[T^{n+1}] = \left[\frac{1}{\widetilde{L}} \int_{-\widetilde{L}/2}^{\widetilde{L}/2}
 d\widetilde{x} \, \left(\frac{\widetilde{T}^{n+1}(\widetilde{x})}{\widetilde{T}^n(\widetilde{x})} 
 - 1\right)^2\right]^{1/2},
\end{equation}
where the integration is performed as indicated in Eqs.~\eqref{eq:rect4} and 
\eqref{eq:rect4coef}.
Figure~\ref{fig:steady} shows that $L_2[T]$ steadily decreases with 
the time $t$ ($t_n = n \Delta t$).

\begin{figure}
\begin{center}
 \begin{tikzpicture}
\fill [black!10!white] (0,0) rectangle (4,4);
\draw [line width=2,dashed] (0,0) rectangle (4,4);
\draw [line width=2] (4,0) -- (4,4);
\draw [line width=2] (0,0) -- (0,4);

\foreach \y in {1,2,...,10} {
    \draw [line width=1.2] (4,\y*0.4-0.4) -- (4.3,\y*0.4);
    \draw [line width=1.2] (-0.3,\y*0.4-0.4) -- (0,\y*0.4);
}
\foreach \y in {1,2,...,10} {
    \draw [line width=1.2] (4.3,\y*0.4-0.4) -- (4,\y*0.4);
    \draw [line width=1.2] (0,\y*0.4-0.4) -- (-0.3,\y*0.4);
}
\draw [dashed] (0.059,0)--(0.059,4);
\draw [dashed] (0.282,0)--(0.282,4);
\draw [dashed] (0.745,0)--(0.745,4);
\draw [dashed] (1.526,0)--(1.526,4);

\draw [dashed] (2.474,0)--(2.474,4);
\draw [dashed] (3.255,0)--(3.255,4);
\draw [dashed] (3.718,0)--(3.718,4);
\draw [dashed] (3.941,0)--(3.941,4);

\node [rotate=90] at (-0.5,2) {$T_{\rm left} = 1 - \Delta T / 2$};
\draw (0.0,0.0) node [anchor=north] {$x=-1/2$};
\node at (2,4.25) {periodic};
\node [rotate=270] at (4.6,2) {$T_{\rm right} = 1 + \Delta T / 2$};
\draw (4,0) node [anchor=north] {$x=1/2$};
\draw (2,0) node [anchor=north] {periodic};
\end{tikzpicture} 
\end{center}
\caption{The simulation setup for the heat transfer problem.
The vertical dashed lines show a sample grid 
employing $S = 4$ points on each half of the channel, 
stretched according to Eq.~\eqref{eq:stretch} with $A = 0.95$.
\label{fig:ht_setup}}
\end{figure}
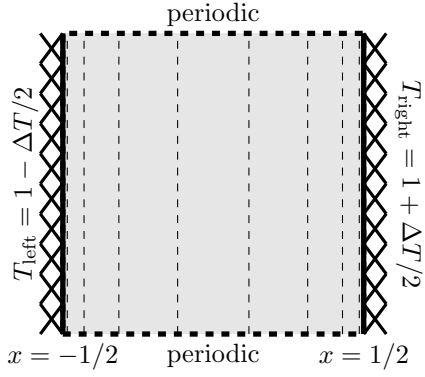

\begin{figure}
\begin{center}
\begin{tabular}{cc}
\includegraphics[angle=0,width=0.48\linewidth]{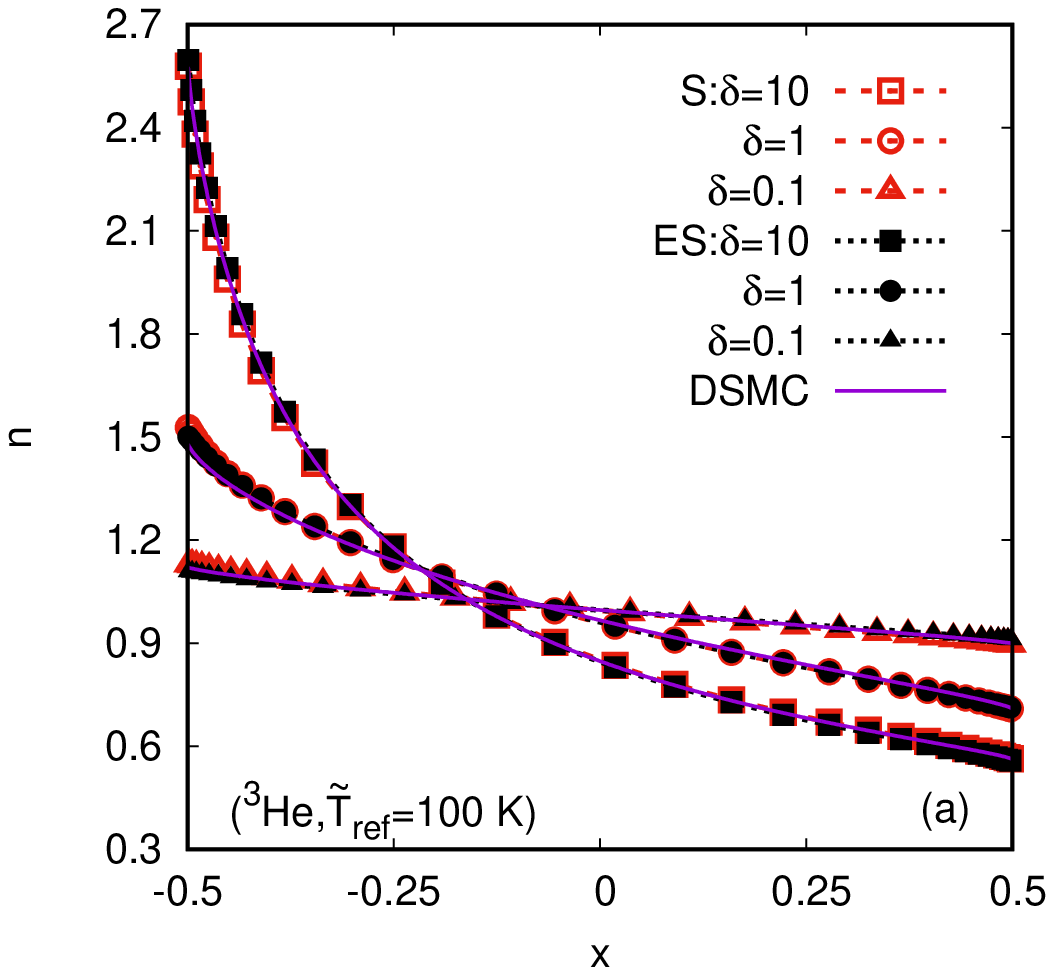} &
\includegraphics[angle=0,width=0.48\linewidth]{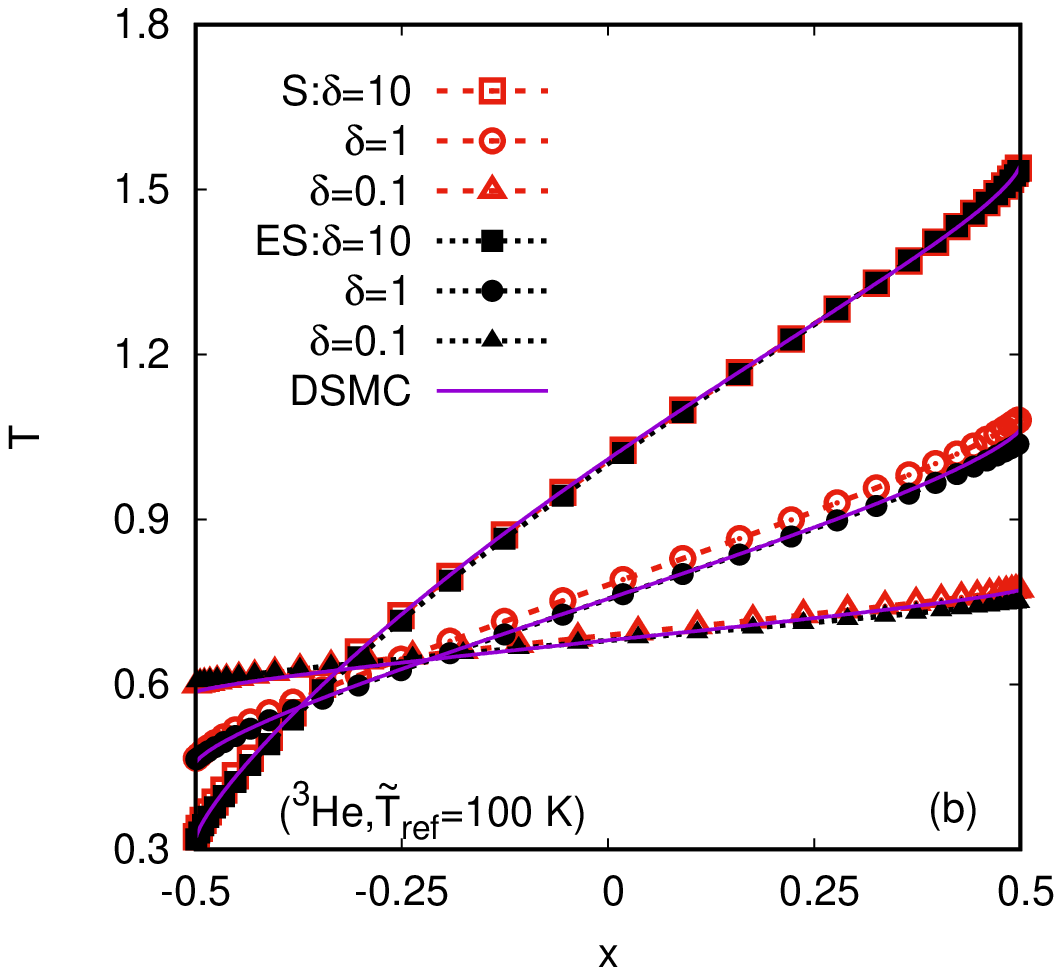} 
\end{tabular}
\end{center}
\caption{Comparison between the S (red dashed lines and empty symbols) 
and ES (black dotted lines and filled symbols) results and the
DSMC (continuous lines) results in the context of the heat transfer 
between stationary plates problem for the profiles of (a) $n$
and (b) $T$ through the channel
($-1/2 \le x\le 1/2$), for ${}^3{\rm He}$ gas constituents.
The reference temperature is set to
$\widetilde{T}_{\rm ref} = 100\ {\rm K}$,
while the temperature difference
between the two walls is $\widetilde{\Delta T} = 1.5 \widetilde{T}_{\rm ref}$.
\label{fig:ht_profiles}}
\end{figure}

\begin{figure}
\begin{center}
\begin{tabular}{cc}
\includegraphics[angle=0,width=0.48\linewidth]{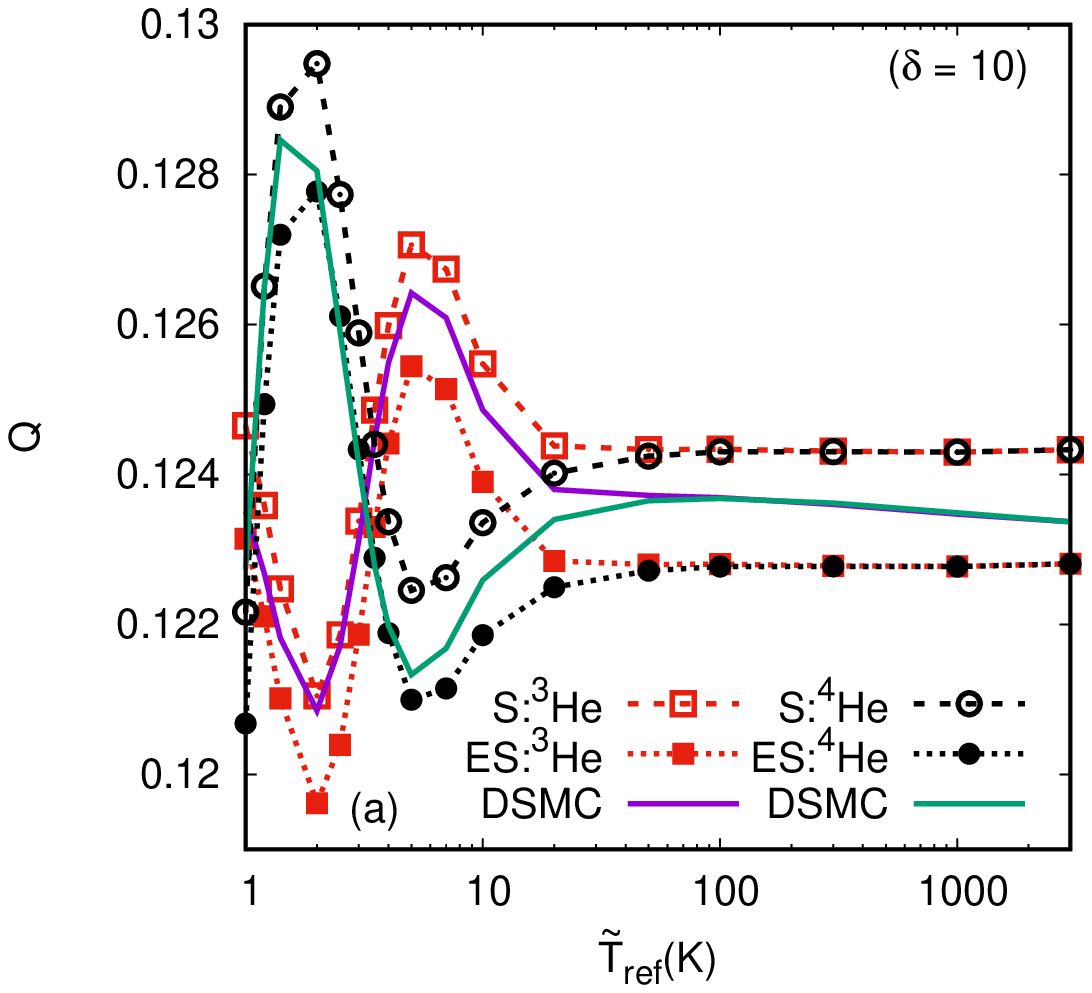} &
\includegraphics[angle=0,width=0.48\linewidth]{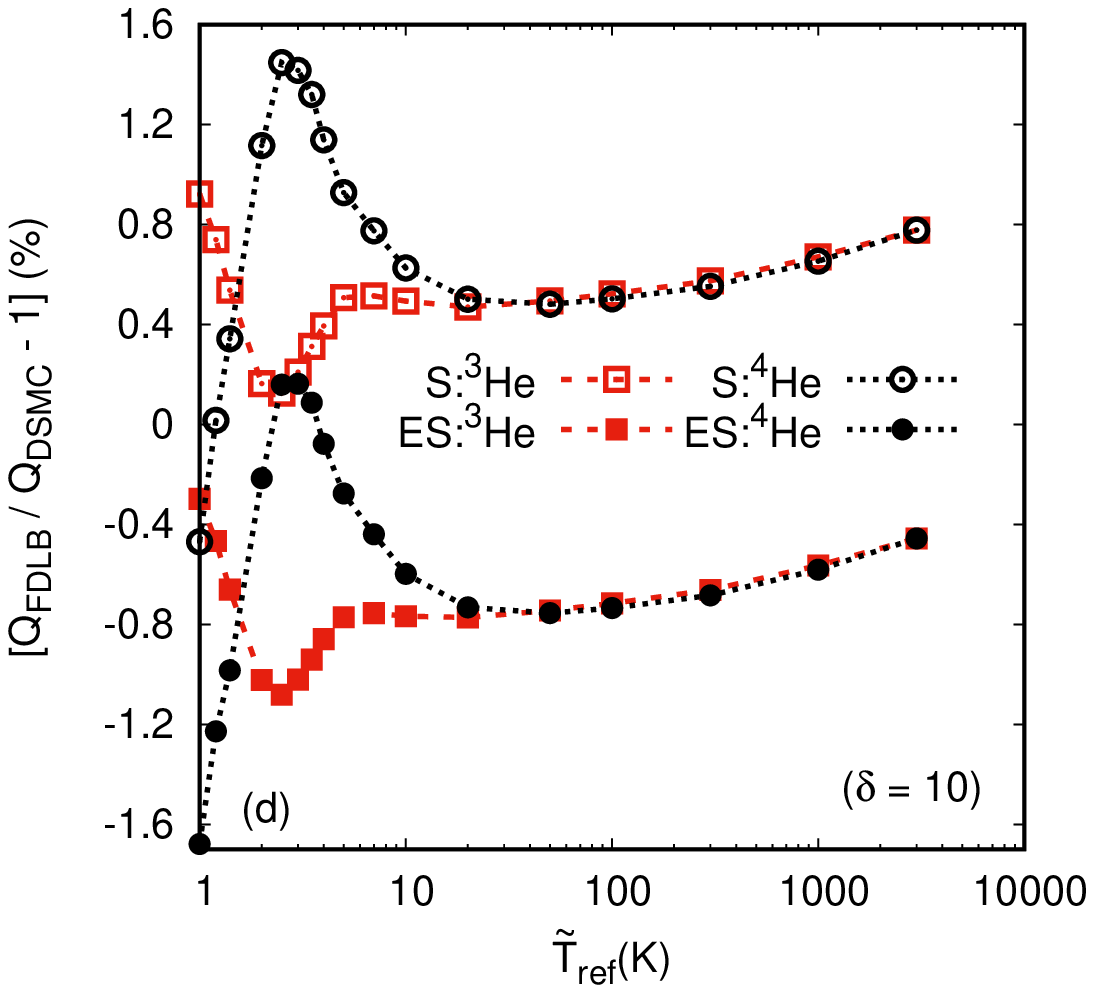} \\
\includegraphics[angle=0,width=0.48\linewidth]{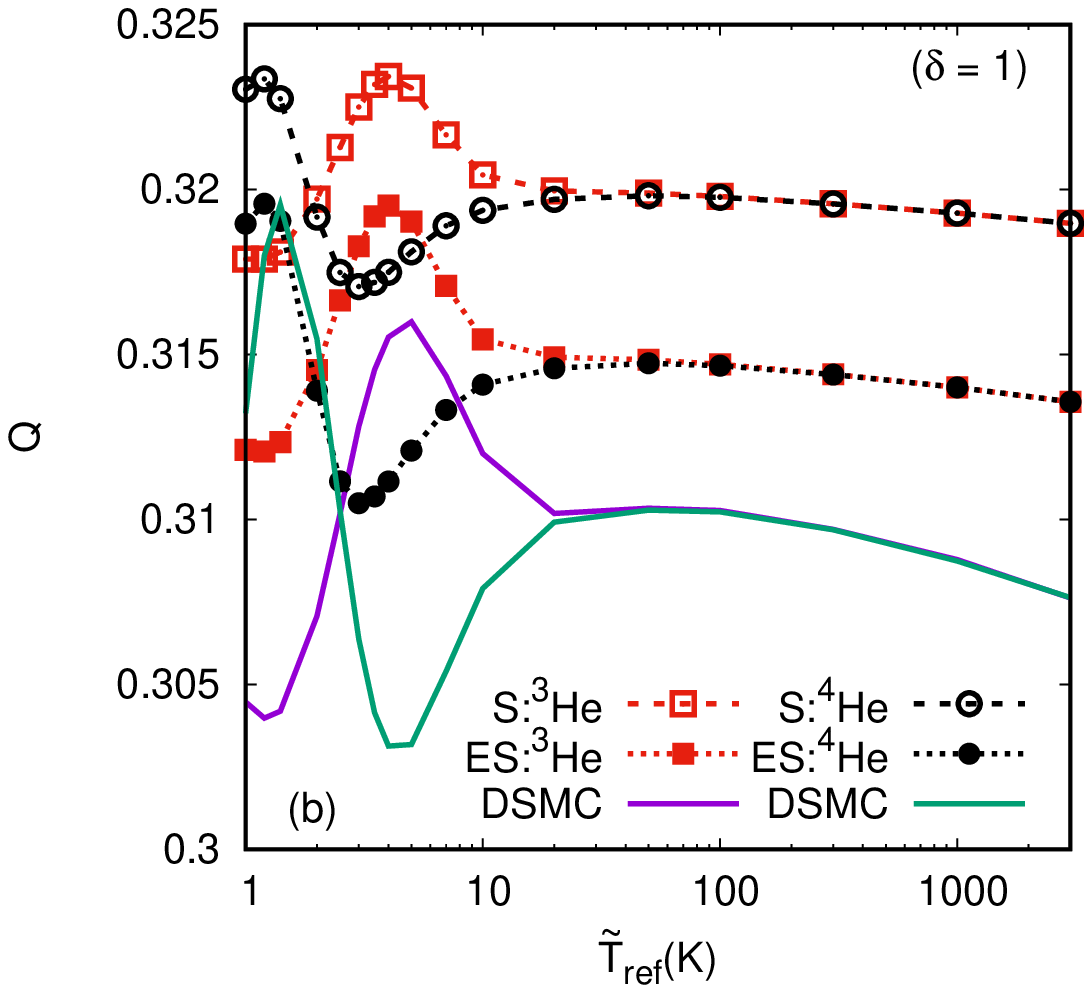} &
\includegraphics[angle=0,width=0.48\linewidth]{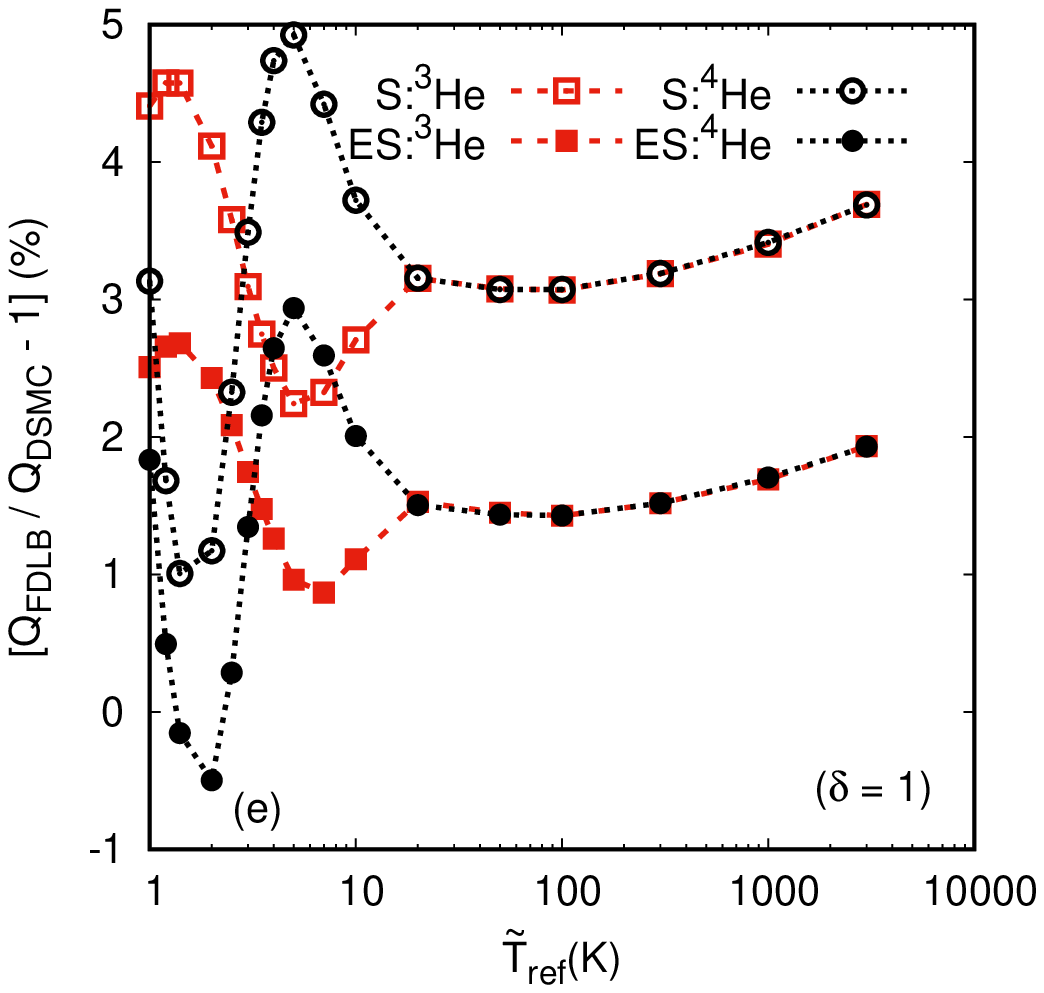} \\
\includegraphics[angle=0,width=0.48\linewidth]{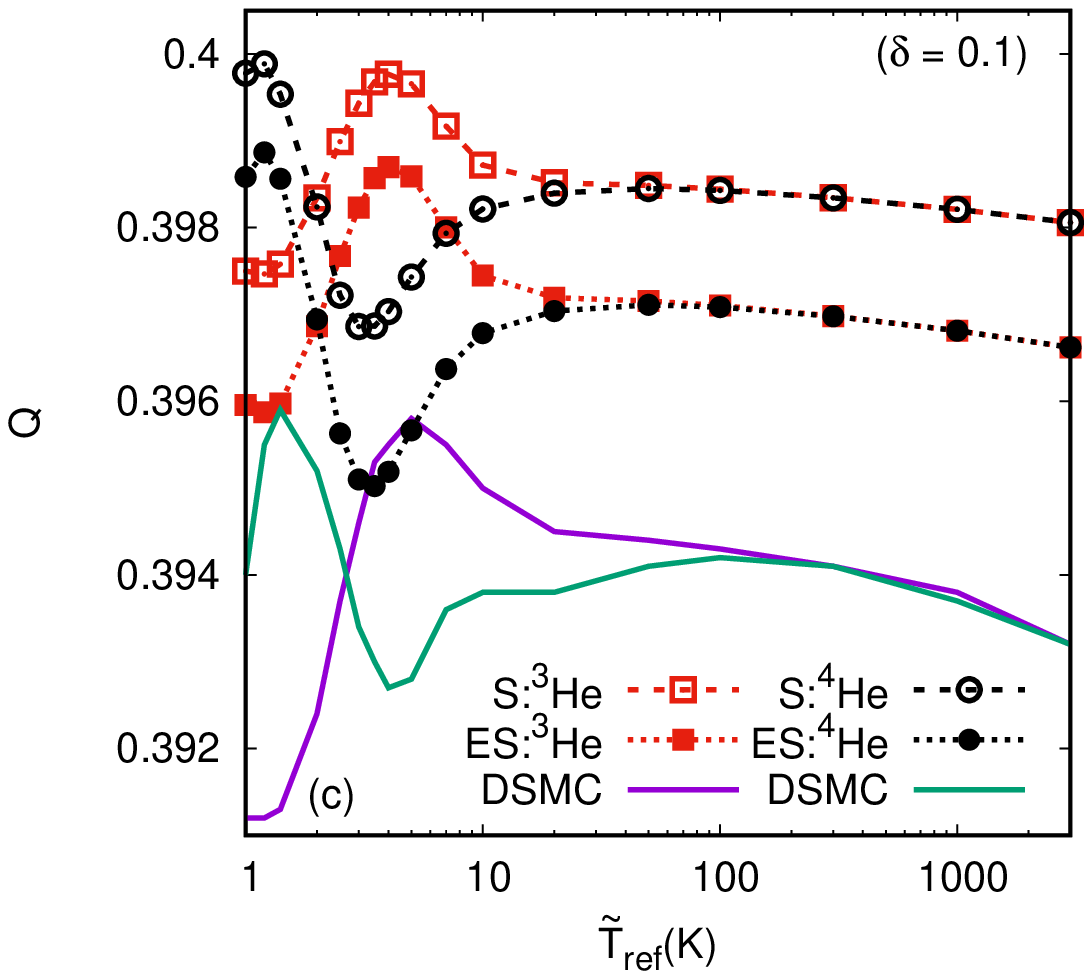} &
\includegraphics[angle=0,width=0.48\linewidth]{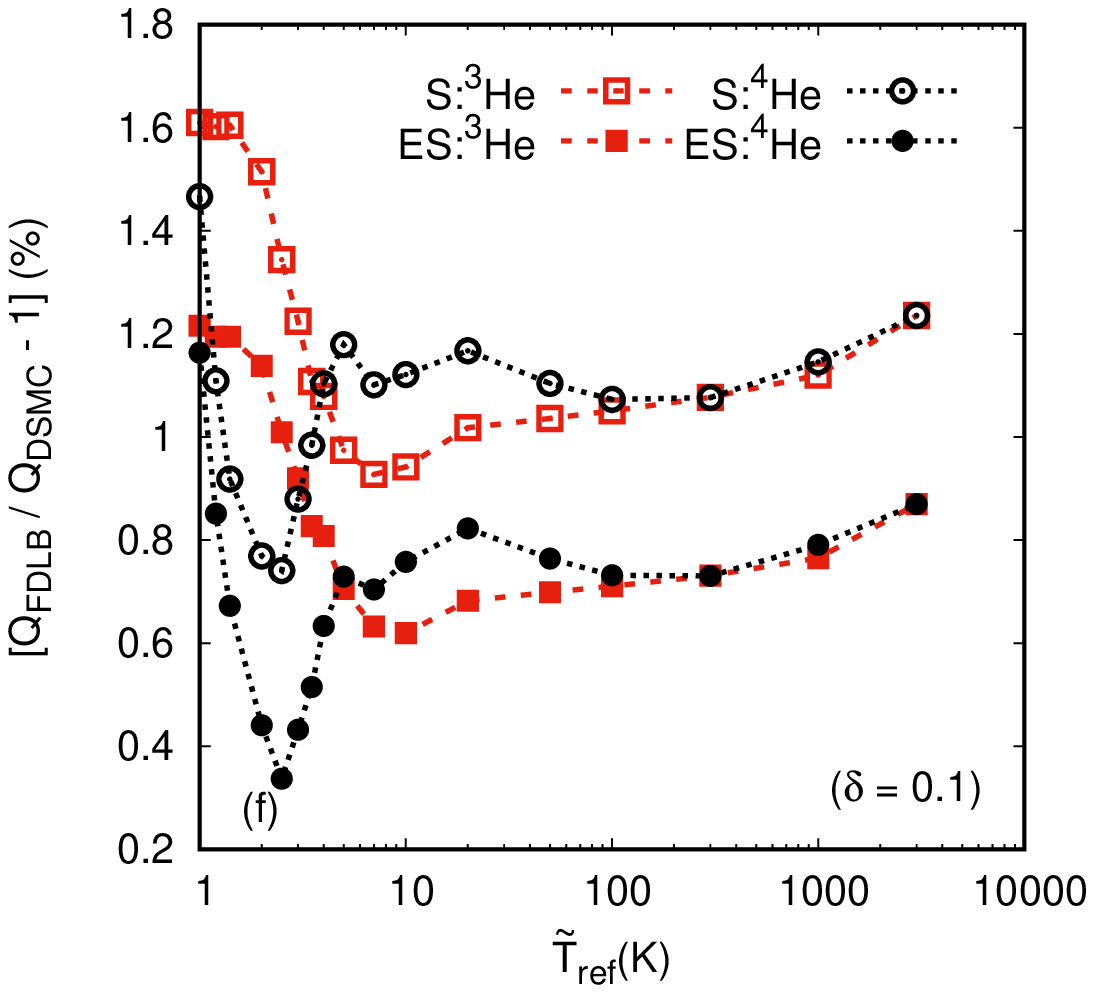}
\end{tabular}
\end{center}
\caption{(Left) Dependence of the constant $Q$, computed 
for the heat transfer 
between stationary plates problem using 
Eq.~\eqref{eq:Q} with $\widetilde{u}_y = 0$,
on the average wall temperature $\widetilde{T}_{\rm ref}$.
(Right) Relative error $Q_{\rm FDLB} / Q_{\rm DSMC} - 1$ 
of the FDLB results with respect to the DSMC 
results. Both ${}^3{\rm He}$ (red dashed lines with squares) and 
${}^4{\rm He}$ (black dotted lines with circles) are considered 
within the S (empty symbols) and 
ES (filled symbols) models
and the results are represented at $\delta = 10$ (top),
$1$ (middle) and $0.1$ (bottom).
\label{fig:ht_Q}}
\end{figure}

\section{Heat transfer}\label{sec:ht}

The first application considered in this paper concerns the heat 
transfer between stationary parallel plates problem. The simulation 
setup is represented schematically in Fig.~\ref{fig:ht_setup}. In our 
simulations, the reference temperature, 
$\widetilde{T}_{\rm ref} = (\widetilde{T}_{\rm left} + 
\widetilde{T}_{\rm right}) / 2$, is varied between 
$1\ {\rm K}$ and $3000\ {\rm K}$. 

Representative profiles of the density $n$ and temperature $T$ are 
shown for ${}^3{\rm He}$ constituents at 
$\widetilde{T}_{\rm ref} = 100\ {\rm K}$ in Fig.~\ref{fig:ht_profiles}.
The DSMC results are shown using solid lines.
The FDLB results obtained with the S model are shown using 
red dashed lines with empty symbols. The FDLB results obtained with 
the ES model are shown using black dotted lines with filled 
symbols. The FDLB data corresponding to $\delta = 10$, $1$ and $0.1$ 
are shown with squares, circles and triangles, respectively.
Very good agreement can be seen between the results obtained using 
the ES model and the DSMC data. There is a visible discrepancy in the 
temperature profile obtained with the Shakohv model at $\delta = 1$.

A more quantitative analysis is performed at the level of 
the quantity $Q$, introduced in Eq.~\eqref{eq:Q}, with 
$\widetilde{u}_y$ set to $0$. Figure~\ref{fig:ht_Q} compares 
the FDLB and DSMC results for $Q$ 
with respect to $\widetilde{T}_{\rm ref}$ for 
$1\ {\rm K} \le \widetilde{T}_{\rm ref} \le 3000\ {\rm K}$, at 
$\delta = 10$ (top line), $1$ (middle line) and $0.1$ (bottom line).
The value of $Q$ is represented in the left column of Fig.~\ref{fig:ht_Q},
while the relative error $Q_{\rm FDLB} / Q_{\rm DSMC} - 1$ is shown
in the right column of Fig.~\ref{fig:ht_Q}.
These results were obtained using
the S (empty symbols) and the ES (filled symbols) models,
for both the
${}^3{\rm He}$ (red lines with squares) and the ${}^4{\rm He}$ 
(black lines with circles) constituents. At $\delta = 10$, the 
S model overestimates the DSMC results.
Contrary to the S model, these
DSMC results are underestimated by the ES model. The relative 
errors are roughly the same in absolute values.
At smaller values of $\delta$, the ES model provides results which 
are more accurate than those obtained using the S model. 
The highest relative discrepancy with respect to the DSMC data 
can be observed at $\delta = 1$, when the relative error 
of the S model reaches almost $5\%$, while for the ES model,
it stays below $3\%$.

\section{Couette flow}\label{sec:couette}

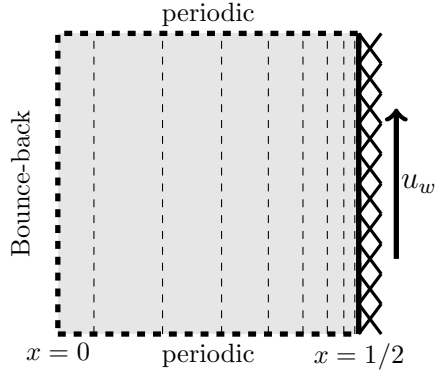
\begin{figure}
\begin{center}
 \begin{tikzpicture}
\fill [black!10!white] (0,0) rectangle (4,4);
\draw [line width=2,dashed] (0,0) rectangle (4,4);
\draw [line width=2] (4,0) -- (4,4);
\draw [line width=2,->] (4.5,1) -- (4.5,3);

\foreach \y in {1,2,...,10} {
    \draw [line width=1.2] (4,\y*0.4-0.4) -- (4.3,\y*0.4);
}
\foreach \y in {1,2,...,10} {
    \draw [line width=1.2] (4.3,\y*0.4-0.4) -- (4,\y*0.4);
}
\draw [dashed] (0.480,0)--(0.480,4);
\draw [dashed] (1.392,0)--(1.392,4);
\draw [dashed] (2.177,0)--(2.177,4);
\draw [dashed] (2.799,0)--(2.799,4);

\draw [dashed] (3.259,0)--(3.259,4);
\draw [dashed] (3.583,0)--(3.583,4);
\draw [dashed] (3.802,0)--(3.802,4);
\draw [dashed] (3.948,0)--(3.948,4);

\node [rotate=90] at (-0.5,2) {Bounce-back};
\draw (0.0,0.0) node [anchor=north] {$x=0$};
\node at (2,4.25) {periodic};
\node at (4.8,2) {\large $u_w$};
\draw (4,0) node [anchor=north] {$x=1/2$};
\draw (2,0) node [anchor=north] {periodic};
\end{tikzpicture} 
\caption{The simulation setup for the Couette flow problem.
The vertical dashed lines show a sample grid 
employing $S = 8$ points, stretched according to 
Eq.~\eqref{eq:stretch} with $A = 0.95$.
\label{fig:couette_setup}}
\end{center}
\end{figure}

\begin{figure}
\begin{center}
\begin{tabular}{cc}
\includegraphics[angle=0,width=0.48\linewidth]{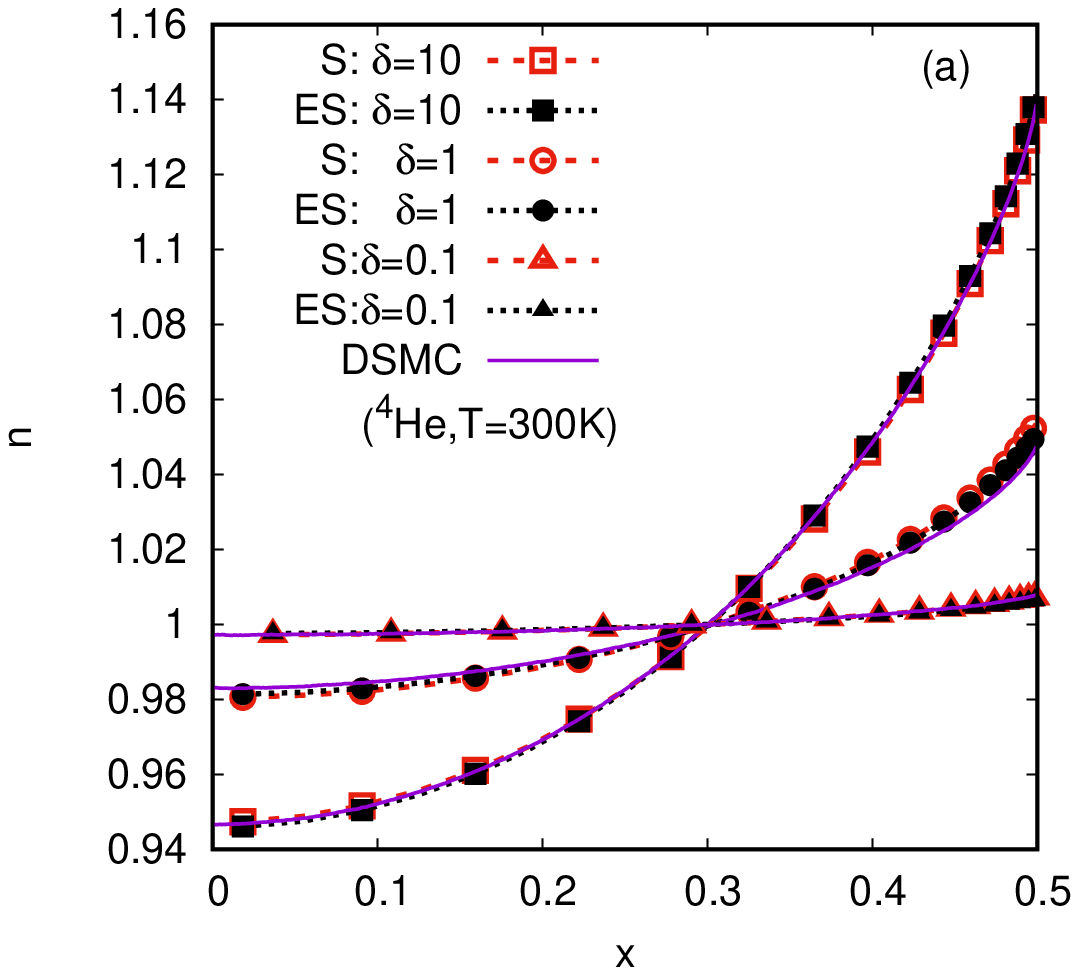} &
\includegraphics[angle=0,width=0.48\linewidth]{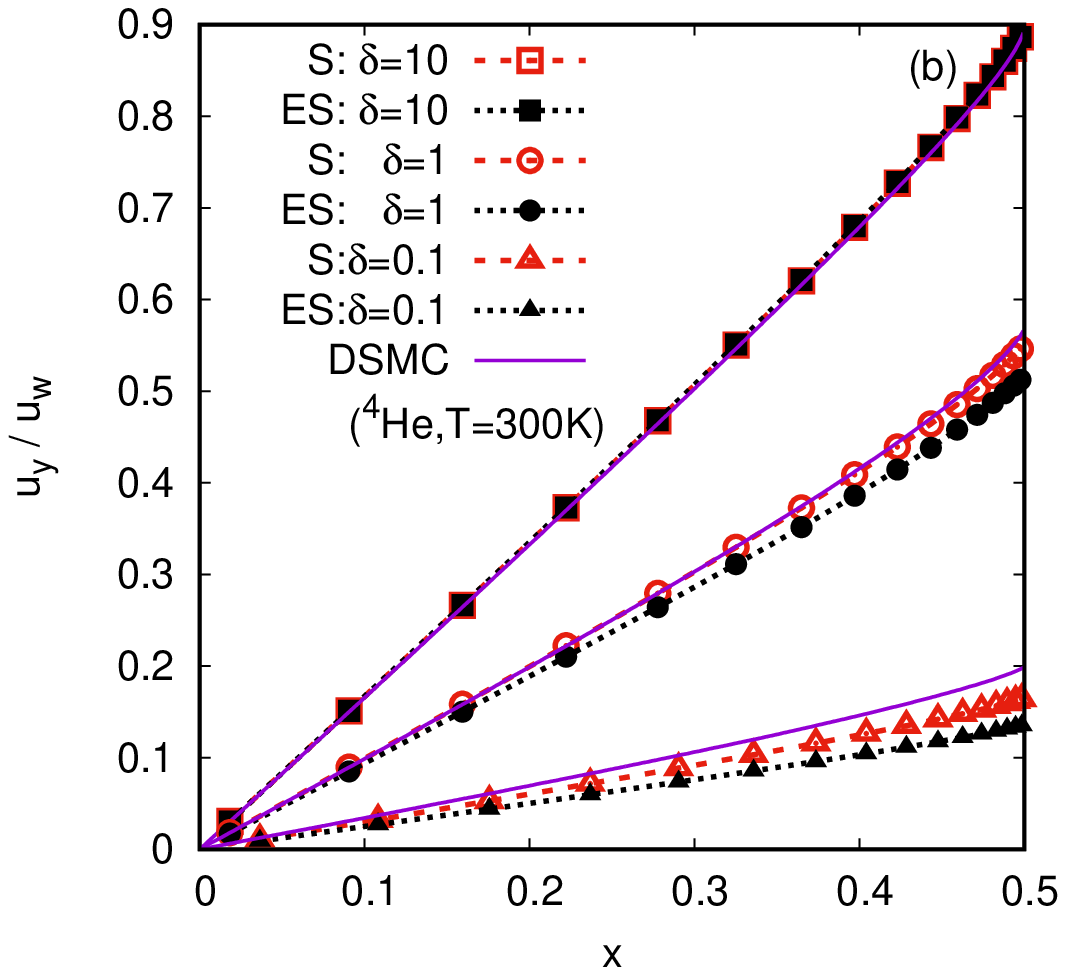} \\
\includegraphics[angle=0,width=0.48\linewidth]{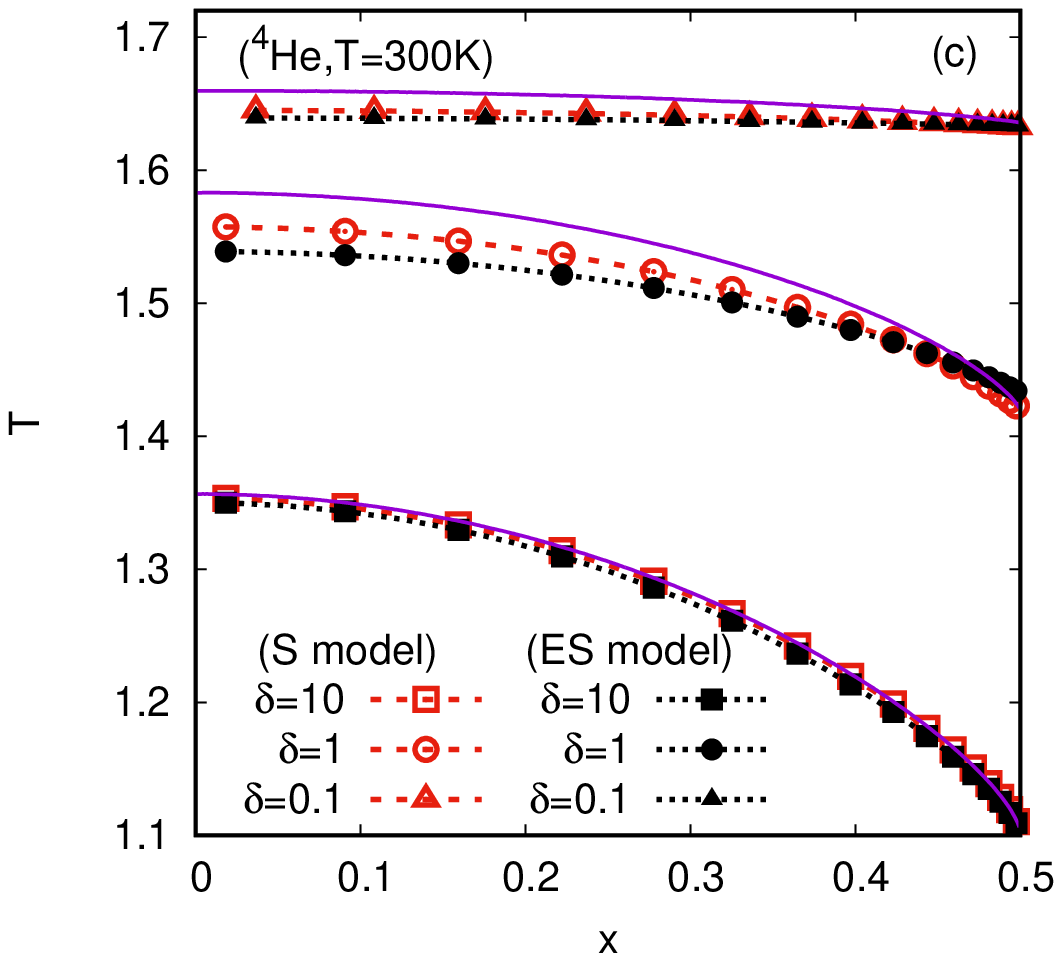} &
\includegraphics[angle=0,width=0.48\linewidth]{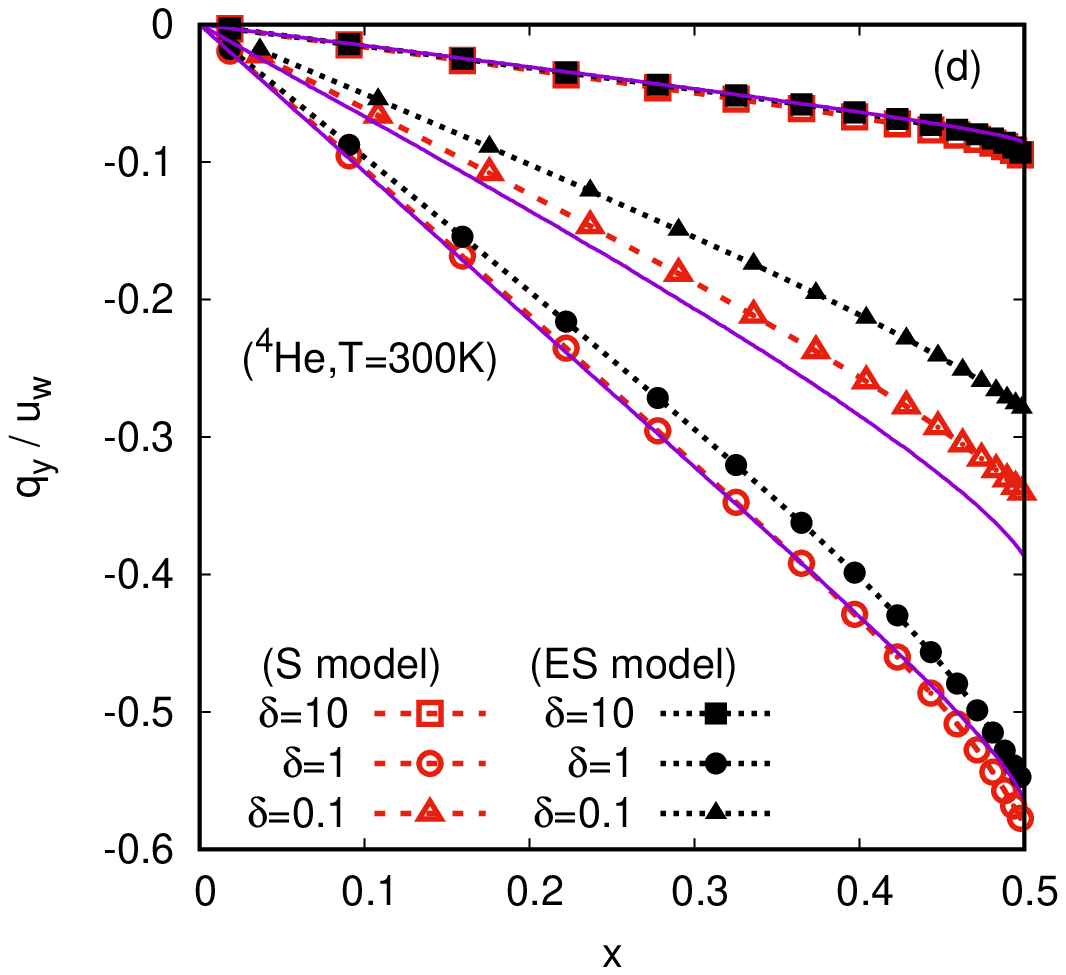} 
\end{tabular}
\end{center}
\caption{Comparison between the FDLB results for the
S model (dashed red lines and empty symbols) 
and ES model (dotted black lines and filled symbols)
and the DSMC results (continuous lines) for the profiles of (a) $n$,
(b) $u_y$, (c) $T$ and (d) $q_y$ through the half-channel
($0 \le x\le 1/2$), for ${}^4{\rm He}$ gas constituents,
in the context of the Couette flow.
The wall temperature is set to
$\widetilde{T}_{\rm ref} = 300\ {\rm K}$,
while the wall velocity is
$u_w = \sqrt{2}$.
\label{fig:couette_profiles}}
\end{figure}

\begin{figure}
\begin{center}
\begin{tabular}{cc}
\includegraphics[angle=0,width=0.48\linewidth]{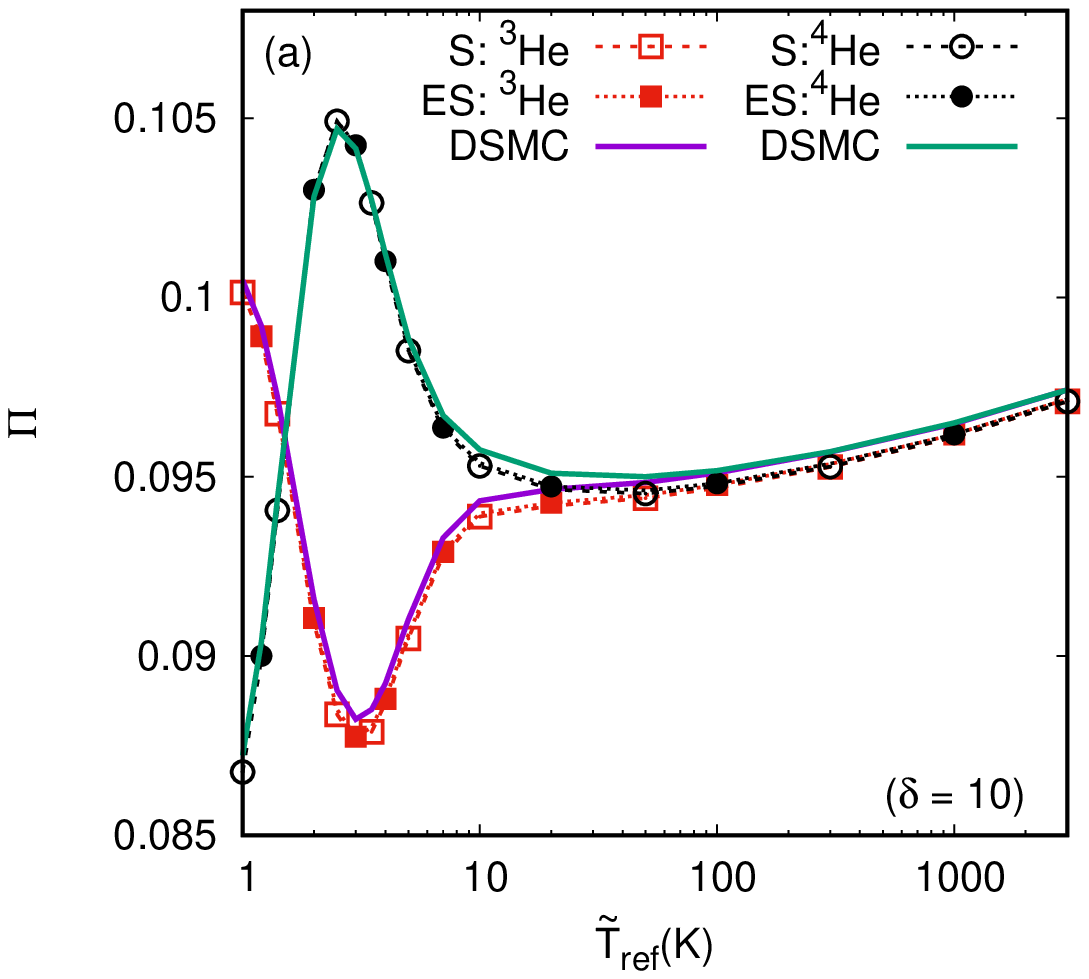} &
\includegraphics[angle=0,width=0.48\linewidth]{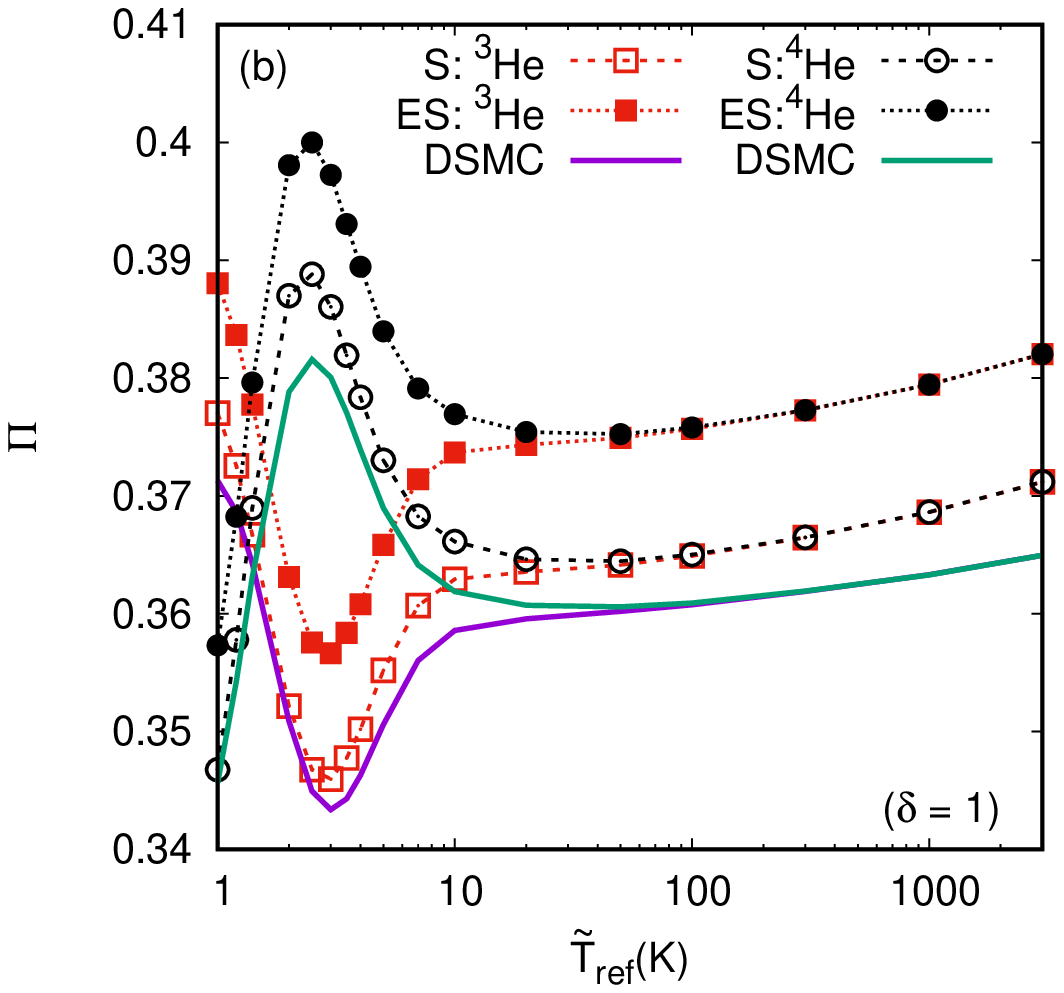} 
\end{tabular}

\begin{tabular}{c}
\includegraphics[angle=0,width=0.48\linewidth]{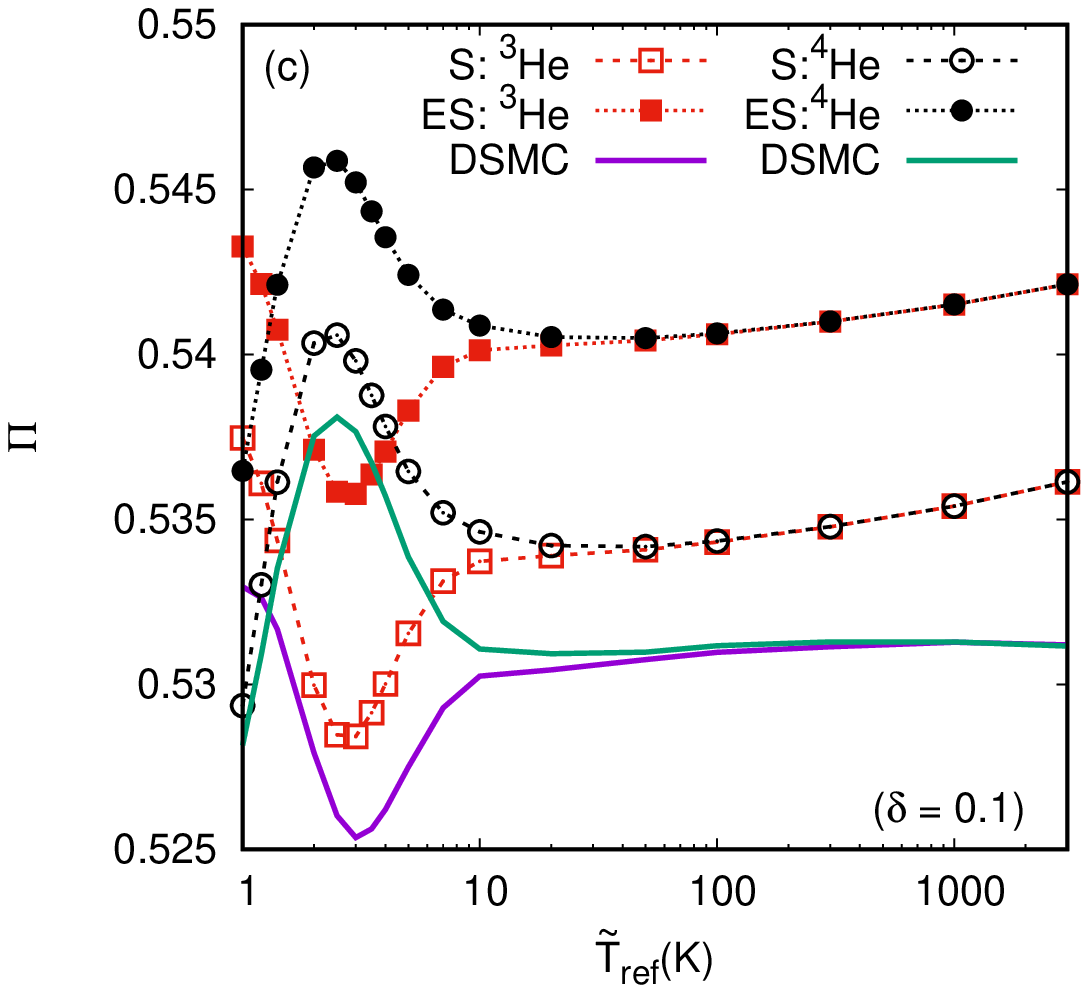}
\end{tabular}
\end{center}
\caption{Dependence of $\Pi$, computed using Eq.~\eqref{eq:Pi_adim}
in the context of the Couette flow,
on the wall temperature $\widetilde{T}_{\rm ref}$ 
for both ${}^3{\rm He}$ (squares) and ${}^4{\rm He}$ (circles), at (a) $\delta = 10$,
(b) $\delta = 1$ and (c) $\delta = 0.1$.
\label{fig:couette_Pi}}
\end{figure}

\begin{figure}
\begin{center}
\begin{tabular}{cc}
\includegraphics[angle=0,width=0.48\linewidth]{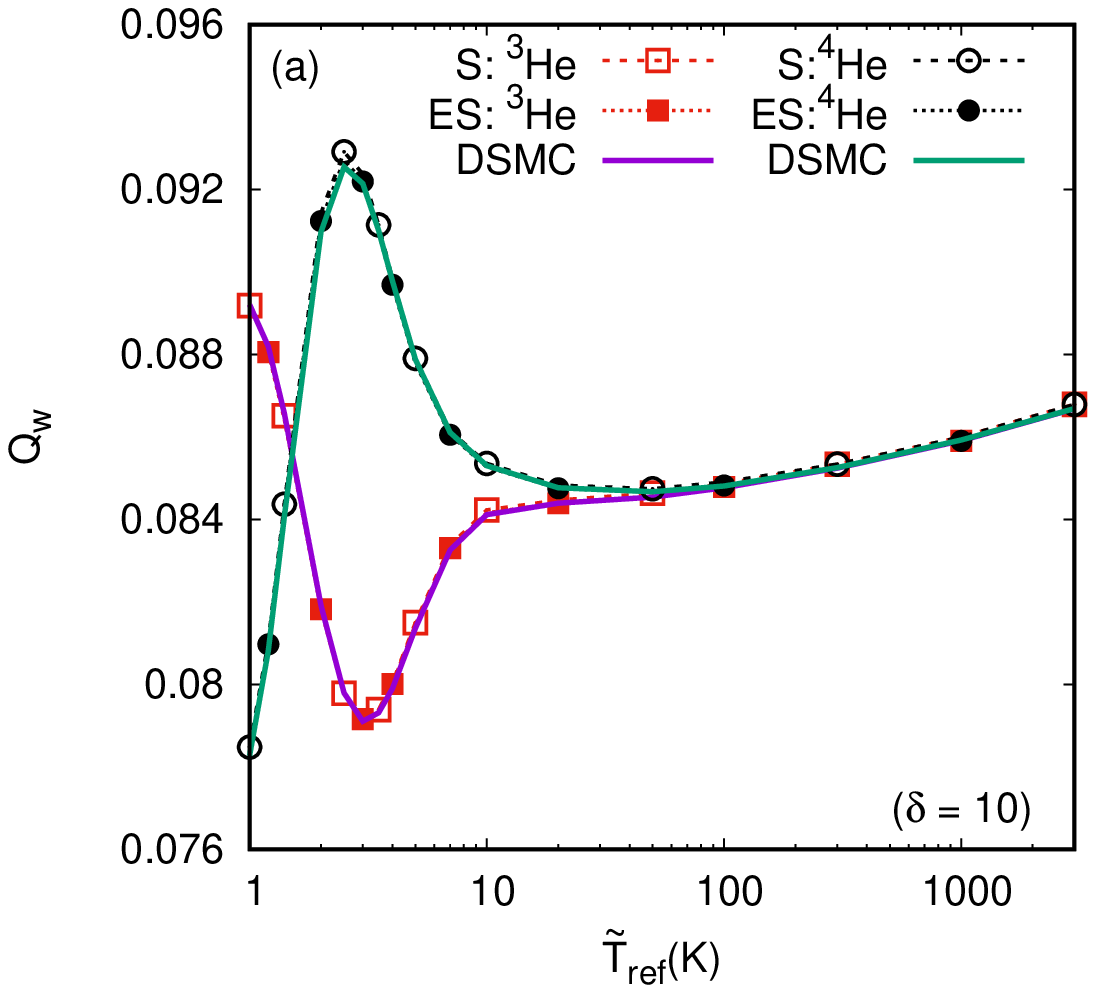} &
\includegraphics[angle=0,width=0.48\linewidth]{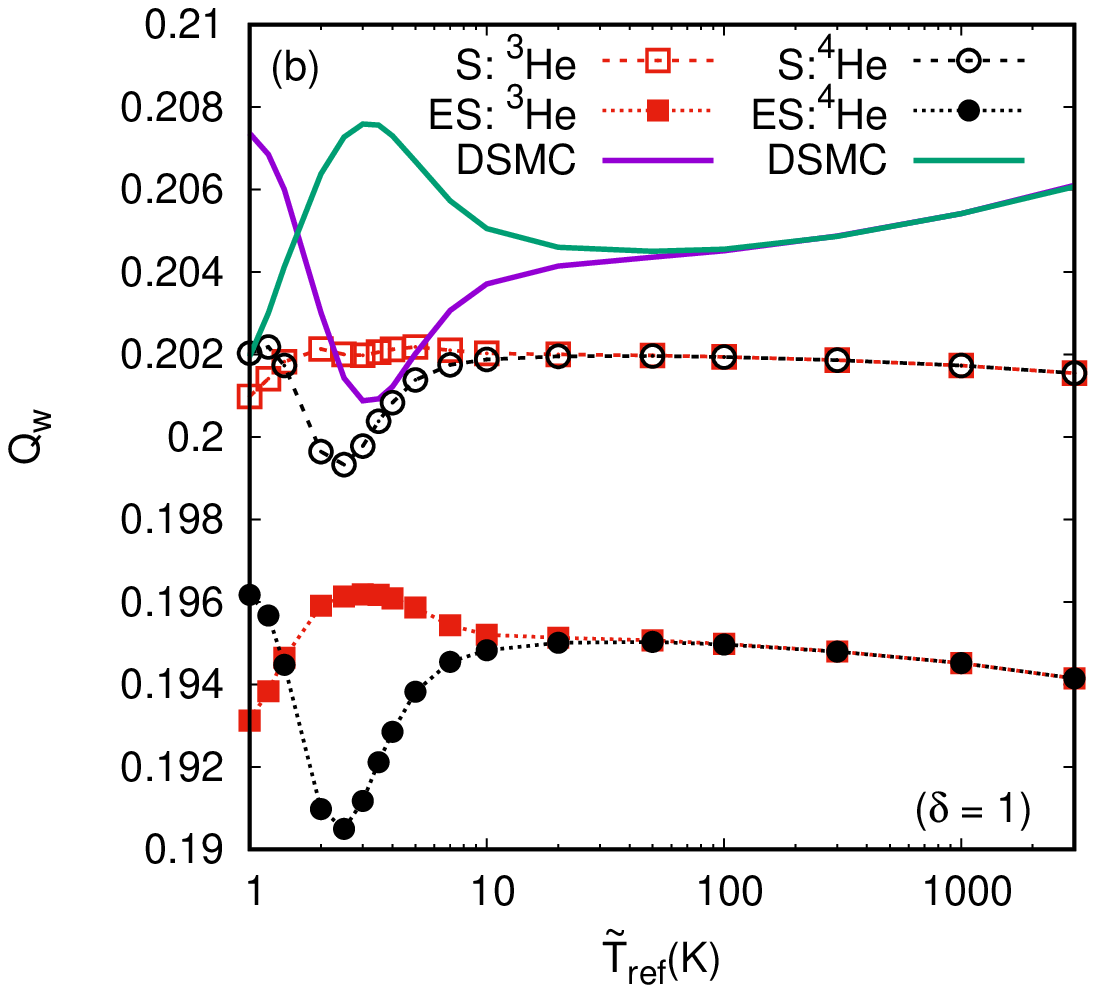} \\
\end{tabular}

\begin{tabular}{c}
\includegraphics[angle=0,width=0.48\linewidth]{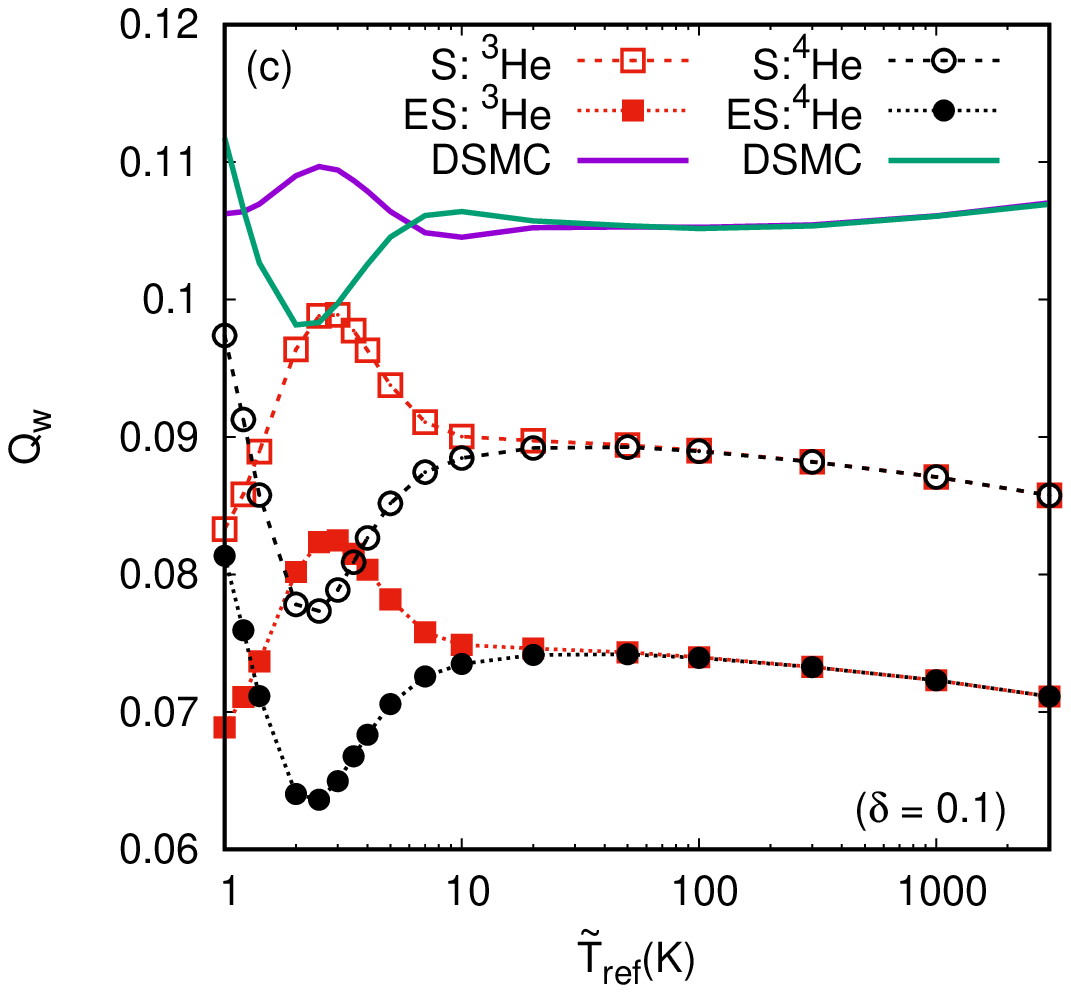}
\end{tabular}
\end{center}
\caption{Dependence of $Q_w$, computed in the context of the Couette flow using 
Eq.~\eqref{eq:Qw}, on the wall temperature $\widetilde{T}_{\rm ref}$ 
for both ${}^3{\rm He}$ (squares) and ${}^4{\rm He}$ (circles), at (a) $\delta = 10$,
(b) $\delta = 1$ and (c) $\delta = 0.1$.
\label{fig:couette_Qw}}
\end{figure}

\begin{figure}
\begin{center}
\begin{tabular}{cc}
\includegraphics[angle=0,width=0.48\linewidth]{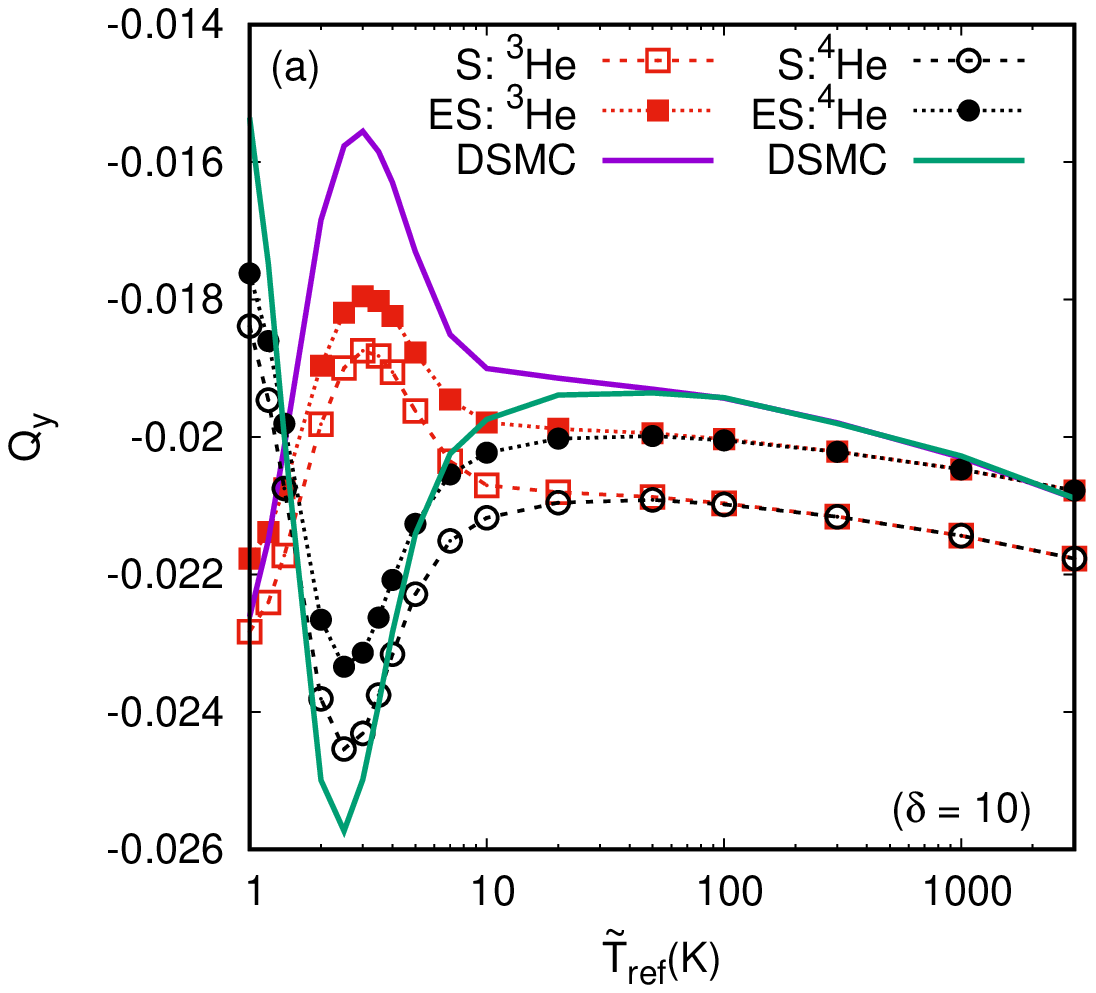} &
\includegraphics[angle=0,width=0.48\linewidth]{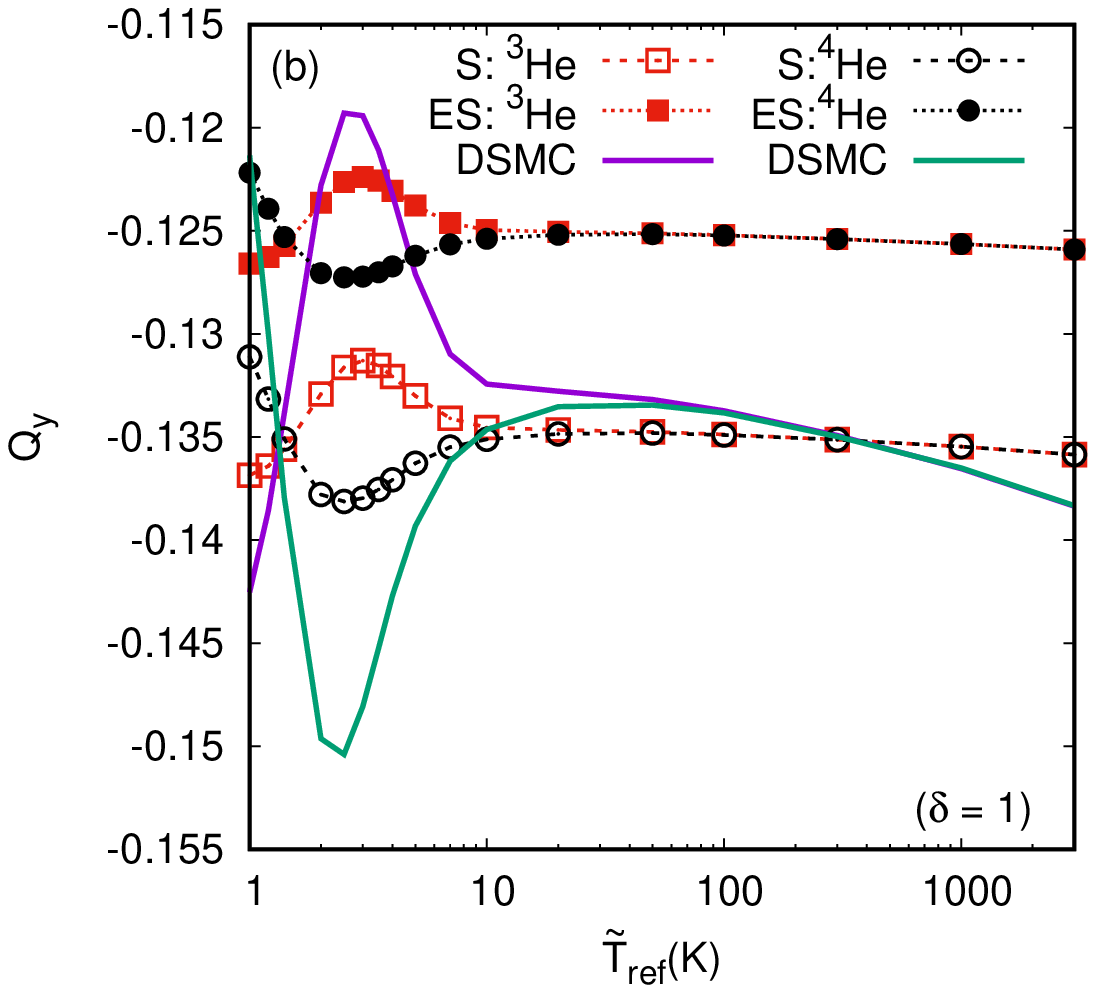} 
\end{tabular}

\begin{tabular}{c}
\includegraphics[angle=0,width=0.48\linewidth]{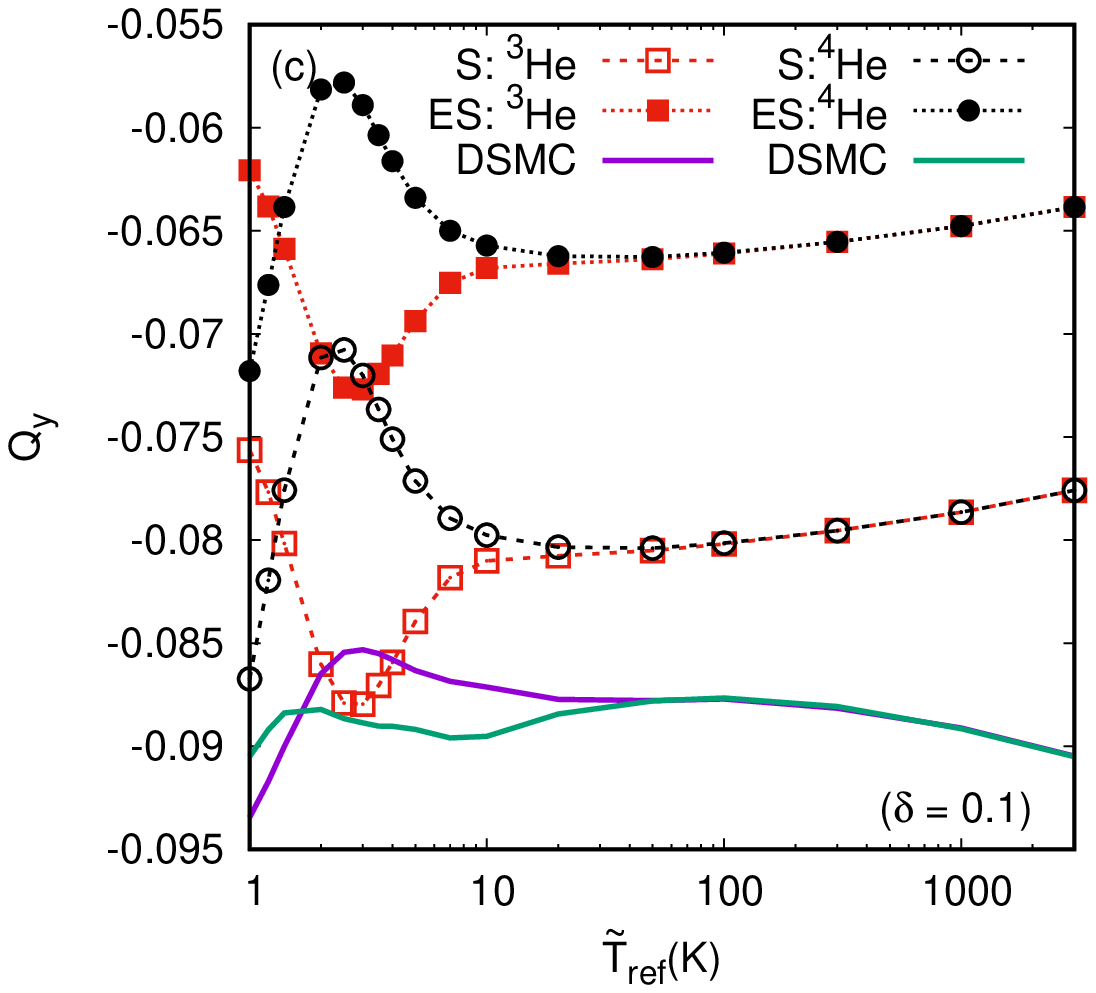}
\end{tabular}
\end{center}
\caption{Dependence of $Q_y$,
computed in the context of the Couette flow using 
Eq.~\eqref{eq:Qy}, on the wall temperature $\widetilde{T}_{\rm ref}$ 
for both ${}^3{\rm He}$ (squares) and ${}^4{\rm He}$ (circles), at (a) $\delta = 10$,
(b) $\delta = 1$ and (c) $\delta = 0.1$.
\label{fig:couette_Qy}}
\end{figure}

\begin{figure}
\begin{center}
\begin{tabular}{cc}
\includegraphics[angle=0,width=0.48\linewidth]{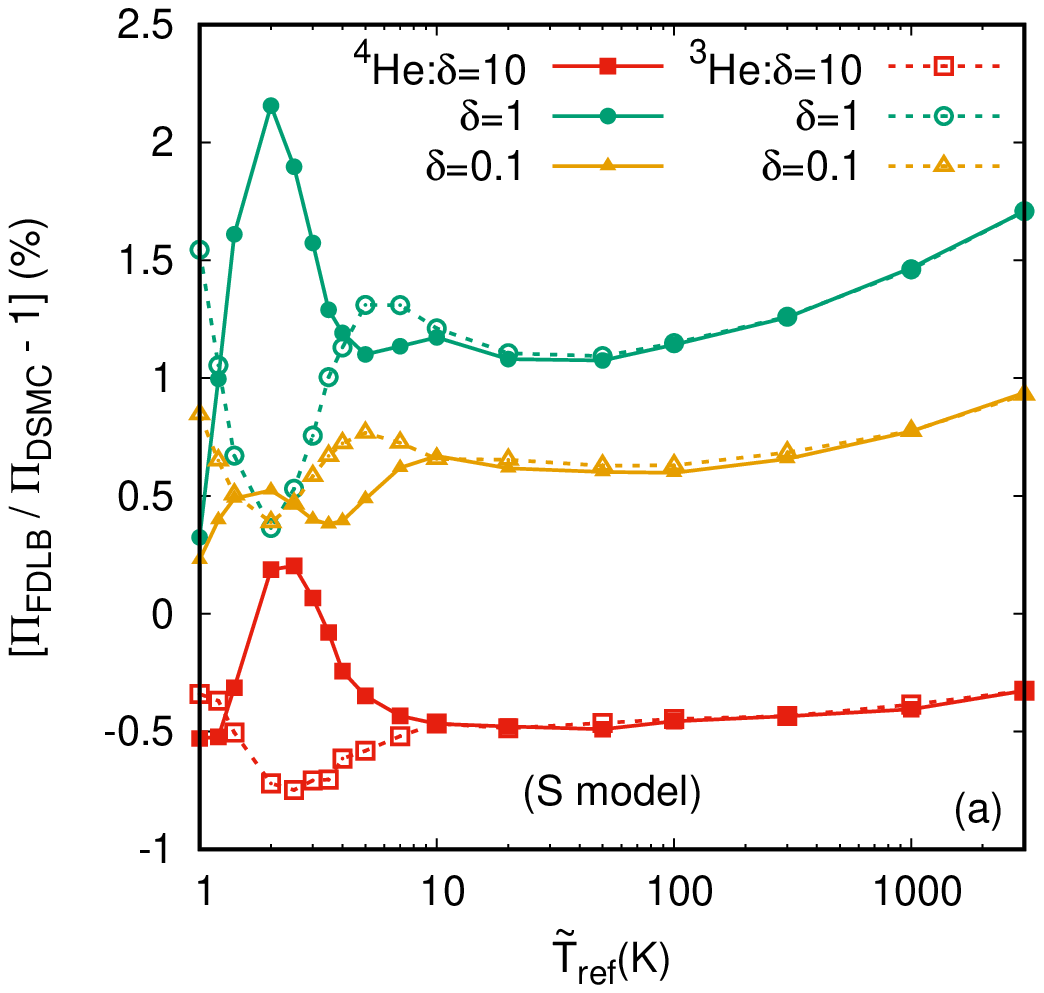} &
\includegraphics[angle=0,width=0.48\linewidth]{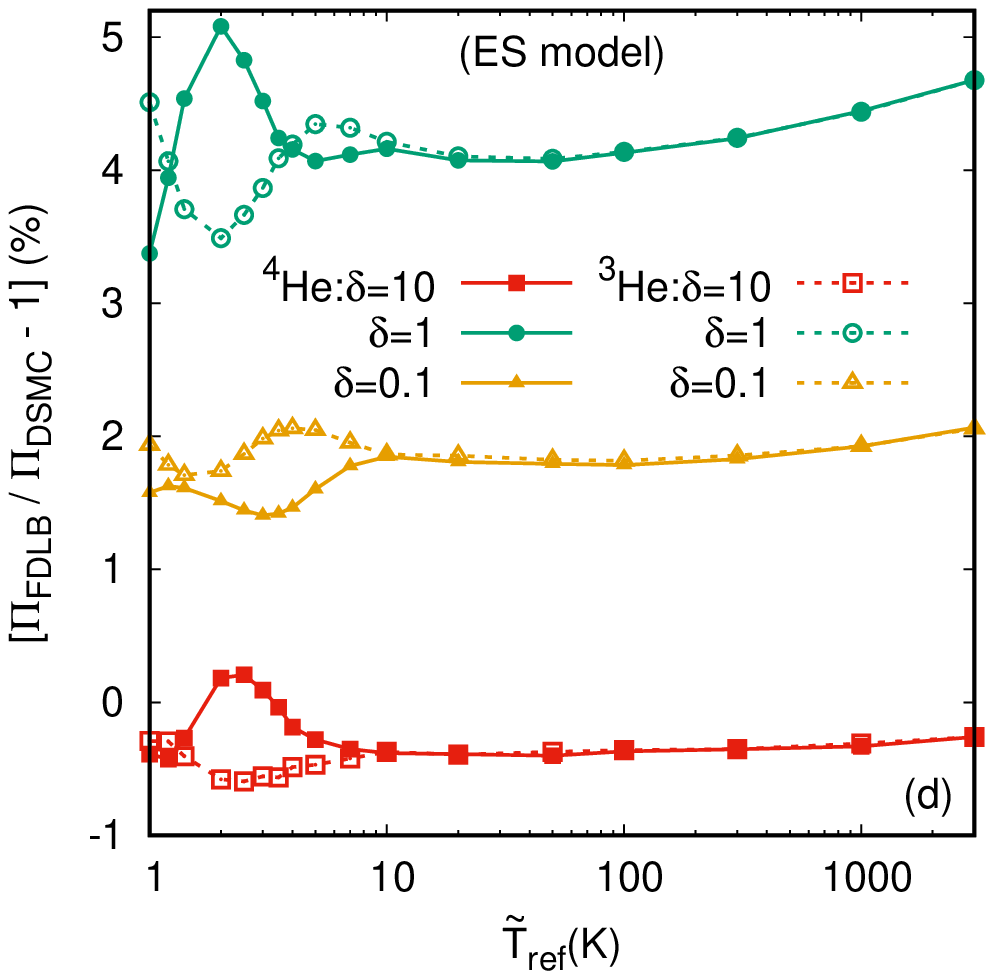} \\
\includegraphics[angle=0,width=0.48\linewidth]{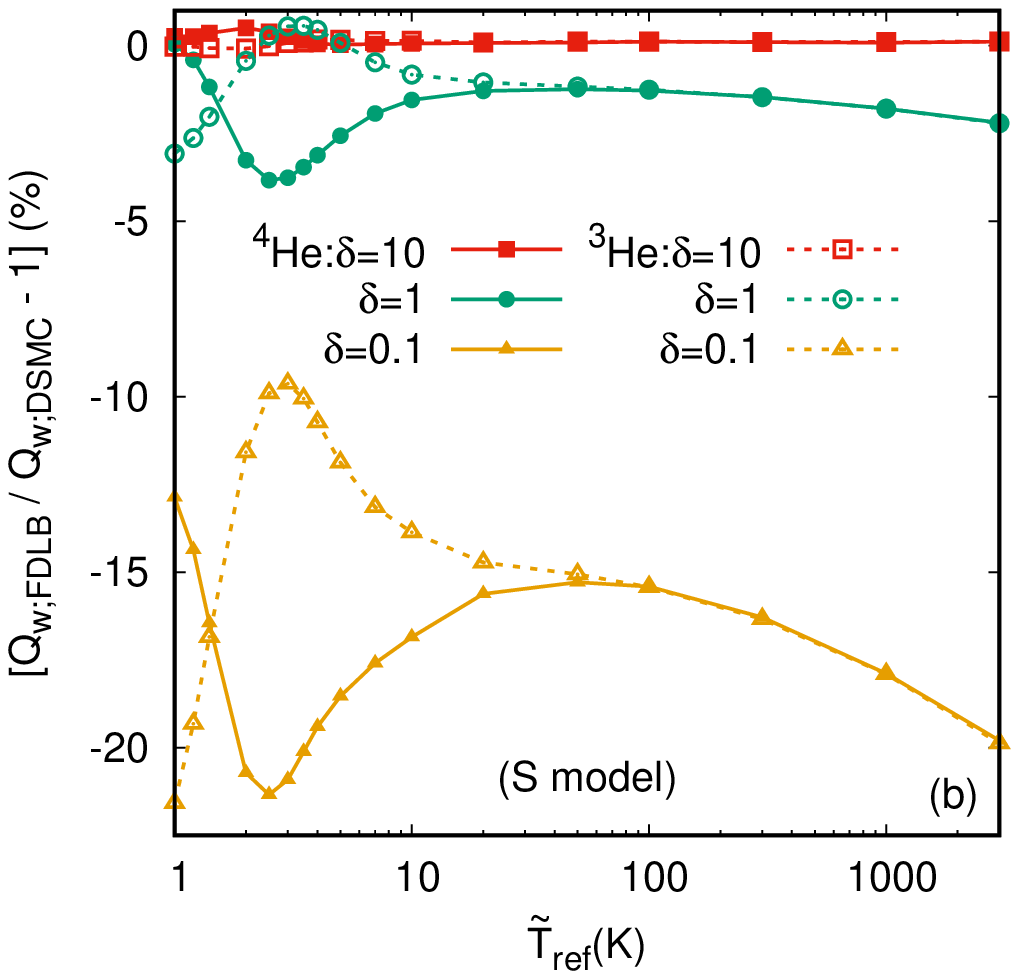} &
\includegraphics[angle=0,width=0.48\linewidth]{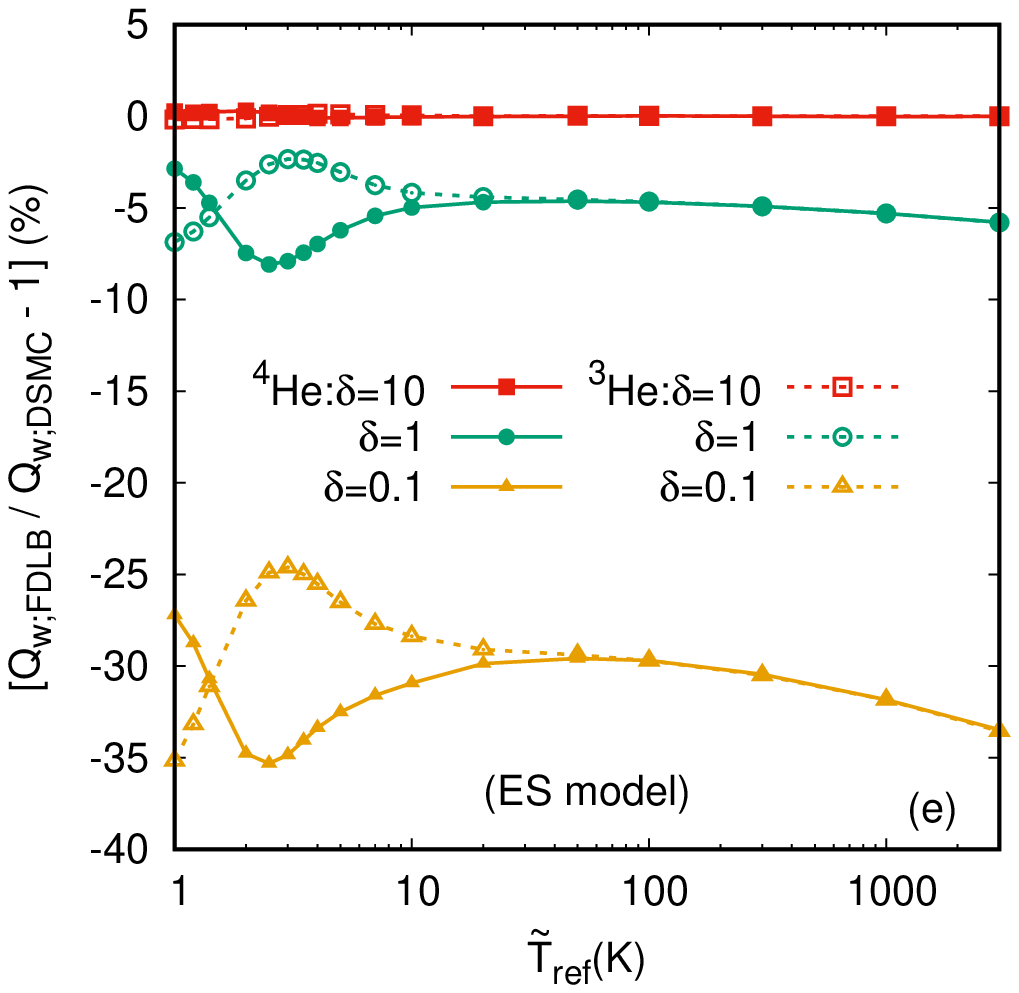} \\
\includegraphics[angle=0,width=0.48\linewidth]{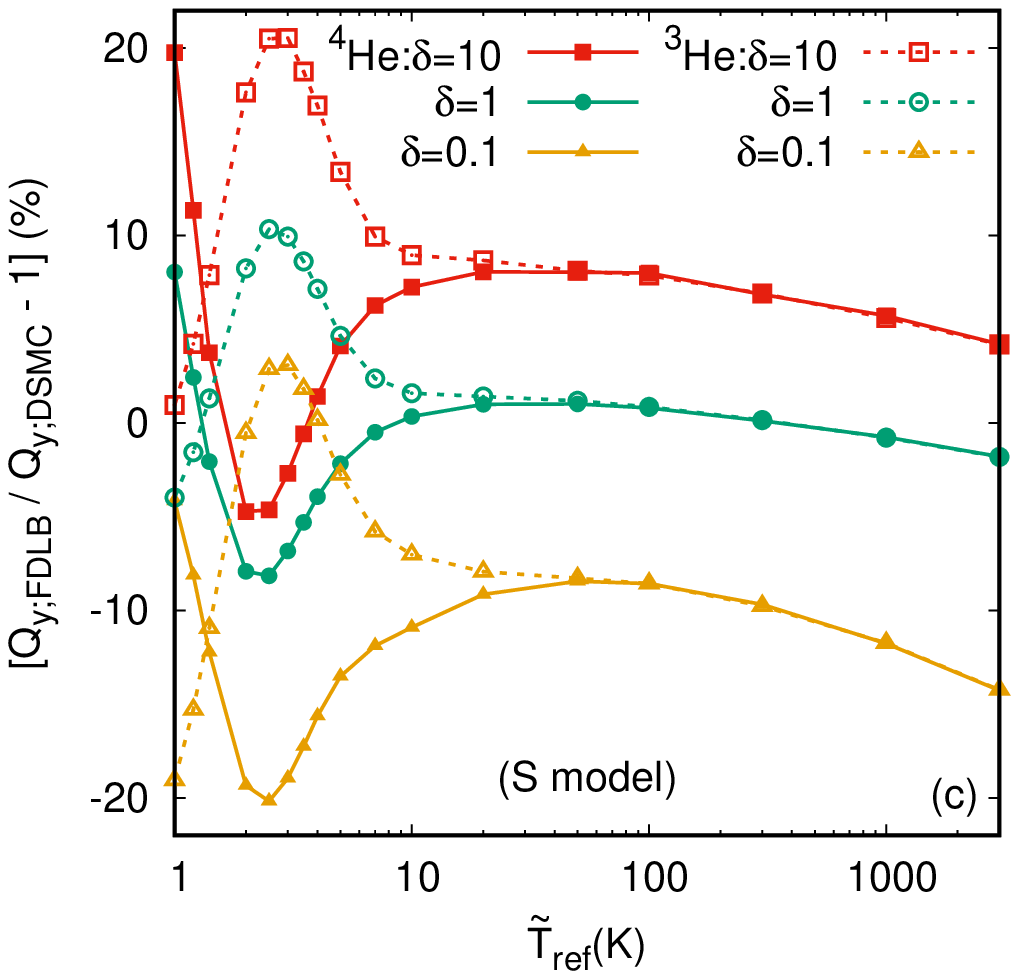} &
\includegraphics[angle=0,width=0.48\linewidth]{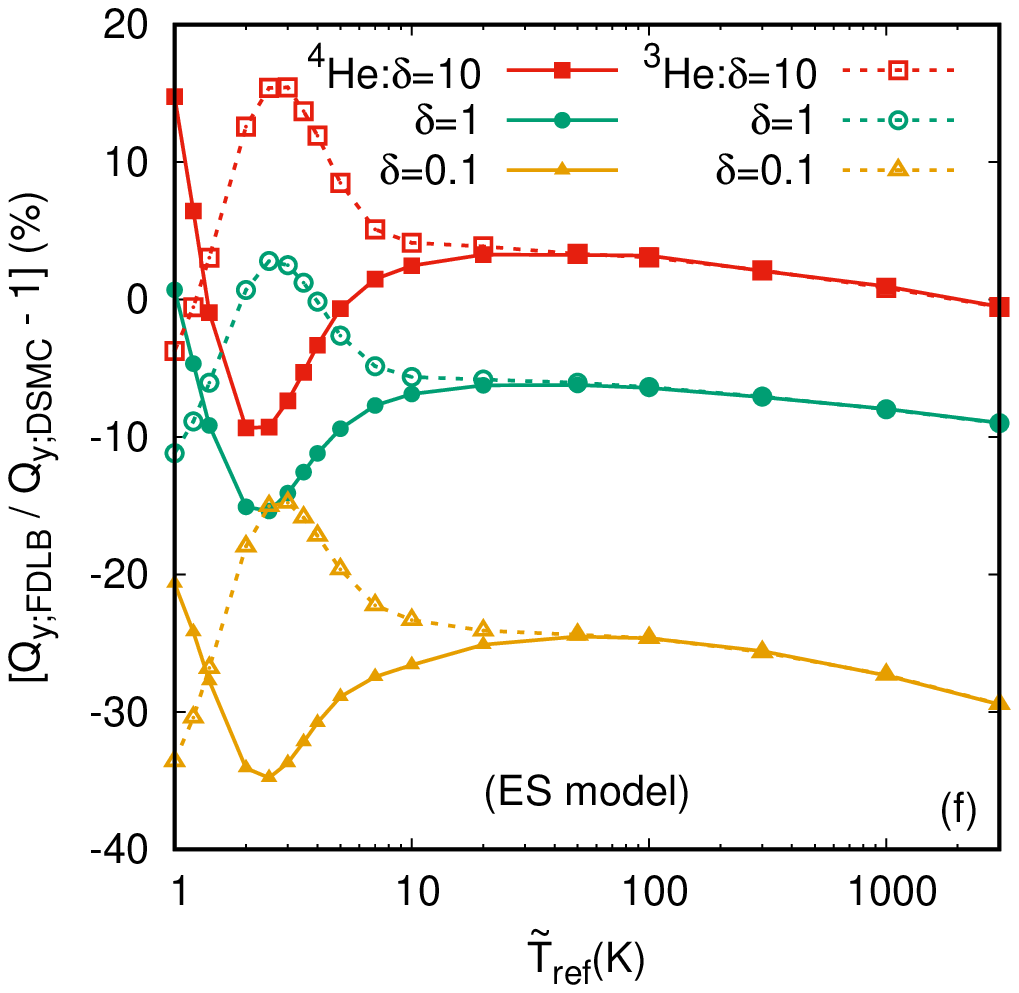} 
\end{tabular}
\end{center}
\caption{Relative errors
$\Pi_{\rm FDLB} / \Pi_{\rm DSMC} - 1$ (top),
$Q_{w;{\rm FDLB}} / Q_{w; {\rm DSMC}} - 1$ (middle) and
$Q_{y; {\rm FDLB}} / Q_{y; {\rm DSMC}} - 1$ (bottom) between the 
DSMC and FDLB results for the S model (left) and ES model (right),
at $\delta = 10$ (squares), $1$ (circles) and $0.1$ (triangles) 
for $1\ {\rm K} \le \widetilde{T}_{\rm ref} \le 3000\ {\rm K}$,
computed in the context of the Couette flow.
\label{fig:couette_err}}
\end{figure}

The second application concerns the Couette flow between 
parallel plates. Due to the symmetry of the flow, only 
the right half of the channel ($0 \le x \le 1/2$) is considered
in the simulation setup, as shown in Fig.~\ref{fig:couette_setup}. 
The walls are kept at constant temperatures $\widetilde{T}_{\rm left} = 
\widetilde{T}_{\rm right} = \widetilde{T}_{\rm ref}$ and 
$\widetilde{T}_{\rm ref}$ is varied bewteen $1\ {\rm K}$ and 
$3000\ {\rm K}$. The wall velocity 
$\widetilde{u}_w = \sqrt{2 \widetilde{K}_B \widetilde{T}_{\rm ref} / 
\widetilde{m}}$ takes the value $u_w = \sqrt{2}$ after 
non-dimensionalization.

Aside from the transversal component $q_x$ of the heat flux, which can
be related at large $\delta$ to the temperature variations with respect 
to the coordinate $x$ via Fourier's law, $q_x = -\kappa \partial_x T$,
the Couette flow exhibits a non-vanishing longitudinal heat flux, $q_y$,
which is a purely microfluidics effect. 
Figure~\ref{fig:couette_profiles} shows a comparison between 
the FDLB results for the S (dashed red lines and empty symbols)
and ES (dotted black lines and filled symbols) models and the 
DSMC results (solid purple lines). The wall temperature is set to
$\widetilde{T}_{\rm ref} = 300\ {\rm K}$ and ${}^4{\rm He}$ 
gas constituents are considered for $\delta = 10$, $1$ and $0.1$.
Both the S and ES models are in good agreement with the DSMC 
data at $\delta = 10$. When $\delta$ decreases, the agreement 
deteriorates, being slightly worse in the case of the 
ES model. Remarkably, the density profiles are well recovered 
with both models at all tested values of $\delta$.

We now consider a more quantitative analysis at the level 
of $\Pi$, $Q_w$ and $Q_y$, computed via Eqs.~\eqref{eq:Pi},
\eqref{eq:Qw} and \eqref{eq:Qy}, respectively.
The variations with the plate temperature $\widetilde{T}_{\rm ref}$ 
of $\Pi$, $Q_w$ and $Q_y$ for 
${}^3{\rm He}$ and ${}^4{\rm He}$
are shown in Figs.~\ref{fig:couette_Pi}, \ref{fig:couette_Qw} and 
\ref{fig:couette_Qy} for 
(a) $\delta = 10$, (b) $\delta = 1$ and (c) $\delta = 0.1$.
Each plot shows curves corresponding to the 
S model (dashed lines with empty symbols), 
ES model (dotted lines with filled symbols) and 
DSMC (solid lines). The data corresponding to 
${}^3{\rm He}$ is shown using red squares, while the 
data for ${}^4{\rm He}$ is shown with black circles.
It can be seen that in general, the agreement between 
the results obtained with the model equations and the 
DSMC results deteriorates as $\delta$ is decreased. 
Contrary to the results obtained in the case of the 
heat transfer problem, the S model gives more accurate 
results compared to the ES model, confirming the results 
reported in Ref.~\cite{meng13jcp}. 
Figure~\ref{fig:couette_err} shows the relative errors
computed with respect to the DSMC results, obtained
with the S (left column) and ES (right column) models.
The results for ${}^4{\rm He}$ are shown with solid 
lines and filled symbols, while those for ${}^3{\rm He}$
are shown with dashed lines and empty symbols.
The data corresponding to $\delta = 10$, $1$ and $0.1$ 
are shown with red squares, green circles and 
amber triangles, respectively. In the case of $\Pi$, the 
relative error of the ES model is roughly twice that of the 
S model. 

It is remarkable that the relative errors for both 
$Q_w$ and $Q_y$ (shown in Figs.~\ref{fig:couette_Qw} and 
\ref{fig:couette_Qy}) reach values around $20\%$
for $\delta =0.1$.
This can be explained since the heat fluxes decrease to $0$ as 
$\delta$ is decreased, while $\Pi$, for which the relative error 
is below $5\%$, attains a finite value as the ballistic regime 
is approached ($\lim_{\delta \rightarrow 0} \Pi = \pi^{-1/2}$). 
Thus, the relative errors for $Q_w$ and $Q_y$ are computed 
by dividing the FDLB values by small numbers.
However, in the case of $Q_y$, the errors are around $20\%$ 
even when $\delta = 10$, whereas for both $Q_w$ and $\Pi$, 
the error at $\delta = 10$ is less than $1\%$. This disagreement 
between the model equations and the DSMC data can be attributed 
to the nature of $Q_y$. Since the longitudinal heat flux, $q_y$, 
is not generated by a temperature gradient (through the 
so-called direct phenomenon), its characteristics must 
depend on higher order transport coefficients, which are 
visible only at the Burnett level \cite{marques00}.
Since the model equations are constructed to ensure consistency 
only at the Navier-Stokes level (corresponding to the first order 
in the Chapman-Enskog expansion),
it is not surprising that such cross phenomena are not 
accurately recovered.

\section{Heat transfer under shear}\label{sec:htsh}

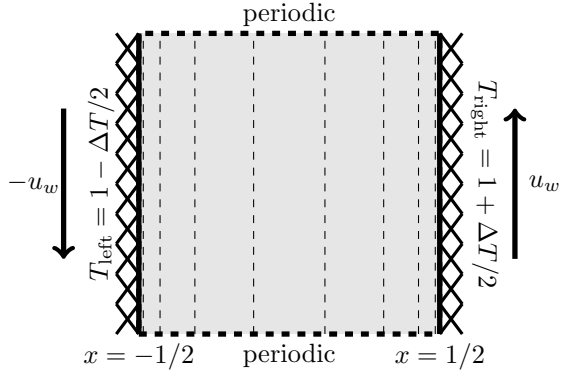
\begin{figure}
\begin{center}
\begin{tikzpicture}
\fill [black!10!white] (0,0) rectangle (4,4);
\draw [line width=2,dashed] (0,0) rectangle (4,4);
\draw [line width=2] (4,0) -- (4,4);
\draw [line width=2] (0,0) -- (0,4);

\draw [line width=2,->] (5,1) -- (5,3);
\draw [line width=2,->] (-1,3) -- (-1,1);

\foreach \y in {1,2,...,10} {
    \draw [line width=1.2] (4,\y*0.4-0.4) -- (4.3,\y*0.4);
    \draw [line width=1.2] (-0.3,\y*0.4-0.4) -- (0,\y*0.4);
}
\foreach \y in {1,2,...,10} {
    \draw [line width=1.2] (4.3,\y*0.4-0.4) -- (4,\y*0.4);
    \draw [line width=1.2] (0,\y*0.4-0.4) -- (-0.3,\y*0.4);
}
\draw [dashed] (0.059,0)--(0.059,4);
\draw [dashed] (0.282,0)--(0.282,4);
\draw [dashed] (0.745,0)--(0.745,4);
\draw [dashed] (1.526,0)--(1.526,4);

\draw [dashed] (2.474,0)--(2.474,4);
\draw [dashed] (3.255,0)--(3.255,4);
\draw [dashed] (3.718,0)--(3.718,4);
\draw [dashed] (3.941,0)--(3.941,4);

\node [rotate=90] at (-0.5,2) {$T_{\rm left} = 1 - \Delta T / 2$};
\draw (0.0,0.0) node [anchor=north] {$x=-1/2$};
\node at (2,4.25) {periodic};
\node [rotate=270] at (4.6,2) {$T_{\rm right} = 1 + \Delta T / 2$};
\node at (5.4,2) {$u_w$};
\node at (-1.4,2) {$-u_w$};
\draw (4,0) node [anchor=north] {$x=1/2$};
\draw (2,0) node [anchor=north] {periodic};
\end{tikzpicture} 
\end{center}
\caption{The simulation setup for the heat transfer under shear 
problem. The vertical dashed lines show a sample grid 
employing $S = 4$ points on each half of the channel, 
stretched according to Eq.~\eqref{eq:stretch} with $A = 0.95$.
\label{fig:htsh_setup}}
\end{figure}

\begin{figure}
\begin{center}
\begin{tabular}{cc}
\includegraphics[angle=0,width=0.48\linewidth]{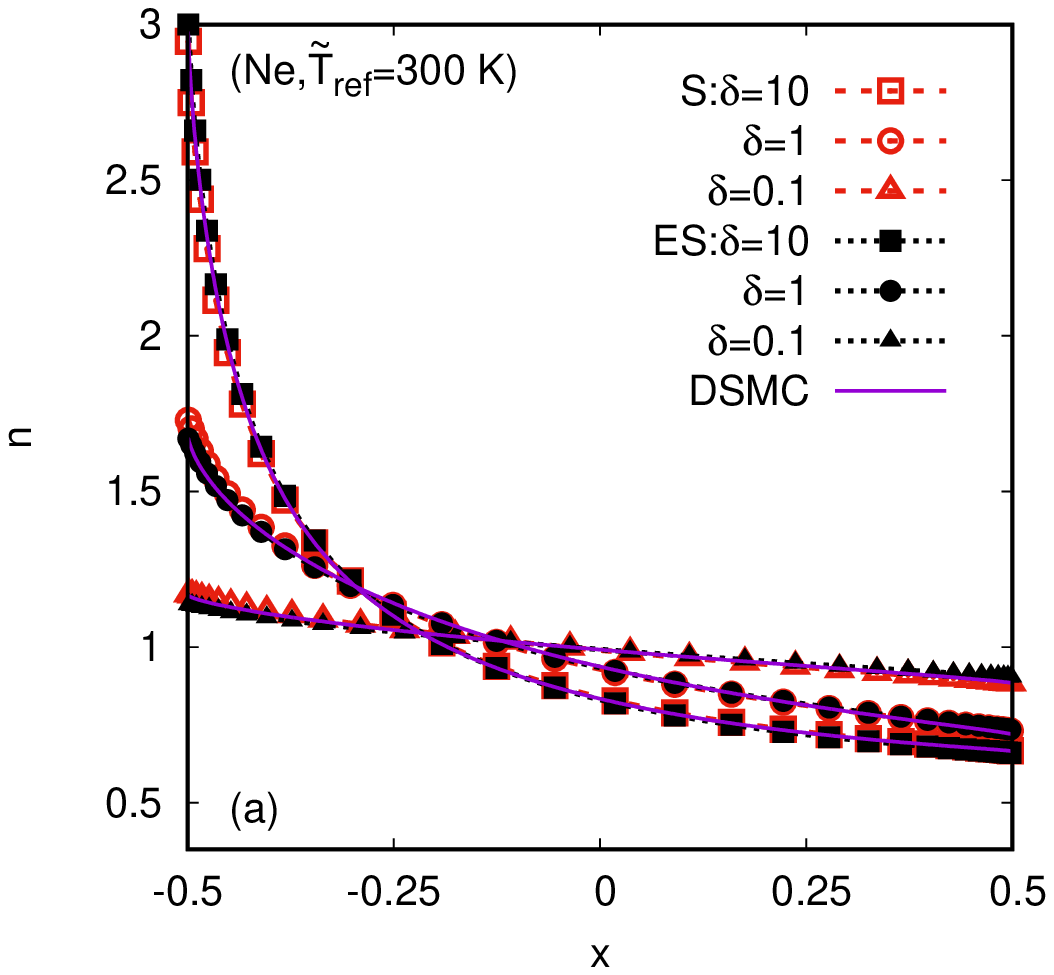} &
\includegraphics[angle=0,width=0.48\linewidth]{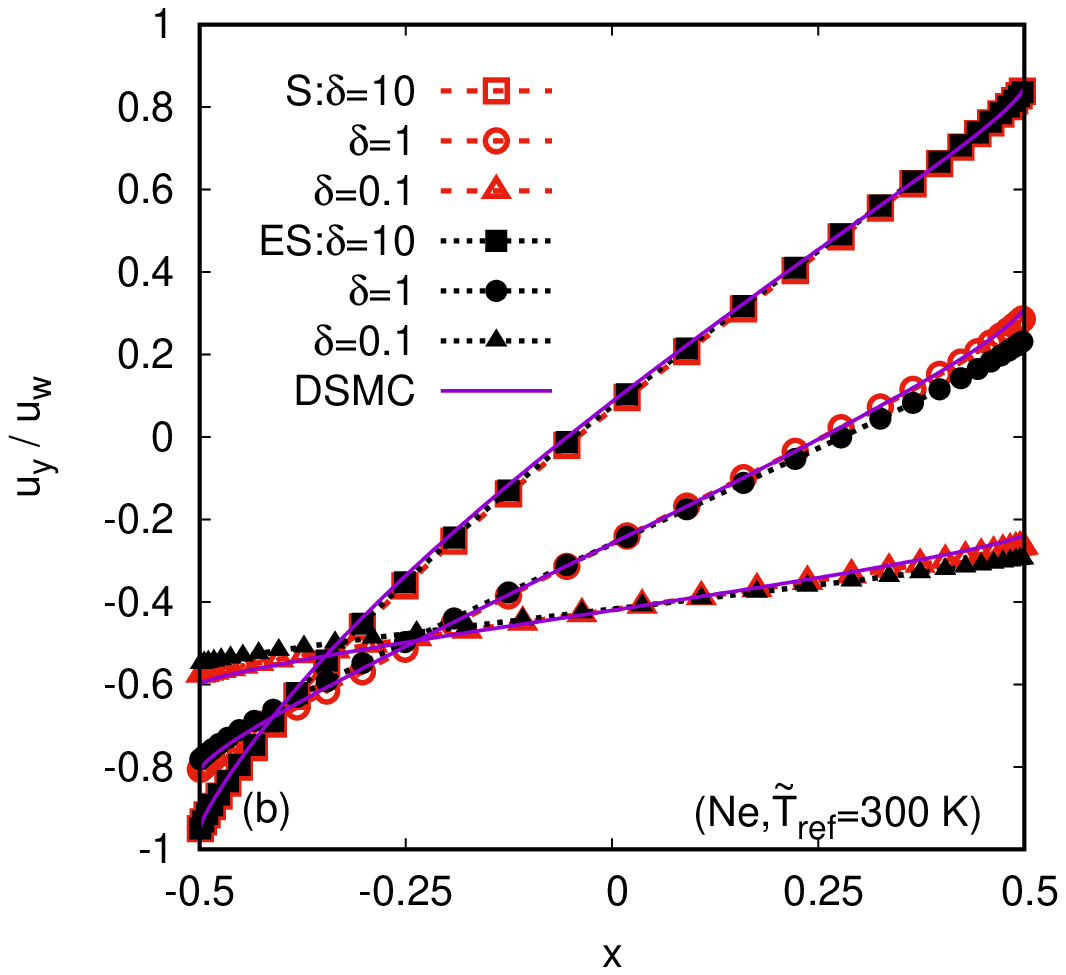} 
\end{tabular}
\begin{center}
\end{center}
\begin{tabular}{c}
\includegraphics[angle=0,width=0.48\linewidth]{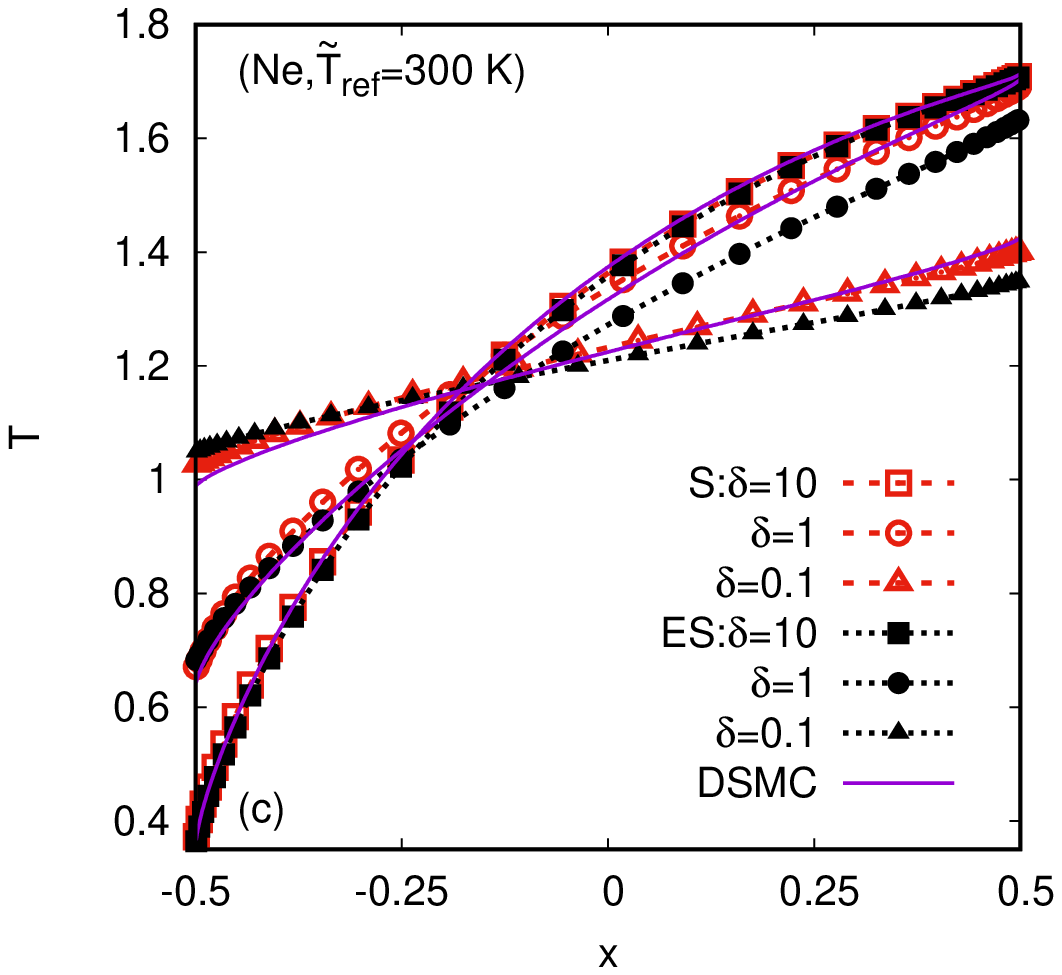} 
\end{tabular}
\end{center}
\caption{Comparison between the FDLB results (dotted lines and points) 
obtained using the S (red empty symbols) and ES (black filled symbols)
models and the DSMC (continuous lines) results for the profiles of $n$ (a),
$u_y$ (b) and $T$ (c) through the channel
($-1/2 \le x\le 1/2$), for ${\rm Ne}$ gas constituents, in the 
context of the heat transfer between moving plates problem.
The reference temperature is set to
$\widetilde{T}_{\rm ref} = 300\ {\rm K}$,
the temperature difference 
between the two walls is $\widetilde{\Delta T} = 1.5 \widetilde{T}_{\rm ref}$
and the wall velocity is $\widetilde{u}_w = \sqrt{2 \widetilde{K}_B 
\widetilde{T}_{\rm ref} / \widetilde{m}}$.
\label{fig:htsh_profiles}}
\end{figure}

\begin{figure}
\begin{tabular}{cc}
\includegraphics[angle=0,width=0.48\linewidth]{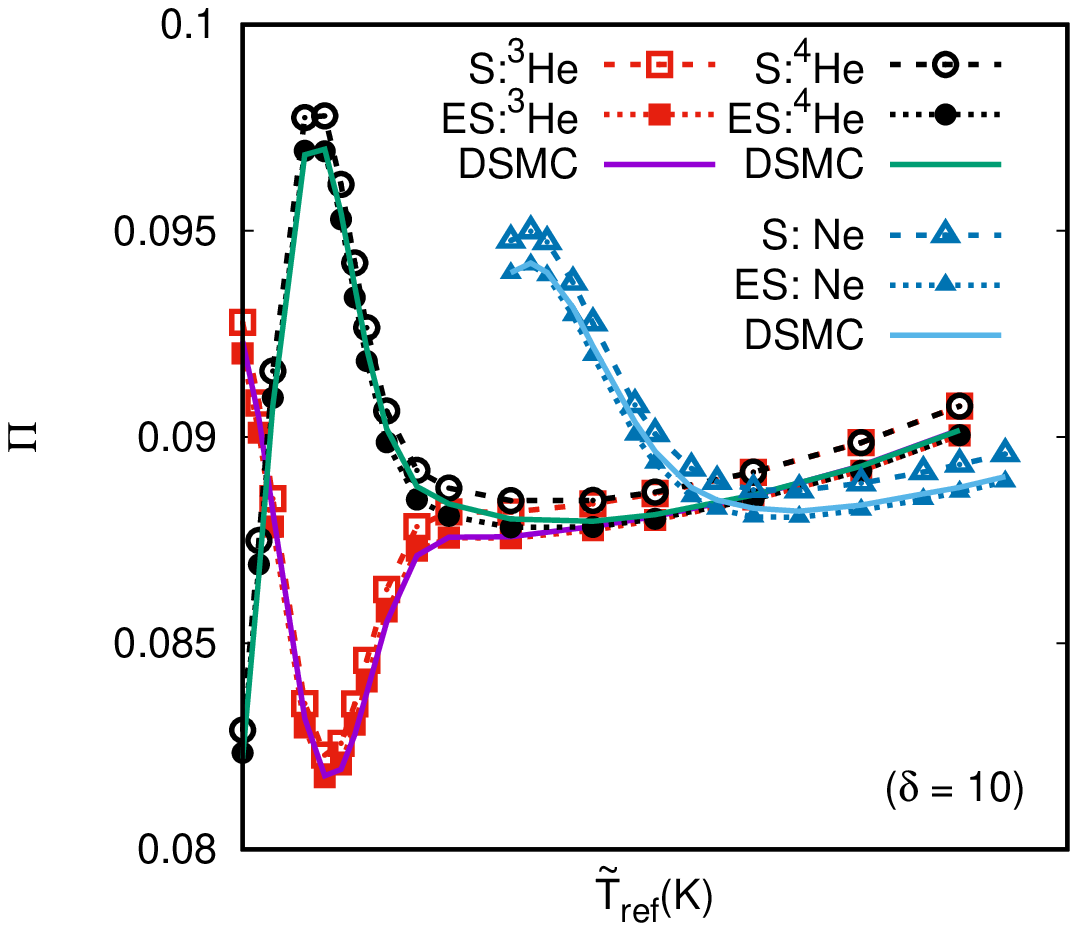} &
\includegraphics[angle=0,width=0.48\linewidth]{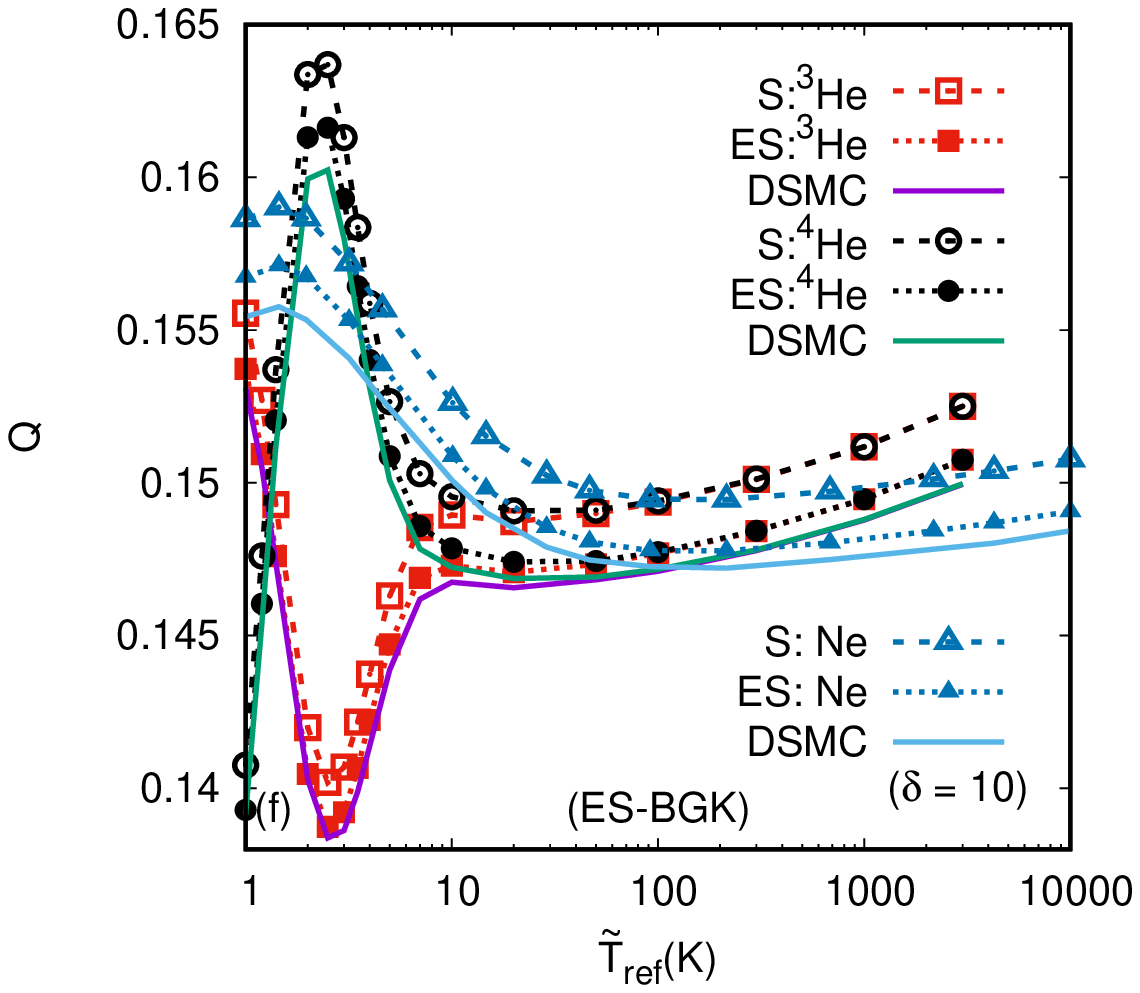} \\
\includegraphics[angle=0,width=0.48\linewidth]{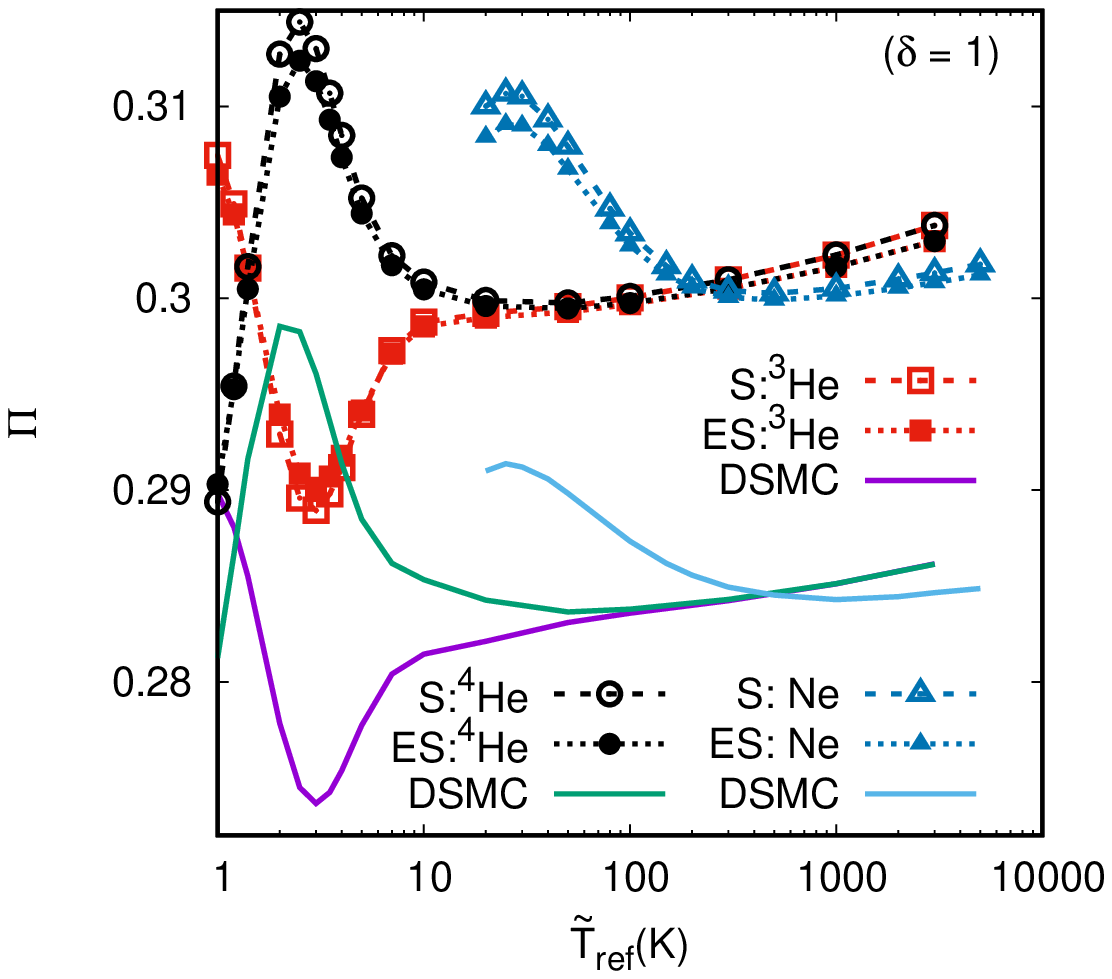} &
\includegraphics[angle=0,width=0.48\linewidth]{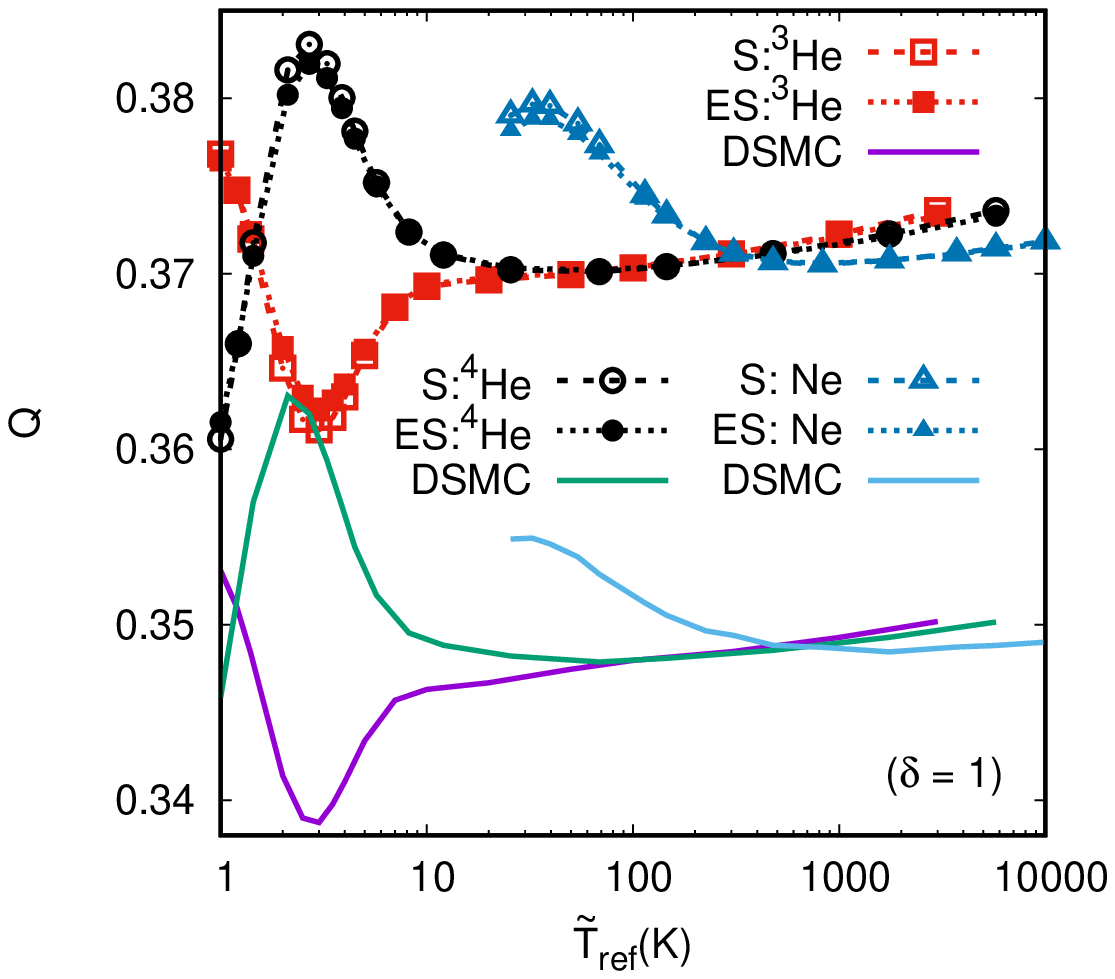} \\
\includegraphics[angle=0,width=0.48\linewidth]{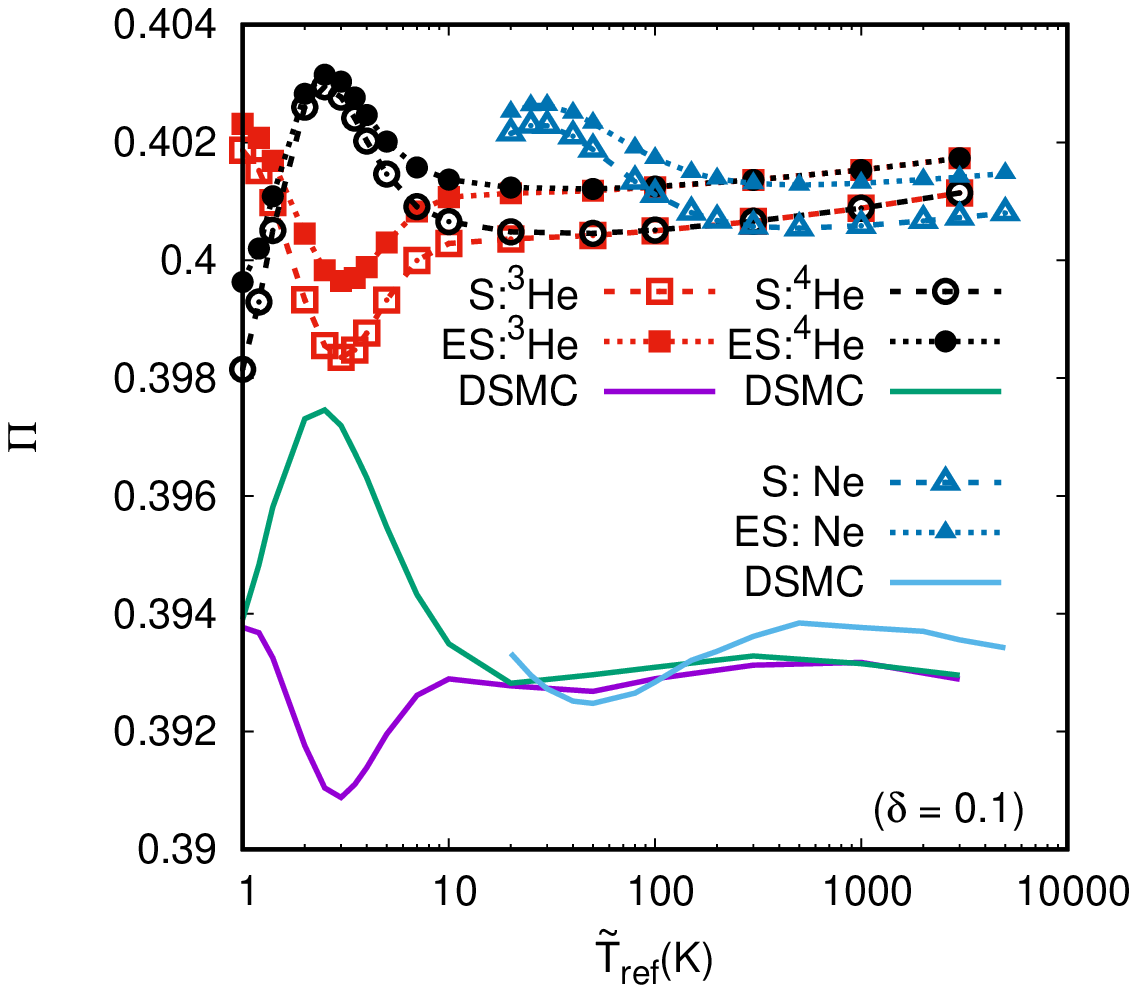} &
\includegraphics[angle=0,width=0.48\linewidth]{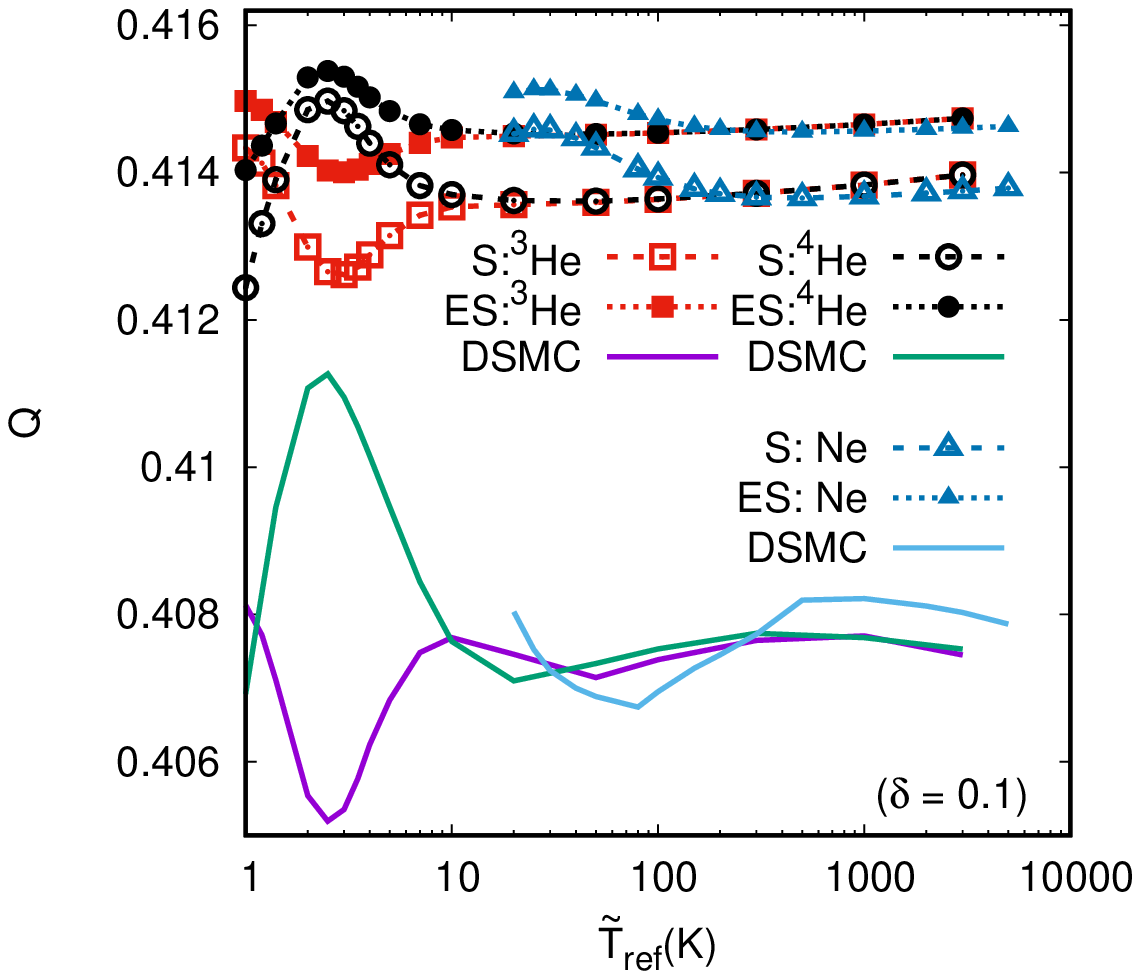}
\end{tabular}
\caption{Dependence of $\Pi$ (left column) and $Q$ (right column), defined in
Eqs.~\eqref{eq:Pi} and \eqref{eq:Q} for the heat transfer between moving 
plates problem, on the average wall temperature
$\widetilde{T}_{\rm ref} = (\widetilde{T}_{\rm left} + \widetilde{T}_{\rm right}) / 2$
for ${}^3{\rm He}$ (red squares), ${}^4{\rm He}$ (black circles) and 
${\rm Ne}$ (blue triangles), at $\delta = 10$ (top line),
$1$ (middle line) and $0.1$ (bottom line). \label{fig:htsh_PiQ}}
\end{figure}

The final example considered in this paper is the heat transfer 
between parallel plates in motion. The simulation setup is represented 
in Fig.~\ref{fig:htsh_setup}.
This example combines the features of the heat transfer 
between stationary plates discussed in Sec.~\ref{sec:ht} and 
those of the Couette flow discussed in Sec.~\ref{sec:couette}.
The reference temperature $\widetilde{T}_{\rm ref} = (\widetilde{T}_{\rm left} + 
\widetilde{T}_{\rm right}) / 2$, is varied between 
$1\ {\rm K}$ and $3000\ {\rm K}$ for ${}^3{\rm He}$ 
and ${}^4{\rm He}$ constituents, while for ${\rm Ne}$,
the range for $\widetilde{T}_{\rm ref}$ is 
$20\ {\rm K} \le \widetilde{T}_{\rm ref} \le 5000\ {\rm K}$. 
As in Sec.~\ref{sec:ht}, the temperature difference 
$\widetilde{\Delta T} = \widetilde{T}_{\rm right} - 
\widetilde{T}_{\rm left}$ obeys Eq.~\eqref{eq:DT}.
Furthermore, the plates have velocities 
$\widetilde{\bm{u}}_{\rm left} = -\widetilde{u}_w \bm{j}$ 
and $\widetilde{\bm{u}}_{\rm right} = \widetilde{u}_w \bm{j}$, 
where $\widetilde{u}_w = \sqrt{2 \widetilde{K}_B \widetilde{T}_{\rm ref} / 
\widetilde{m}}$, such that the Mach number is given by Eq.~\eqref{eq:Ma}.

Figure~\ref{fig:htsh_profiles} shows the 
profiles of the density (a), velocity (b) and 
temperature (c) for the case of ${\rm Ne}$ constituents 
at $T = 300\ {\rm K}$.
In general, good agreement can be seen between the results 
corresponding to the model equations and the DSMC results.
A larger discrepancy can be seen between the ES model 
and the DSMC results, especially in the temperature profile 
at $\delta = 1$ and $0.1$.

\begin{figure}
\begin{center}
\begin{tabular}{cc}
\includegraphics[angle=0,width=0.48\linewidth]{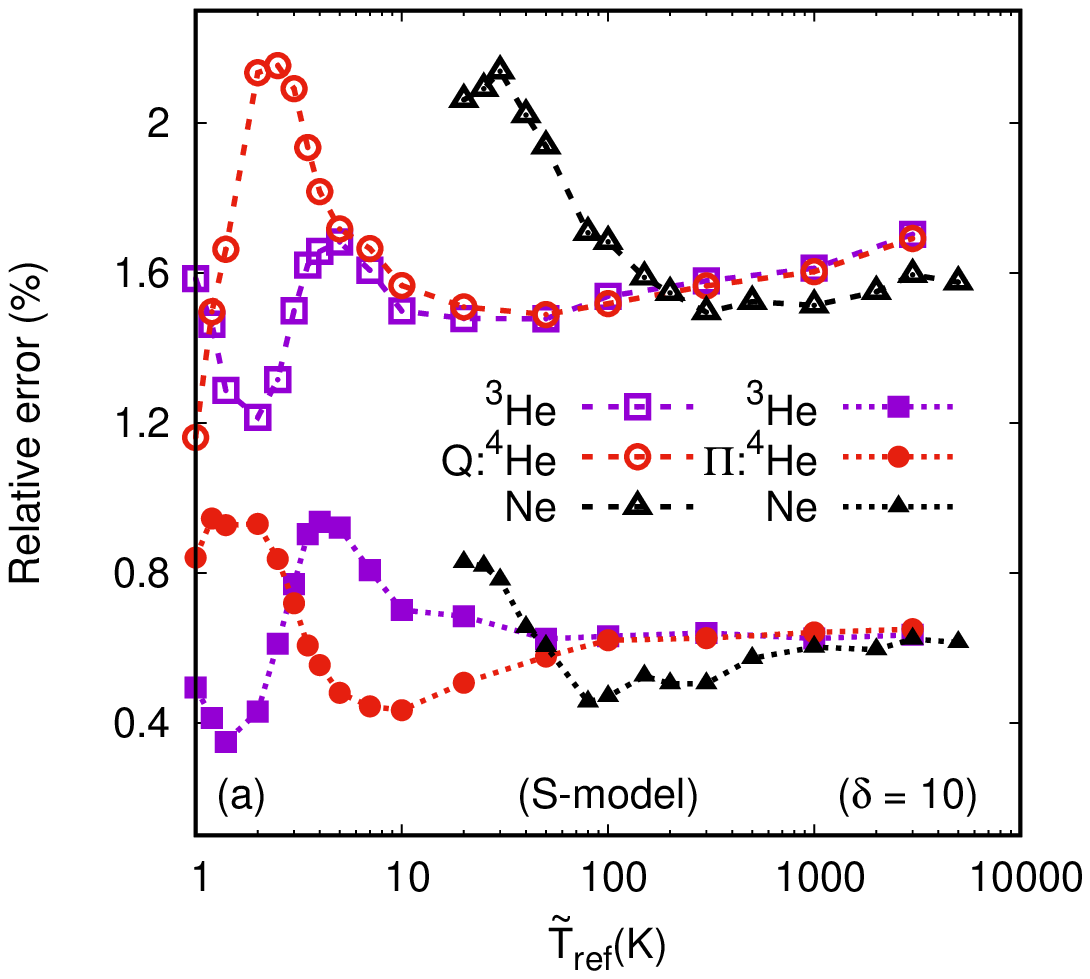} &
\includegraphics[angle=0,width=0.48\linewidth]{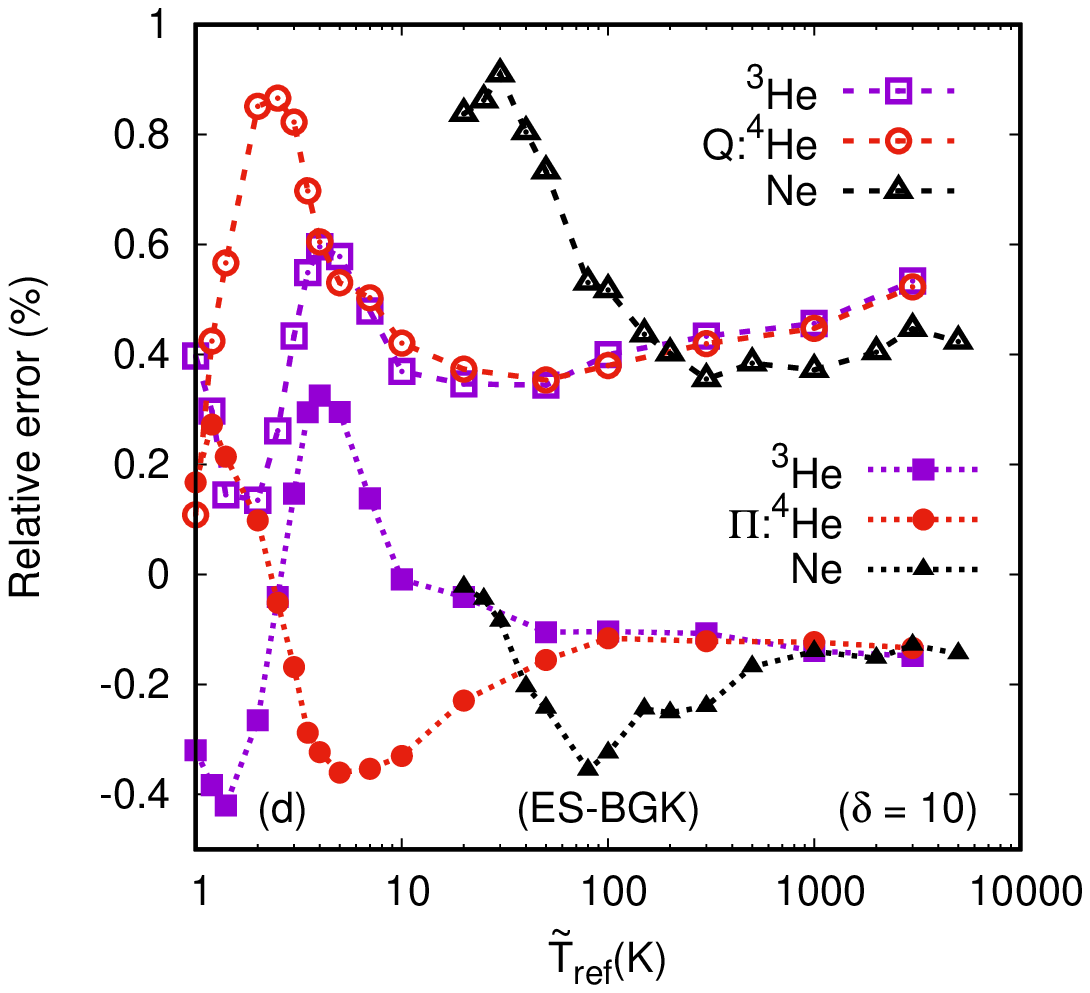} \\
\includegraphics[angle=0,width=0.48\linewidth]{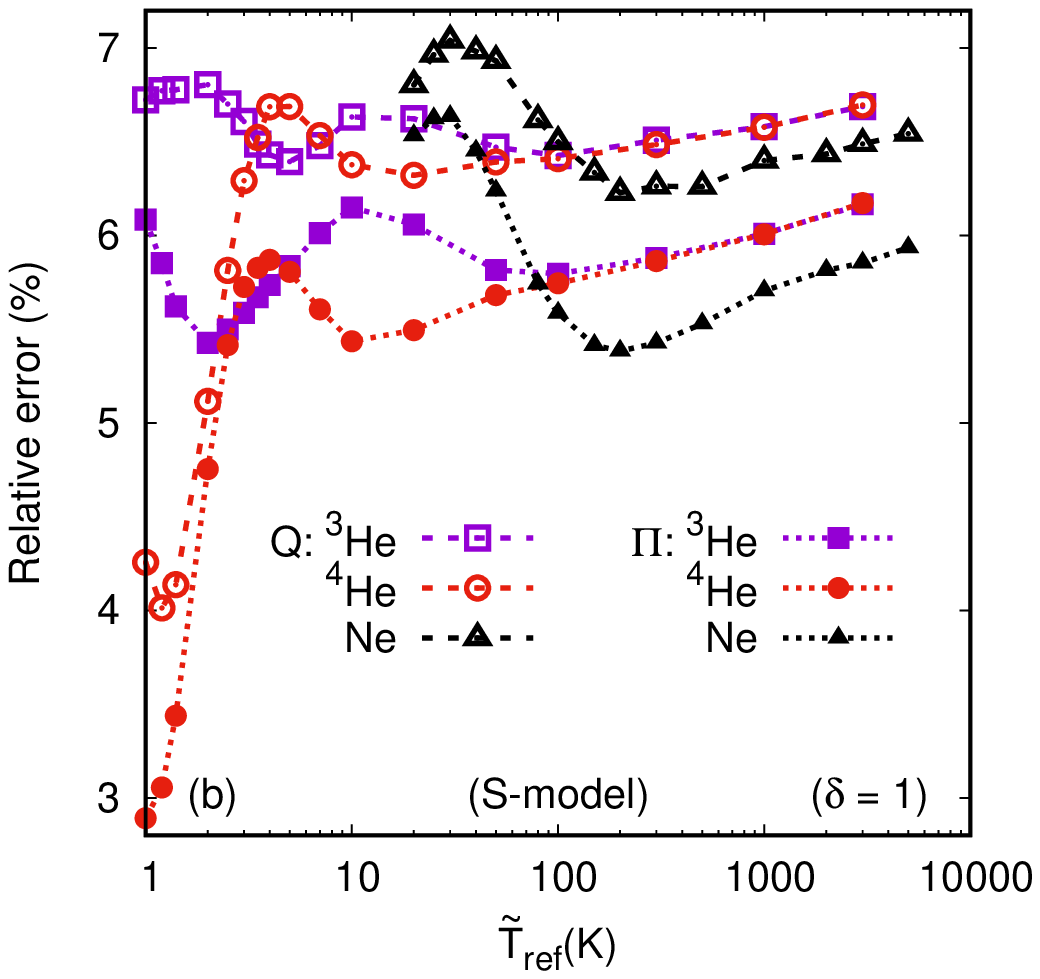} &
\includegraphics[angle=0,width=0.48\linewidth]{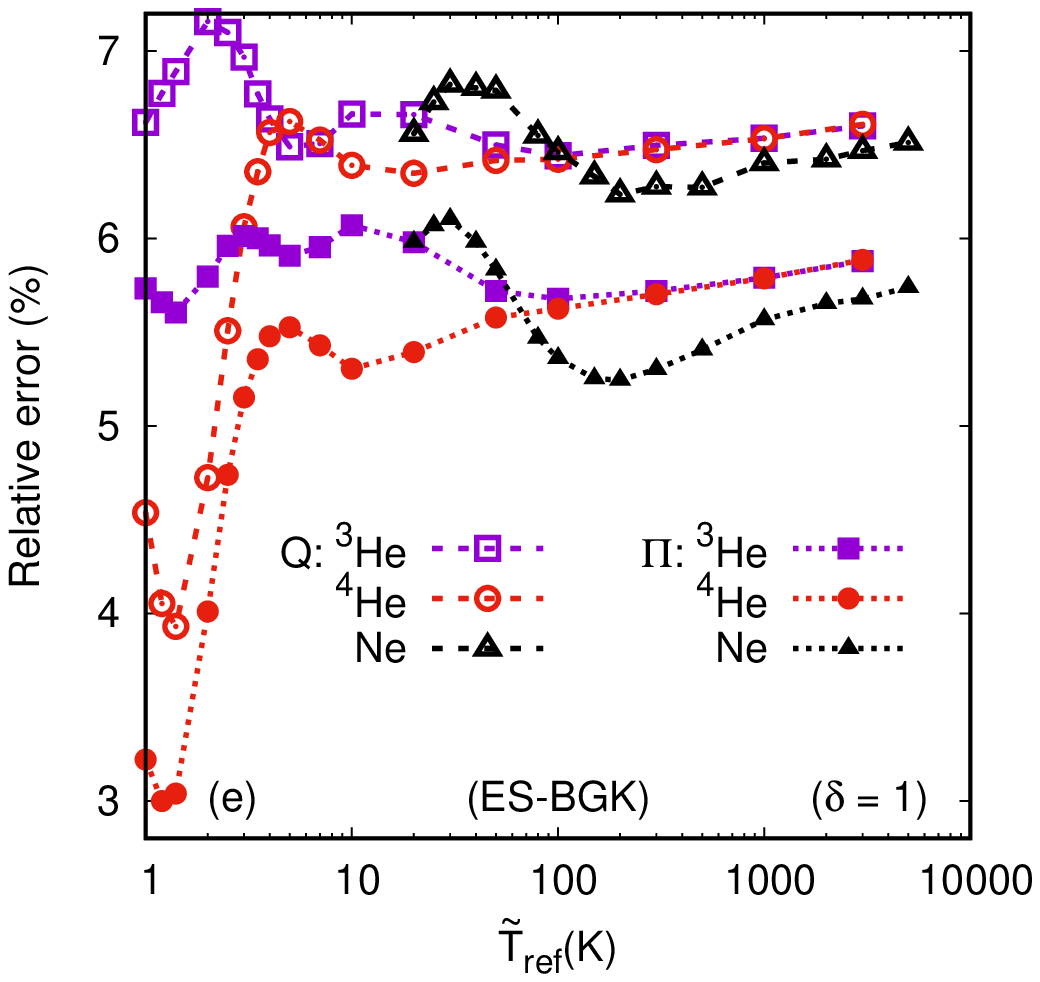} \\
\includegraphics[angle=0,width=0.48\linewidth]{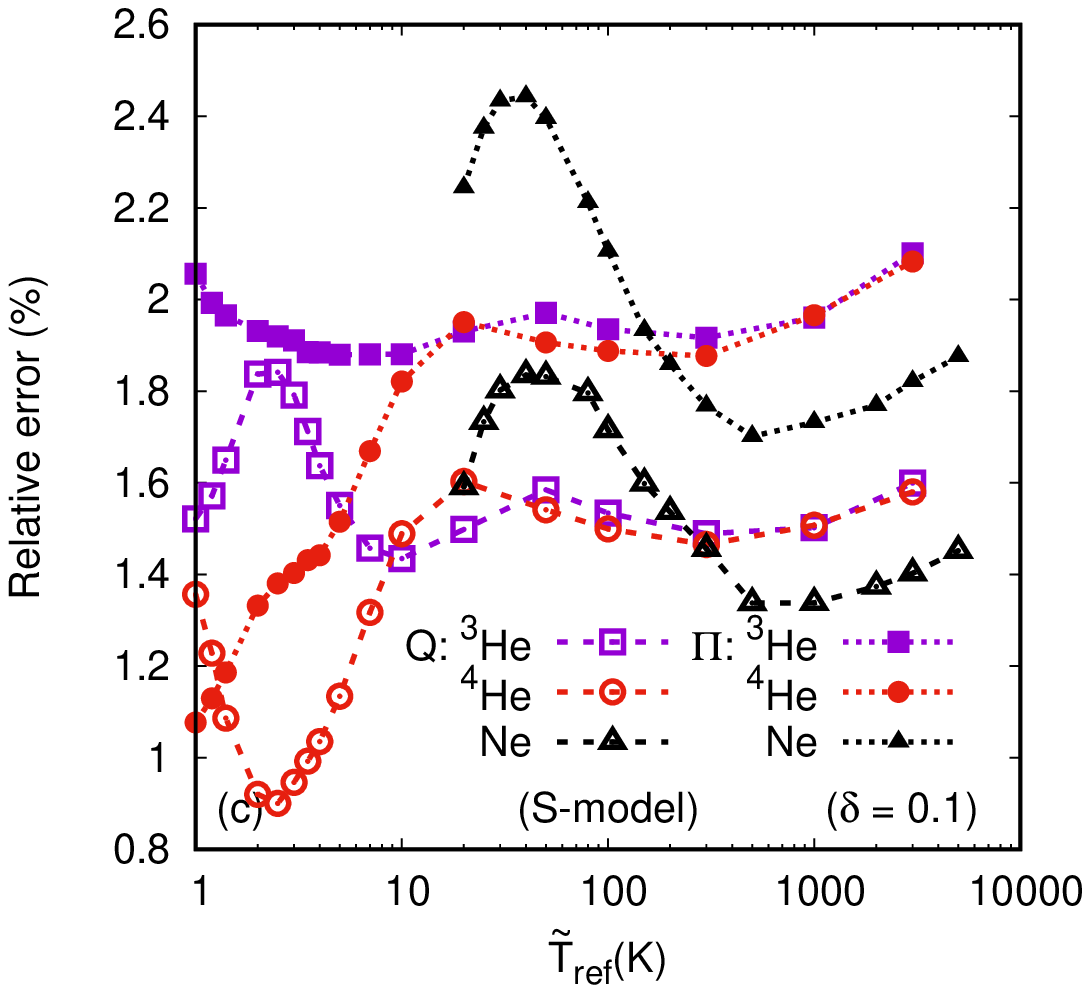} &
\includegraphics[angle=0,width=0.48\linewidth]{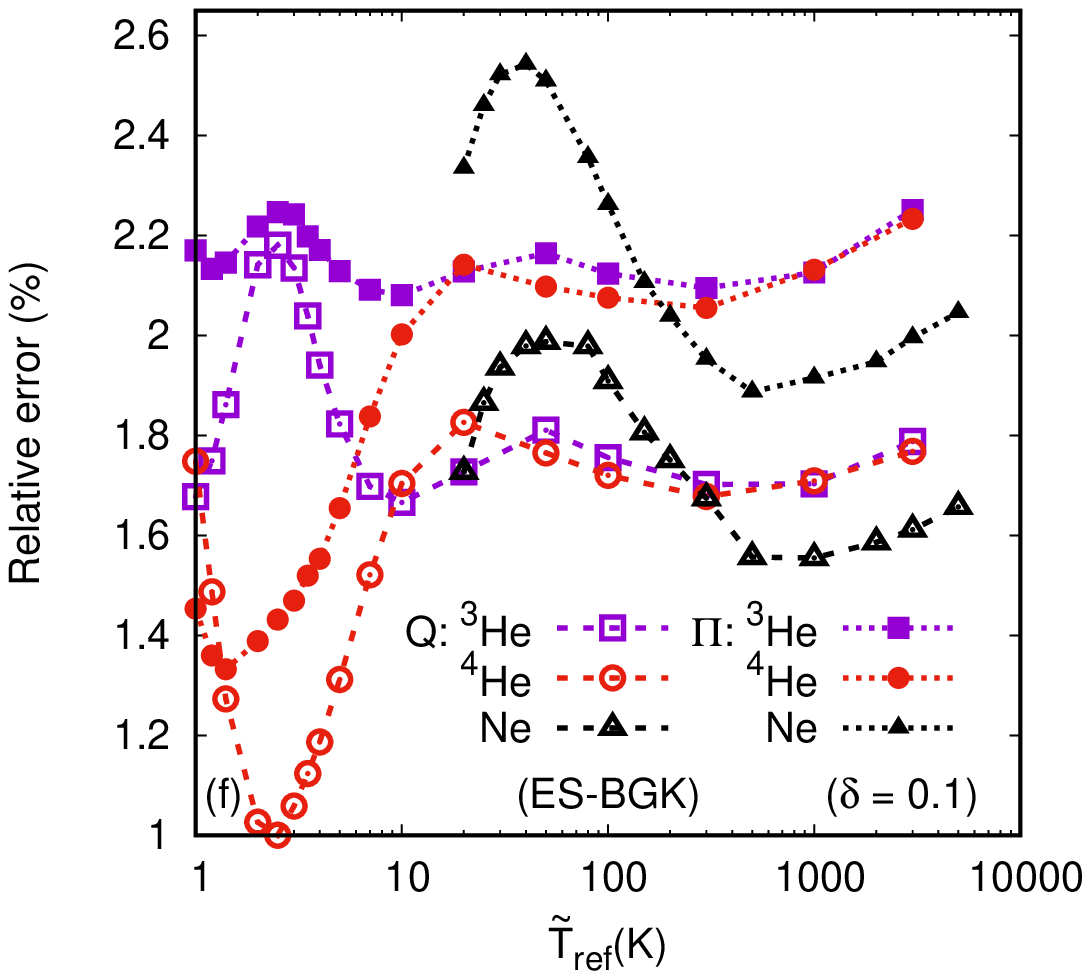}
\end{tabular}
\end{center}
\caption{Dependence of the relative errors 
$Q_{\rm FDLB} / Q_{\rm DSMC} - 1$ (dashed lines and empty symbols) and 
$\Pi_{\rm FDLB} / \Pi_{\rm DSMC} - 1$ (dotted lines and filled symbols),
expressed in percentages,
where the FDLB results are obtained using the S (left column) and 
ES (right column) models, for the heat transfer between moving 
plates problem, on the average wall temperature
$T_{\rm ref} = (T_{\rm left} + T_{\rm right}) / 2$
for ${}^3{\rm He}$ (squares), ${}^4{\rm He}$ (circles) and 
${\rm Ne}$ (triangles), at $\delta = 10$ (top line),
$1$ (middle line) and $0.1$ (bottom line). \label{fig:htsh_err}}
\end{figure}

A quantitative analysis can be made at the level of
the nondimensional quantities $\Pi$ and $Q$, computed
using Eqs.~\eqref{eq:Pi} and \eqref{eq:Q}.
Figure~\ref{fig:htsh_PiQ} shows a comparison between the FDLB results for 
the S (dashed lines with empty symbols) and the ES (dotted 
lines with filled symbols) models and the DSMC results (solid lines), obtained 
for ${}^3{\rm He}$ (squares), ${}^4{\rm He}$ (circles),
and ${\rm Ne}$ (triangles) constituents.
Figure~\ref{fig:htsh_err} shows the relative errors in $Q$ (dashed lines 
and empty symbols) and $\Pi$ (dotted lines and filled symbols)
computed for the S model (left column) and ES model (right column) with 
respect to the DSMC results for ${}^3{\rm He}$ (squares), 
${}^4{\rm He}$ (circles) and ${\rm Ne}$ (triangles). At $\delta = 10$ (top line), 
the results obtained using the ES model seem to be in better agreement 
with the DSMC results than those obtained using the S model. At 
$\delta = 1$ (middle line) and $0.1$ (bottom line), the two models give results with similar
accuracy.
As noticed in the case of the heat transfer between stationary 
plates and in the case of the direct phenomena in the Couette flow,
the relative erros are highest at $\delta = 1$, where they take values 
between $6-8\%$ (about $1\%$ higher for $Q$ than for $\Pi$).

\section{Conclusion} \label{sec:conc}

In this paper, we presented a systematic comparison between the results obtained 
using the Boltzmann equation with the Shakhov (S) and Ellipsoidal-BGK (ES) 
models for the collision term and those obtained using the direct simulation 
Monte Carlo (DSMC) method for three benchmark channel flows between 
parallel plates, namely: heat transfer between static walls,
Couette flow and heat transfer under shear. 
The results were obtained numerically in the nonlinear 
regime [${\rm Ma} \simeq 2.19$ for the case when the parallel plates are 
moving and $2 (\widetilde{T}_{\rm right} - \widetilde{T}_{\rm left}) / 
(\widetilde{T}_{\rm right} + \widetilde{T}_{\rm left}) = 1.5$ for the 
heat transfer problems], by 
considering ${}^3{\rm He}$ and ${}^4{\rm He}$ 
constituents interacting via {\it ab initio} potentials. 
We also considered ${\rm Ne}$ constituents for the heat transfer under shear problem.

In the kinetic theory setup, the connection with the DSMC simulations
was established at the level of the transport coefficients 
(dynamic viscosity $\widetilde{\mu}$ and heat conductivity $\widetilde{\kappa}$).
For ${}^3{\rm He}$ and ${}^4{\rm He}$, the range of values for the reference 
temperature $\widetilde{T}_{\rm ref} = 
(\widetilde{T}_{\rm right} + \widetilde{T}_{\rm left}) / 2$ 
was $1\ {\rm K} \le \widetilde{T}_{\rm ref} \le 3000\ {\rm K}$, 
while for the ${\rm Ne}$ constituents, it was 
$20\ {\rm K} \le \widetilde{T}_{\rm ref} \le 5000\ {\rm K}$.
We considered three values for the rarefaction parameter, namely 
$\delta = 10$ (slip flow regime), $\delta = 1$ (transition regime) 
and $\delta = 0.1$ (early free molecular flow regime).

We first conducted a qualitative comparison at the level of the profiles 
of the density, temperature, velocity and heat flux. 
In all cases considered, the density profile was well recovered 
with both kinetic models, for all values of the rarefation parameter. 
In the context of the heat transfer problem, the results obtained using 
the ES model were in better agreement with the DSMC results for the 
temperature profile.
In the Couette and heat transfer with shear problems, the S model 
seemed to give results which were closer to the DSMC predictions 
for all quantities (temperature, velocity and heat flux).

We next considered a quantitative comparison of the performance of the 
kinetic models with respect to the DSMC data by comparing the numerical 
values for non-dimensional quantities derived from the 
longitudinal heat flux (in the case of heat transfer between stationary 
and moving plates, denoted $Q$), 
shear stress (in the case of Couette flow and heat transfer between moving 
plates, denoted $\Pi$), as well as the half-channel heat flow rate,
$Q_y$, and heat transfer rate through the boundary, $Q_w$ (in the 
case of the Couette flow). Among these quantities, we can distinguish 
two categories. The first category (containing $Q$, $\Pi$ and $Q_w$) 
refers to quantities related to ``direct phenomena,'' which are 
driven by, e.g., shear rate $\partial_x u_y$ for $\Pi$ and temperature gradient 
$\partial_x T$ for $Q$, as predicted by the Navier-Stokes-Fourier 
theory. The second category (containing $Q_y$) refers to 
quantities related to ``cross phenomena,'' visible at the 
level of the Burnett equations, in which the usual 
thermodynamic forces driving the non-equilibrium quantity 
are absent (i.e., non-vanishing $q_y$ when $\partial_y T = 0$).

For the quantities in the first category (corresponding to direct 
phenomena), the agreement between the kinetic models and the DSMC 
results was within a few percent at $\delta = 10$, which confirms 
the validity of these models in the slip flow regime. 
At $\delta = 1$, the errors seem to be bounded within $8\%$ for 
both models, with the ES model giving better results in the 
heat transfer between stationary plates problem, while the S model 
performs better for the Couette flow and heat transfer under shear 
problems. When $\delta = 0.1$, 
the free molecular flow regime is approached. For the quantities 
that attain a finite value in this regime ($Q$ in the heat transfer 
problems and $\Pi$ in the Couette flow problem), the relative 
errors drop compared to $\delta = 1$, to within $2\%-3\%$. 
On the contrary, the relative errors for the heat flux $Q_w$ 
measured at the wall in the Couette flow grow to around
$20\%$ for the S model and $30\%$ for the ES model. This can 
be attributed to the fact that $Q_w$ decreases towards $0$ as the 
free molecular flow regime is approached, such that the relative 
errors are computed by dividing the results obtained within the model 
equations by a small quantity.

When considering the quantity $Q_y$ from the second category, 
which is generated through the cross-phenomena, 
the results of the kinetic models had relative errors of the 
order of $20\%$ even at $\delta = 10$, highlighting that 
the model equations do not accurately take into account such 
phenomena. At $\delta = 1$, the relative errors decrease 
to around $10\%$ for the S model and $15\%$ for the 
ES model. At $\delta = 0.1$, they increase again to around $20\%$ 
and $35\%$ for the S and ES models, respectively.
As was the case for $Q_w$, the large values of the relative errors 
of $Q_y$ encountered 
at $\delta = 10$ and $\delta = 0.1$ may be caused by the 
fact that $Q_y$ vanishes in the inviscid 
($\delta \rightarrow \infty$)
and free molecular flow ($\delta \rightarrow 0$) regimes.

In conclusion, our results demonstrate that even in the strongly non-linear 
regime, the model equations can give reasonably accurate results,
with errors of up to $10\%$ for quantities related to direct phenomena 
throughout the rarefaction spectrum (provided they remain finite in the 
free molecular flow regime), while the errors for the cross phenomena-related quantities 
seem to be within $35\%$. Due to the computational efficiency of the 
finite difference lattice Boltzmann (FDLB)
algorithm employed in this paper, solving the 
kinetic model equations can provide a cheap and reasonably accurate 
solution for the flow properties in the case of realistic monatomic 
gases under rarefied conditions.

{\bf Acknowledgments.} VEA gratefully acknowledges the generous support of
the Romanian-U.S. Fulbright Commission through The Fulbright Senior Postdoctoral
Program for Visiting Scholars 2017-2018, Grant number 678/2018. 
FS acknowledges the Brazilian Agency CNPq, Brazil, 
for the support of his research, grant 304831/2018-2.
VEA is grateful to Professor P. Dellar (Oxford University, UK) for preliminary 
discussions regarding the projection of the Shakhov collision term onto 
orthogonal polynomials.
The numerical simulations were performed on the Turing High Performance 
Computing cluster and the Computing cluster at the Computer Science Department 
of the Old Dominion University (Norfolk, VA, USA).
The computer simulations reported in this paper were done using the Portable 
Extensible Toolkit for Scientific Computation (PETSc 3.6) developed at Argonne 
National Laboratory, Argonne, Illinois \cite{petsc-user-ref,petsc-efficient}.

\appendix

\section{FDLB models for the hydrodynamic regime}
\label{app:hydro}

When $\delta \gtrsim 100$, the flow enters the hydrodynamic regime, where 
it can be approximately described by the Navier-Stokes equations \cite{kundu16},
\begin{align}
 \frac{D \rho}{Dt} + \rho \nabla \cdot \bm{u} =& 0,\nonumber\\
 \rho \frac{D u_i}{Dt} + \partial_i P =& -\partial_j \sigma_{ij},\nonumber\\
 \rho \frac{De}{Dt} + P \nabla \cdot \bm{u} =& -\sigma_{ij} \partial_i u_j -
 \nabla \cdot \bm{q},
 \label{eq:NS}
\end{align}
where $D/Dt = \partial_t + \bm{u} \cdot \nabla$ is the convective (material) derivative,
$\sigma_{ij} = T_{ij} - P \delta_{ij}$ is the shear stress tensor, $\bm{q}$ is the heat flux,
$e = \frac{3}{2m} T$ is the specific energy for an ideal monatomic gas. 
The Newtonian fluid model and Fourier's law 
give the following constitutive equations for $\tau_{ij}$ and $\bm{q}$:
\begin{equation}
 \tau_{ij} = -\mu \left( \partial_i u_j + \partial_j u_i - 
 \frac{2}{3} \delta_{ij} \nabla \cdot \bm{u}\right), \qquad 
 \bm{q} = -\kappa \nabla T,
 \label{eq:NS_const}
\end{equation}
where $\mu$ and $\kappa$ are the dynamic viscosity and heat conductivity, respectively.

According to the Chapman-Enskog expansion, briefly discussed in \ref{app:hydro:CE},
low order moments of the reduced distribution functions $\phi$ and $\chi$ 
are required to ensure the relations in Eq.~\eqref{eq:NS_const}. The evolution and 
stationary state properties of these moments can be obtained by employing similarly 
low order quadratures (i.e., $Q_x^\phi = 5$ and $Q_y^\phi = 4$ for the $\phi$ 
distribution; and $Q_x^\chi = 3$ and $Q_y^\chi = 2$ for the $\chi$ distribution).

As the quadrature order is lowered and $\delta$ is increased, the recovery of the 
conservation equations becomes increasingly challenging when the distributions 
are evaluated directly (i.e., using the hybrid method described in Sec.~\ref{sec:FDLB:feq}). In the traditional lattice Boltzmann framework, the key 
to employing the low order quadratures is to project the local equilibrium 
distribution on a set of orthogonal polynomials, which is subsequently 
truncated at an order $N_x^*$ ($* \in \{\phi, \chi\}$). 

\subsection{Chapman-Enskog analysis}\label{app:hydro:CE}

To derive the hydrodynamic regime from the kinetic model equation, 
\begin{equation}
 \frac{\partial f}{\partial t} + 
 \frac{\bm{p}}{m} \cdot \nabla f = -\frac{1}{\tau_*} 
 (f - f_*),
\end{equation}
the fluid can be assumed to be very close to isotropic thermal equilibrium, described 
by $f = f_* = \feq$, where $\feq$ is the Maxwell-Boltzmann distribution. 
The deviations of $f$ and $f_*$ from $\feq$,
denoted by $\delta f = f - \feq$ and $\delta f_* = f_* - \feq$, can be assumed 
to be of the same order as the relaxation time $\tau_*$, which is considered to 
be small. To first order with respect to $\tau_*$, the deviation $\delta f$ can 
be written as
\begin{equation}
 \delta f = \delta f_* - \tau_* \left(\frac{\partial \feq}{\partial t} 
 + \frac{\bm{p}}{m} \cdot \nabla \feq\right),
 \label{eq:CE:deltaf}
\end{equation}
where $\feq$ is determined by the density $\rho$, velocity $\bm{u}$ and 
temperature $T$. 
The constitutive relations for $\sigma_{ij} = T_{ij} - P \delta_{ij}$ and $q_i$ 
given in Eq.~\eqref{eq:NS_const} can be obtained by taking the second and third 
order moments of Eq.~\eqref{eq:CE:deltaf} with respect to the momentum space:
\begin{align}
 \sigma_{ij} =& \sigma^*_{ij} - \tau_* 
 \left( \frac{\partial }{\partial t} \int d^3p\, \frac{p_i p_j}{m} \feq + 
 \partial_k \int d^3p \, \frac{p_i p_j p_k}{m^2} \feq\right),\nonumber\\
 q_i + u_j \sigma_{ij} =& q_i^* + u_j \sigma^*_{ij} - \tau_* \left(
 \frac{\partial }{\partial t} \int d^3p\, \frac{\bm{p}^2 p_i}{2m^2} \feq + 
 \partial_k \int d^3p \, \frac{\bm{p}^2 p_i p_k}{2m^3} \feq\right),
 \label{eq:CE:sq_aux}
\end{align}
where $\sigma^*_{ij}$ and $q_i^*$ are obtained by taking moments of $\delta f_*$,
\begin{equation}
 \int d^3p \, \frac{p_i p_j}{m} \delta f_* = \sigma^*_{ij}, \qquad
 \int d^3p \, \frac{\bm{p}^2 p_i}{2m^2} \delta f_* = q_i^* + \sigma^*_{ij} u_j.
 \label{eq:CE:intfs}
\end{equation}
For completeness, the details of the Chapman-Enskog procedure for the 
S and ES models employed in this paper are briefly presented. The integrals 
of $\feq$ entering Eq.~\eqref{eq:CE:sq_aux} are
\begin{align}
 \int d^3p\, \frac{p_i p_j}{m} \feq =& P \delta_{ij} + \rho u_i u_j, \nonumber\\
 \int d^3p\, \frac{p_i p_j p_k}{m^2} \feq =& P (\delta_{ij} u_k + \delta_{jk} u_i + 
 \delta_{ki} u_j) + \rho u_i u_j u_k, \nonumber\\ 
 \int d^3p\, \frac{\bm{p}^2 p_i}{2m^2} \feq =& \frac{5 P}{2} u_i + 
 \frac{\rho \bm{u}^2}{2} u_i, \nonumber\\ 
 \int d^3p\, \frac{\bm{p}^2 p_i p_k}{2m^3} \feq =& P\left(\frac{5 T}{2m} + 
 \frac{\bm{u}^2}{2}\right) 
 \delta_{ik} + \left(\frac{7P}{2} + 
 \frac{\rho \bm{u}^2}{2}\right) u_i u_k.
 \label{eq:CE:intfeq}
\end{align}
Using the above relations and replacing the time derivatives using the 
Euler (inviscid) form of the Navier-Stokes equations \eqref{eq:NS},
\begin{equation}
 \frac{D\rho}{Dt} + \rho \nabla \cdot \bm{u} = O(\tau_*), \qquad 
 \rho \frac{D\bm{u}}{Dt} + \nabla P = O(\tau_*), \qquad 
 \rho \frac{De}{Dt} + P \nabla \cdot \bm{u} = O(\tau_*),
\end{equation}
it can be seen that
\begin{align}
 \sigma_{ij} =& \sigma^*_{ij} - \tau_* P
 \left(\partial_i u_j + \partial_j u_i - 
 \frac{2}{3} \nabla \cdot\bm{u} \delta_{ij}\right),&
 \bm{q} =& \bm{q}_* - \frac{5\tau_* P}{2m} \nabla T.
 \label{eq:CE:sqstar_aux}
\end{align}

In the Shakhov (S) and ellipsoidal (ES) models, when $f_*$ are given by $f_{\rm S}$ 
and $f_{\rm ES}$ introduced in Eqs.~\eqref{eq:sdef} and \eqref{eq:fES},
$\sigma^*_{ij}$ and $\bm{q}_*$ are given by:
\begin{align}
 \sigma^{\rm S}_{ij} =& 0, &
 \bm{q}_{\rm S} =& (1 - {\rm Pr}) \bm{q},\nonumber\\
 \sigma^{\rm ES}_{ij} =& -\frac{1 - {\rm Pr}}{\rm Pr}\sigma_{ij}, & 
 \bm{q}_{\rm ES} =& 0.
 \label{eq:CE:sqstar}
\end{align}
Substituting Eq.~\eqref{eq:CE:sqstar} into Eq.~\eqref{eq:CE:sqstar_aux} and 
comparing the result to Eq.~\eqref{eq:NS_const}, the relations given in 
Eqs.~\eqref{eq:Stcoeff} and \eqref{eq:EStcoeff} between the relaxation times, 
$\tau_{\rm S}$ and $\tau_{\rm ES}$, and the transport coefficients $\mu$ and 
$\kappa$ can be obtained.


The recovery of the constitutive equations \eqref{eq:NS_const} 
is conditioned by the correct recovery of the integrals 
\eqref{eq:CE:intfs} and \eqref{eq:CE:intfeq} of $f_*$ and $\feq$.

In the case when the flow is trivial along the $z$ 
direction ($d = 2$), the $p_z$ degree of freedom 
can be integrated automatically and Eqs.~\eqref{eq:CE:intfeq}
and \eqref{eq:CE:intfs} can be written in terms of the reduced 
distributions $\phi$ and $\chi$ introduced in Eq.~\eqref{eq:fred_def}.
The highest moments required for the reduced equilibrium distributions 
$\phi_*$ and $\chi_*$ are derived from the fourth order moment on the last 
line of Eq.~\eqref{eq:CE:intfeq}. Substituting $\feq \rightarrow f_*$, 
the following equation is obtained:
\begin{equation}
 \int d^3p\, \frac{\bm{p}^2 p_i p_k}{2m^3} f_* =
 \int dp_x\,dp_y\left[ 
 \frac{(p_x^2 + p_y^2)p_i p_k}{2m^3} \phi_* + 
 \frac{p_i p_k}{2m^2} \chi_* \right].
\end{equation}
It can be seen that the above integrals require the correct recovery 
of the moments with respect to $p_x$ of order $4$ for $\phi_*$ and 
of order $2$ for $\chi_*$. This can be achieved using the half-range 
Gauss-Hermite quadrature employing $Q_x^\phi = 5$ and $Q_x^\chi = 3$ 
points on each of the $p_x > 0$ and $p_x < 0$ semiaxes. Furthermore,
$\phi_*$ and $\chi_*$ must be replaced by a truncated expansion with 
respect to the half-range Hermite polynomials of orders $Q_x^* - 1$. 
More details for each of the collision models are given below.

\subsection{Maxwell-Boltzmann distribution}\label{app:hydro:MB}

The projection of the Maxwell-Boltzmann distribution $\feq$ 
with respect to the half-range Gauss-Hermite polynomials  
was derived in Ref.~\cite{ambrus16jcp}. Here, we only summarise 
the details. Considering the factorisation 
$\feq = n g_x g_y g_z$ introduced in Eq.~\eqref{eq:feq},
the expansion of $\feq$ can be performed at the level of each 
$g_i$ factor individually. Specifically, the $g_z$ factor is integrated 
out when introducing the reduced distributions. The $g_y$ factor 
is expanded up to order $N_y^*$ ($* \in \{\phi, \chi\}$) 
with respect to the full-range Hermite polynomials, as 
summarised in Eqs.~\eqref{eq:gN}--\eqref{eq:Gell} and 
\eqref{eq:S2D_Gell}--\eqref{eq:S2D_Gell_UT} 
in the contexts of the ES and S models.
For the Maxwell-Boltzmann distribution, Eqs.~\eqref{eq:gN} and 
\eqref{eq:Gell} can be used by substituting $\zeta_y$ and $T_y$ 
by $u_y$ and $T$, respectively.

The $g_x$ factor is expanded with respect to the half-range Hermite 
polynomials up to orders $N_x^\phi = Q_x^\phi - 1 = 4$ and 
$N_x^\chi = Q_x^\chi - 1 = 2$:
\begin{equation}
 g^{(N_x^*)}(p_x, u_x, T) = 
 \frac{\omega(\overline{p}_x)}{p_{0,x}} 
 \sum_{r = 0}^{N_x^*} \hh_r(|\overline{p}_x|) 
 [\theta(p_x) \mathcal{G}^+_r(u_x, T) + 
 \theta(-p_x) \mathcal{G}^-_r(u_x, T)],\label{eq:gNhalf}
\end{equation}
where $\theta(x)$ is the Heaviside step function.
The coefficients $\mathcal{G}^{\pm}_r$ are 
given by
\begin{equation}
 \mathcal{G}^\pm_r(u_x, T) = 
 \frac{1}{2} \sum_{s = 0}^r (\pm 1)^s \hh_{r,s} 
 \left(\frac{m T}{2p_{0,x}^2}\right)^{s/2} 
 \left[(1 \pm {\rm erf}\ \zeta) P_s^+(\zeta) \pm 
 \frac{2}{\sqrt{\pi}} e^{-\zeta^2} P_s^*(\zeta)\right],
 \label{eq:halfG}
\end{equation}
where $\zeta \equiv \zeta(u_x, T) = u_x \sqrt{m / 2T}$ and
$\hh_{r,s}$ represent the coefficients of $x^s$ in the 
expression for $\hh_{r}(x)$, as shown in Eq.~\eqref{eq:hh_exp}.
The polynomial $P_s^*(\zeta)$,
\begin{equation}
 P_s^*(\zeta) = \sum_{j = 0}^{s - 1} \binom{s}{j} 
 P_j^+(\zeta) P_{s-j-1}^-(\zeta),
\end{equation}
is defined with the help of the polynomials $P_s^\pm(\zeta)$, 
which satisfy:
\begin{equation}
 P_s^\pm(\zeta) = e^{\mp \zeta^2} \frac{d^s}{d\zeta^s} e^{\pm \zeta^2}.
\end{equation}

\subsection{$d = 1$: ES model} \label{app:hydro:d1ES}

The mass equilibrium distribution $\phi_{\rm ES}$, given by
Eq.~\eqref{eq:phiES1D}, can be rewritten in the language of the 
previous subsection in the following form:
\begin{equation}
 \phi_{\rm ES} = n g(p_x, u_x, T \mathcal{B}_{xx}),
 \label{eq:app_phiES1D_aux}
\end{equation}
such that its truncated version is
\begin{align}
 \phi_{\rm ES}^{(N_x^\phi)} =& \frac{\omega(\overline{p}_x)}{p_{0,x}} 
 \sum_{r = 0}^{N_x^\phi} \hh_r(|\overline{p}_x|) 
 [\theta(p_x) \Phi^+_{{\rm ES};r}(u_x, T) + 
 \theta(-p_x) \Phi^-_{{\rm ES};r}(u_x, T)],\nonumber\\
 \Phi^\pm_{{\rm ES};r}(u_x, T) =& n \mathcal{G}^\pm_r(u_x, T \mathcal{B}_{xx}),
 \label{eq:app_phiES1D}
\end{align}
where $\mathcal{B}_{xx}$ is introduced in Eq.~\eqref{eq:phiES1D_Bxx}.
The expansion coefficients $\Phi_{{\rm ES};r}^\pm(u_x, T)$ can be written entirely 
in terms of the expansion coefficients $\mathcal{G}^\pm_r$ corresponding 
to the Maxwell-Boltzmann distribution.
The energy equilibrium distribution satisfying  
$\chi_{\rm ES} = 2 T_{\rm red} \phi_{\rm ES}$ admits a similar decomposition:
\begin{align}
 \chi_{\rm ES}^{(N_x^\chi)} =& \frac{\omega(\overline{p}_x)}{p_{0,x}} 
 \sum_{r = 0}^{N_x^\chi} \hh_r(|\overline{p}_x|) 
 [\theta(p_x) X^+_{{\rm ES};r}(u_x, T) + 
 \theta(-p_x) X^-_{{\rm ES};r}(u_x, T)], \nonumber\\ 
 X_{{\rm ES};r}(u_x, T) =& 2 P_{\rm red} \mathcal{G}^\pm_r(u_x, T \mathcal{B}_{xx}).
 \label{eq:app_chiES1D}
\end{align}
where $P_{\rm red} = n T_{\rm red} = \frac{3}{2} P - \frac{1}{2} T_{xx}$.

\subsection{$d = 1$: S model} \label{app:hydro:d1S}

In the S model, the distributions $\phi_{\rm S}$ and $\chi_{\rm S}$, given 
in Eq.~\eqref{eq:phiS1D}, can be expanded as:
\begin{align}
 \begin{pmatrix}
  \phi_{\rm S} \\ \chi_{\rm S}
 \end{pmatrix} =& \frac{\omega(\overline{p}_x)}{p_{0,x}} 
 \sum_{r = 0}^{N_x^*} \hh_r(|\overline{p}_x|) 
 \left[\theta(p_x) 
 \begin{pmatrix}
  \Phi^+_{{\rm S};r}\\
  X^+_{{\rm S};r}
 \end{pmatrix} + \theta(-p_x) 
 \begin{pmatrix}
  \Phi^-_{{\rm S};r}\\
  X^-_{{\rm S};r}
 \end{pmatrix}\right], \nonumber\\
 \Phi^\pm_{{\rm S};r}(u_x, T) =& n\left[\mathcal{G}^\pm_r(u_x, T) + 
 \frac{1 - {\rm Pr}}{5 n T^2} \mathcal{G}^{\phi;\pm}_{{\rm S};r}(u_x, T)\right],\nonumber\\
 X^\pm_{{\rm S};r}(u_x, T) =& 2nT \left[\mathcal{G}^\pm_r(u_x, T) + 
 \frac{1 - {\rm Pr}}{5 n T^2} \mathcal{G}^{\chi;\pm}_{{\rm S};r}(u_x, T)\right],
 \label{eq:app_phiS1D_aux}
\end{align}
where $\mathcal{G}^{\pm}_r(u_x, T)$ is given in Eq.~\eqref{eq:halfG},
while $\mathcal{G}^{*;\pm}_{{\rm S};r}(u_x, T)$ can be found as follows:
\begin{align}
 \begin{pmatrix}
  \mathcal{G}^{\phi;\pm}_{{\rm S};r} \\
  \mathcal{G}^{\chi;\pm}_{{\rm S};r}
 \end{pmatrix} =& q_x \int_0^\infty dp_x\, 
 g_x(\pm p_x, u_x, T) \left[\frac{(\pm p_x - mu_x)^2}{m T} - 
 \begin{pmatrix}
  3 \\ 1
 \end{pmatrix}\right] \nonumber\\
 & \times (\pm p_x - mu_x) \hh(\overline{p}_x).
\end{align}
Employing the expansion for $g_x(p_x, u_x, T)$ in Eq.~\eqref{eq:gNhalf},
the coefficients $\mathcal{G}^{*;\pm}_{{\rm S}; r}(u_x, T)$ can be obtained 
as follows:
\begin{equation}
 \mathcal{G}^{*;\pm}_{{\rm S};r}(u_x, T) = q_x \sum_{r' = 0}^{\infty} 
 \mathcal{G}^\pm_{r'}(u_x, T)
 \int_0^\infty d\overline{p}_x \, \omega(\overline{p}_x) \hh_{r'}(\overline{p}_x)
 \sum_{k = 0}^3 (\pm 1)^k A^*_k [\overline{p}_x^k \hh(\overline{p}_x)],
 \label{eq:app_phiS1D_aux2}
\end{equation}
where the coefficients $A^*_k$ are given by
\begin{align}
 \begin{pmatrix}
  A^\phi_0 \\ A^\chi_0
 \end{pmatrix} =& \left[
 \begin{pmatrix}
  3 \\ 1
 \end{pmatrix} - \frac{mu_x^2}{T}\right] mu_x, &
 \begin{pmatrix}
  A^\phi_1 \\ A^\chi_1
 \end{pmatrix} =& 3\frac{m u_x^2 p_{0,x}}{T} - 
 \begin{pmatrix}
  3 \\ 1
 \end{pmatrix} p_{0,x}, \nonumber\\
 A_2^\phi =& A_2^\chi = -\frac{3 p_{0,x}^2 u_x}{T}, &
 A_3^\phi =& A_3^\chi = \frac{p_{0,x}^3}{mT}.
\end{align}

In order to evaluate Eq.~\eqref{eq:app_phiS1D_aux2} using the orthogonality relation 
for the half-range Hermite polynomials \cite{ambrus16jcp},
\begin{equation}
 \int_0^\infty d\overline{p}_x\, \omega(\overline{p}_x) \hh_r(\overline{p}_x) 
 \hh_{r'}(\overline{p}_x) = \delta_{r,r'},
\end{equation}
the following recurrence relation can be employed to eliminate the factors of $\overline{p}_x$:
\begin{equation}
 \overline{p}_x \hh_r(\overline{p}_x) = \frac{1}{a_r} \hh_{r+1}(\overline{p}_x) - 
 \frac{b_r}{a_r} \hh_r(\overline{p}_x) - \frac{c_r}{a_r} \hh_{r-1}(\overline{p}_x),
 \label{eq:hh_recur}
\end{equation}
where the recurrence coefficients $a_r$, $b_r$ and $c_r$ can be obtained by 
the procedure described in Ref.~\cite{ambrus16jcp}. There is an easy relation 
allowing $c_r$ to be eliminated in favour of $a_r$:
\begin{equation}
 c_r = -\frac{a_r}{a_{r-1}}.
\end{equation}
The recurrence in Eq.~\eqref{eq:hh_recur} can be used to obtain the following relation:
\begin{equation}
 \overline{p}_x^n \hh_r(\overline{p}_x) = 
 \sum_{r' = -n}^n B^{(n)}_{r,r'} \hh_{r'}(\overline{p}_x),
 \label{eq:hh_recurr_B}
\end{equation}
where it is understood that $B^{(n)}_{r,r'} = 0$ when $r + r' < 0$. 
The coefficients $B^{(n)}_{r,r'}$ depend only on the properties of the 
half-range Hermite polynomials and can thus be computed automatically at 
runtime. Their explicit values are given for $0 \le n \le 3$ at the end 
of this subsection. 

After applying the recurrence relations to eliminate all factors of $\overline{p}_x$, 
Eq.~\eqref{eq:app_phiS1D_aux2} becomes
\begin{equation}
 \mathcal{G}^{*;\pm}_{{\rm S};r}(u_x, T) = q_x \sum_{i = 0}^3 (\pm 1)^i A_i^* 
 \sum_{r' = -i}^i B^{(i)}_{r;r'} \mathcal{G}^\pm_{r + r'}(u_x, T).
\end{equation}
The coefficients $B^{(i)}_{r;r'}$ entering the above expression 
can be computed as follows. For $i = 0$, we have
\begin{equation}
 B^{(0)}_{r;0} = 1. 
\end{equation}
At $i = 1$, there are three non-vanishing coefficients:
\begin{equation}
 B^{(1)}_{r;-1} = \frac{1}{a_{r-1}}, \qquad 
 B^{(1)}_{r;0} = -\frac{b_r}{a_r}, \qquad 
 B^{(1)}_{r;1} = \frac{1}{a_r}.
\end{equation}
We remind the reader that $B^{(i)}_{r;r'} = 0$ whenever $r + r' < 0$, e.g. 
$B^{(1)}_{0;-1} = 0$. At $i = 2$, we find
\begin{gather}
 B^{(2)}_{r;-2} = \frac{1}{a_{r-2} a_{r-1}}, \qquad 
 B^{(2)}_{r;-1} = -\frac{1}{a_{r-1}}\left(
 \frac{b_r}{a_r} + \frac{b_{r-1}}{a_{r-1}}\right), \qquad 
 B^{(2)}_{r;0} = \frac{1 + b_r^2}{a_r^2} + \frac{1}{a_{r-1}^2}, \nonumber\\
 B^{(2)}_{r;1} = -\frac{1}{a_r}\left(\frac{b_{r+1}}{a_{r+1}} + \frac{b_r}{a_r}\right), \qquad 
 B^{(2)}_{r;2} = \frac{1}{a_r a_{r+1}}. 
\end{gather}
Finally, $B^{(3)}_{r;r'}$ is given by
\begin{gather}
 B^{(3)}_{r;-3} = \frac{1}{a_{r-3} a_{r-2} a_{r-1}}, \qquad 
 B^{(3)}_{r;-2} = -\frac{1}{a_{r-2} a_{r-1}}\left(\frac{b_{r-2}}{a_{r-2}} + 
 \frac{b_{r-1}}{a_{r-1}} + \frac{b_r}{a_r}\right), \nonumber\\
 B^{(3)}_{r;-1} = \frac{1}{a_{r-1}}\left(\frac{1}{a^2_{r-2}} + 
 \frac{1 + b_{r-1}^2}{a_{r-1}^2} + 
 \frac{b_{r-1} b_r}{a_{r-1} a_r} + \frac{1 + b_r^2}{a_r^2}\right),\nonumber\\
 B^{(3)}_{r;0} = -\left(\frac{b_{r-1}}{a_{r-1}^3} + \frac{2 b_r}{a_{r-1}^2 a_r} + 
 \frac{b_r(2 + b_r^2)}{a_r^3} + \frac{b_{r+1}}{a_r^2 a_{r+1}}\right),\nonumber\\
 B^{(3)}_{r;1} = \frac{1}{a_r}\left(\frac{1}{a_{r-1}^2} + 
 \frac{1 + b_r^2}{a_r^2} + \frac{b_r b_{r+1}}{a_r a_{r+1}} + 
 \frac{1 + b_{r+1}^2}{a_{r+1}^2}\right),\nonumber\\
 B^{(3)}_{r;2} = -\frac{1}{a_r a_{r+1}} \left(
 \frac{b_r}{a_r} + \frac{b_{r+1}}{a_{r+1}} + \frac{b_{r+2}}{a_{r+2}}\right),\qquad
 B^{(3)}_{r;3} = \frac{1}{a_r a_{r+1} a_{r+2}}.
\end{gather}

\subsection{$d = 2$: ES model} \label{app:hydro:d2ES}

When $d = 2$ and $\phi_{\rm ES}$ is given by Eq.~\eqref{eq:phiES2D},
we seek the expansion coefficients $\Phi^{\pm,H}_{{\rm ES};r,\ell}$
and $X^{\pm,H}_{{\rm ES};r,\ell}$ defined through
\begin{align}
 \phi_{\rm ES}^{(N_x^\phi, N_y^\phi)} =& 
 \frac{\omega(\overline{p}_x) \omega(\overline{p}_y)}
 {p_{0,x} p_{0,y}} \sum_{r = 0}^{N_x^\phi} \sum_{\ell = 0}^{N_y^\phi} 
 \frac{1}{\ell!} \hh_r(|\overline{p}_x|) H_\ell(\overline{p}_y) 
[\theta(p_x) \Phi^{+,H}_{{\rm ES};r,\ell} + 
\theta(-p_x) \Phi^{-,H}_{{\rm ES};r,\ell}],\nonumber\\
 \chi_{\rm ES}^{(N_x^\chi, N_y^\chi)} =& 
 \frac{\omega(\overline{p}_x) \omega(\overline{p}_y)}
 {p_{0,x} p_{0,y}} \sum_{r = 0}^{N_x^\chi} \sum_{\ell = 0}^{N_y^\chi} 
 \frac{1}{\ell!} \hh_r(|\overline{p}_x|) H_\ell(\overline{p}_y) 
 [\theta(p_x) X^{+,H}_{{\rm ES};r,\ell} + \theta(-p_x) X^{-,H}_{{\rm ES};r,\ell}].
 \label{eq:d2ES}
\end{align}
Due to the relation $\chi_{\rm ES} = T_{\rm red} \phi_{\rm ES}$, where 
$T_{\rm red} = P_{\rm red} / n$ and $P_{\rm red}$ is defined in 
Eq.~\eqref{eq:ES2DPred}, the coefficients $X^{\pm,H}_{{\rm ES};r, \ell}$ and 
$\Phi^{\pm,H}_{{\rm ES};r, \ell}$ can be related via
\begin{equation}
 X^{\pm,H}_{{\rm ES};r, \ell} = T_{\rm red} \Phi^{\pm,H}_{{\rm ES};r, \ell}.
\end{equation}

Inverting Eq.~\eqref{eq:d2ES} and using Eq.~\eqref{eq:phiES2D} to replace 
$\phi_{\rm ES}$, we find
\begin{equation}
 \Phi^{\pm, H}_{{\rm ES};r, \ell} = 
 n \sum_{r' = 0}^\infty  \mathcal{G}^\pm_{r'}(u_x, T \mathcal{B}_{xx}) 
 \int_0^\infty dp_x\,\frac{\omega(\overline{p}_x)}{p_{0,x}} 
 \mathcal{G}^H_\ell(\zeta_y^\pm, T_y) \hh_r(\overline{p}_x) \hh_{r'}(\overline{p}_x),
 \label{eq:d2ES_Phirl}
\end{equation}
where the expansions in Eqs.~\eqref{eq:gNhalf} and \eqref{eq:gN} were used 
to replace the functions $g(p_x, u_x, T \mathcal{B}_{xx})$ and 
$g(p_y, \zeta_y, T_y)$, respectively. 
The summation range for $r'$ was extended to $\infty$ to allow coefficients 
$\mathcal{G}^\pm_{r'}$ of orders $r' > N_x^\phi$ to be taken into account.
The superscript $\pm$ in $\zeta_y^\pm$ 
indicates the sign of $p_x$ in Eq.~\eqref{eq:phiES2D_zetaT}, i.e.
\begin{equation}
 \zeta_y^\pm = u_y - 
 \frac{\mathcal{B}_{xy}}{\mathcal{B}_{xx}} u_x \pm 
 \frac{p_{0,x} \mathcal{B}_{xy}}{m \mathcal{B}_{xx}} \overline{p}_x.
\end{equation}
The coefficients $\mathcal{G}^H_\ell(\zeta^\pm_y, T_y)$ are given explicitly 
for $0 \le \ell \le 3$ in Eq.~\eqref{eq:Gell}. For larger values of $\ell$,
the following formula can be employed \cite{ambrus16jcp}:
\begin{equation}
 \mathcal{G}^H_\ell(\zeta^\pm_y, T_y) = \sum_{s = 0}^{\lfloor \ell / 2 \rfloor}
 \frac{\ell!}{2^s s! (\ell - 2s)!} \mathfrak{U}^{\ell-2s}(\zeta^\pm_y)
 \mathfrak{I}^s(T_y).
\end{equation}
It can be seen that $\mathcal{G}^H_\ell(\zeta_y^\pm, T_y)$ 
is a polynomial of order $\ell$ with respect to $\zeta^\pm_y$, and therefore 
with respect to $\overline{p}_x$. Thus, it can be expanded as follows:
\begin{equation}
 \mathcal{G}^H_\ell(\zeta^\pm_y, T_y) = \sum_{k = 0}^\ell (\pm 1)^k \mathcal{G}^H_{\ell;k} 
 \overline{p}_x^k.
 \label{eq:d2ESGellk}
\end{equation}
Using now the recurrsion relation for the half-range Hermite polynomials 
given in Eq.~\eqref{eq:hh_recurr_B}, it can be shown that 
$\Phi^{\pm, H}_{{\rm ES}; r, \ell}$ reduces to
\begin{equation}
 \Phi^{\pm, H}_{{\rm ES}; r, \ell} = n \sum_{k = 0}^\ell (\pm 1)^k \mathcal{G}_{\ell;k}^H 
 \sum_{r = -k}^k B^{(k)}_{r,r'} \mathcal{G}^\pm_{r+r'}(u_x, T \mathcal{B}_{xx}).
\end{equation}
The first few functions $\mathcal{G}^H_{\ell;k}$ can be obtained 
by inspection:
\begin{gather}
 \mathcal{G}^H_{0;0} = 1, \qquad 
 \mathcal{G}^H_{1;0} = \mathcal{U}_y, \qquad
 \mathcal{G}^H_{1;1} = \mathcal{B}, \nonumber\\
 \mathcal{G}^H_{2;0} = \mathcal{U}_y^2 + \mathfrak{I}(T_y), \qquad 
 \mathcal{G}^H_{2;1} = 2 \mathcal{B} \mathcal{U}_y, \qquad
 \mathcal{G}^H_{2;2} = \mathcal{B}^2, \nonumber\\
 \mathcal{G}^H_{3;0} = \mathcal{U}_y^3 + 3 \mathcal{U}_y \mathfrak{I}(T_y), \qquad 
 \mathcal{G}^H_{3;1} = 3 \mathcal{B}[\mathcal{U}_y^2 + \mathfrak{I}(T_y)], \nonumber\\
 \mathcal{G}^H_{3;2} = 3 \mathcal{B}^2 \mathcal{U}_y, \qquad
 \mathcal{G}^H_{3;3} = \mathcal{B}^3,
\end{gather}
where $\mathfrak{I}(T_y) = \frac{m T_y}{p_{0,y}^2} - 1$ 
and $T_y = \frac{{\rm det} \mathcal{B}}{\mathcal{B}_{xx}} T$ are defined 
in Eqs.~\eqref{eq:Gell_UT} and \eqref{eq:phiES2D_zetaT}, respectively, while 
$\mathcal{U}_y$ and $\mathcal{B}$ are introduced below:
\begin{equation}
 \mathcal{U}_y = \frac{m}{p_{0,y}}\left(u_y - 
 u_x \frac{\mathcal{B}_{xy}}{\mathcal{B}_{xx}}\right), \qquad 
 \mathcal{B} = \frac{p_{0,x} \mathcal{B}_{xy}}{p_{0,y} \mathcal{B}_{xx}}.
\end{equation}
For larger values of $\ell$, the following formula can be used:
\begin{equation}
 \mathcal{G}^H_{\ell;k} = \frac{\mathcal{B}^k}{k!} 
 \sum_{s = 0}^{\lfloor \frac{\ell - k}{2}\rfloor} 
 \frac{\ell!}{2^s s! (\ell - 2s - k)!} 
 \mathfrak{I}^s(T_y) 
 \mathcal{U}_y^{\ell - 2s -k}.
\end{equation}

\subsection{$d = 2$: S model} \label{app:hydro:d2S}

For the $d = 2$ case, the same strategy employed in Subsec.~\ref{app:hydro:d1S} can 
be employed. Expanding $\phi_{\rm S}$ and $\chi_{\rm S}$ via
\begin{multline}
 \begin{pmatrix}
  \phi_{\rm S} \\ \chi_{\rm S}
 \end{pmatrix} = \frac{\omega(\overline{p}_x) \omega(\overline{p}_y)}{p_{0,x} p_{0,y}} 
 \sum_{r = 0}^{N_x^*} \hh_r(|\overline{p}_x|) \sum_{\ell = 0}^{N_y^*} 
 \frac{1}{\ell!} H_\ell(\overline{p}_y)\\
 \times \left[\theta(p_x) 
 \begin{pmatrix}
  \Phi^{+;H}_{{\rm S};r,\ell}\\
  X^+_{{\rm S};r,\ell}
 \end{pmatrix} + \theta(-p_x) 
 \begin{pmatrix}
  \Phi^{-;H}_{{\rm S};r,\ell}\\
  X^-_{{\rm S};r,\ell}
 \end{pmatrix}\right], 
\end{multline}
the coefficients $\Phi^\pm_{{\rm S};r,\ell}$ and $X^\pm_{{\rm S};r,\ell}$ can 
be obtained by taking into account Eq.~\eqref{eq:S2D_Gell}:
\begin{equation}
\begin{pmatrix}
  \Phi_{{\rm S};r,\ell}^{\pm,H} \\ X_{{\rm S};r,\ell}^{\pm,H}
 \end{pmatrix} = 
 \begin{pmatrix}
  n \\ P
 \end{pmatrix}
 \left[\mathcal{G}^H_\ell(u_y, T) \mathcal{G}^\pm_r(u_x, T) + 
 \frac{1 - {\rm Pr}}{5 n T^2} 
 \begin{pmatrix}
  \mathfrak{G}^{\phi;\pm,H}_{{\rm S};r,\ell} \\
  \mathfrak{G}^{\chi;\pm,H}_{{\rm S};r,\ell}
 \end{pmatrix}\right].
 \label{eq:S2D_Grell}
\end{equation}
The coefficients $\mathfrak{G}^{*;\pm,H}_{{\rm S};r,\ell}$ are obtained by 
computing the following integrals:
\begin{equation}
 \mathfrak{G}^{*;\pm,H}_{{\rm S};r,\ell} = 
 \sum_{r' = 0}^\infty \mathcal{G}^\pm_{r'}(u_x, T)
 \int_0^\infty dp_x\,\frac{\omega(\overline{p}_x)}{p_{0,x}}
 \mathfrak{G}_{S;\ell}^{*;H}(\pm \overline{p}_x) \hh_r(\overline{p}_x) 
 \hh_{r'}(\overline{p}_x),
 \label{eq:S2D_Grell_int}
\end{equation}
where $\mathfrak{G}_{S;\ell}^{*;H}(\pm \overline{p}_x)$ was introduced in 
Eq.~\eqref{eq:S2D_Gell}. As in Eq.~\eqref{eq:d2ES_Phirl}, 
the summation with respect to $r'$ was extended to $\infty$. Using now an expansion 
of $\mathfrak{G}_{S;\ell}^{*;H}(\pm \overline{p}_x)$ similar to that 
introduced in Eq.~\eqref{eq:d2ESGellk},
\begin{equation}
 \mathfrak{G}_{S;\ell}^{*;H}(\pm \overline{p}_x) = \sum_{k = 0}^3 
 (\pm 1)^k \mathfrak{G}_{S;\ell;k}^{*;H} \overline{p}_x^k,
\end{equation}
the integral in Eq.~\eqref{eq:S2D_Grell_int} can be performed 
using the recurrence relation in Eq.~\eqref{eq:hh_recurr_B}:
\begin{equation}
 \mathfrak{G}^{*;\pm,H}_{{\rm S};r,\ell} = \sum_{k = 0}^3 (\pm 1)^k 
 \mathfrak{G}^{*;H}_{S;\ell;k} \sum_{r' = -k}^k 
 B^{(k)}_{r,r'} \mathcal{G}^\pm_{r+r'}(u_x, T).
\end{equation}

We close this subsection by giving explicitly the expressions for 
$\mathfrak{G}^{*;H}_{S;\ell;k}$ for the values of $\ell$ and $k$ that are 
relevant to this paper. From Eq.~\eqref{eq:S2D_Gellfrak}, it can be seen that 
$p_x$ enters $\mathfrak{G}^{*;H}_{S;\ell}$ only through the terms 
$\mathcal{I}^*_s$, defined in Eq.~\eqref{eq:S2D_Is} and given explicitly 
for $0 \le s \le 3$ in Eq.~\eqref{eq:S2D_Is_explicit}. Thus, the expression for 
$\mathfrak{G}^{*;H}_{S;\ell;k}$ is the same as that for 
$\mathfrak{G}^{*;H}_{S;\ell}$, but with $\mathcal{I}^*_s$ replaced 
by $\mathcal{I}^*_{s;k}$, where $\mathcal{I}^*_{s;k}$ represents 
the coefficient of $\overline{p}_x^k$ in $\mathcal{I}^*_{s}$.
Specifically, we find
\begin{align}
 \begin{pmatrix}
  \mathcal{I}^{\phi}_{0;0} \\
  \mathcal{I}^{\chi}_{0;0}
 \end{pmatrix} =& m u_x q_x \left[
 \begin{pmatrix}
  3 \\ 1
 \end{pmatrix} -\frac{mu_x^2}{T} \right], &
 \begin{pmatrix}
  \mathcal{I}^{\phi}_{0;1} \\
  \mathcal{I}^{\chi}_{0;1}
 \end{pmatrix} =& p_{0,x} q_x \left[\frac{3mu_x^2}{T} - 
 \begin{pmatrix}
  3 \\ 1
 \end{pmatrix}\right], \nonumber\\
 \mathcal{I}^{*}_{0;2} =& -\frac{3 p_{0,x}^2}{T} u_x q_x, &
 \mathcal{I}^{*}_{0;3} =& \frac{p_{0,x}^3}{mT} q_x,\nonumber\\
 \begin{pmatrix}
  \mathcal{I}^{\phi}_{1;0} \\
  \mathcal{I}^{\chi}_{1;0}
 \end{pmatrix} =& m T q_y \left[\frac{mu_x^2}{T} + 
 \begin{pmatrix}
  -1 \\ 1
 \end{pmatrix}\right], & 
 \mathcal{I}^{*}_{1;1}=& -2 m u_x p_{0,x} q_y,\nonumber\\
 \mathcal{I}^{*}_{1;2}=& p^2_{0,x} q_y, &
 \mathcal{I}^{*}_{1;3}=& 0,\nonumber \\
 \begin{pmatrix}
  \mathcal{I}^{\phi}_{2;0} \\
  \mathcal{I}^{\chi}_{2;0}
 \end{pmatrix} =& -m^2 u_x q_x T \left[
 \begin{pmatrix}
  -1 \\ 1
 \end{pmatrix} + \frac{mu_x^2}{T} \right], &
 \begin{pmatrix}
  \mathcal{I}^{\phi}_{2;1} \\
  \mathcal{I}^{\chi}_{2;1}
 \end{pmatrix} =& m p_{0,x} q_x T \left[\frac{3mu_x^2}{T} + 
 \begin{pmatrix}
  -1 \\ 1
 \end{pmatrix}\right], \nonumber\\
 \mathcal{I}^{*}_{2;2} =& -3 m p_{0,x}^2 u_x q_x, &
 \mathcal{I}^{*}_{2;3} =& p_{0,x}^3 q_x,\nonumber\\
 \begin{pmatrix}
  \mathcal{I}^{\phi}_{3;0} \\
  \mathcal{I}^{\chi}_{3;0}
 \end{pmatrix} =& 3 (mT)^2 q_y \left[\frac{mu_x^2}{T} + 
 \begin{pmatrix}
  1 \\ 3
 \end{pmatrix}\right], &
 \mathcal{I}^{*}_{3;1}=& -6 m^2 T p_{0,x} u_x q_y,\nonumber\\
 \mathcal{I}^{*}_{3;2}=& 3m T p^2_{0,x} q_y, &
 \mathcal{I}^{*}_{3;3}=& 0.
\end{align}

\subsection{Numerical results}\label{app:hydro:res}

\begin{figure}
\begin{center}
\begin{tabular}{cc}
 \includegraphics[width=0.48\linewidth]{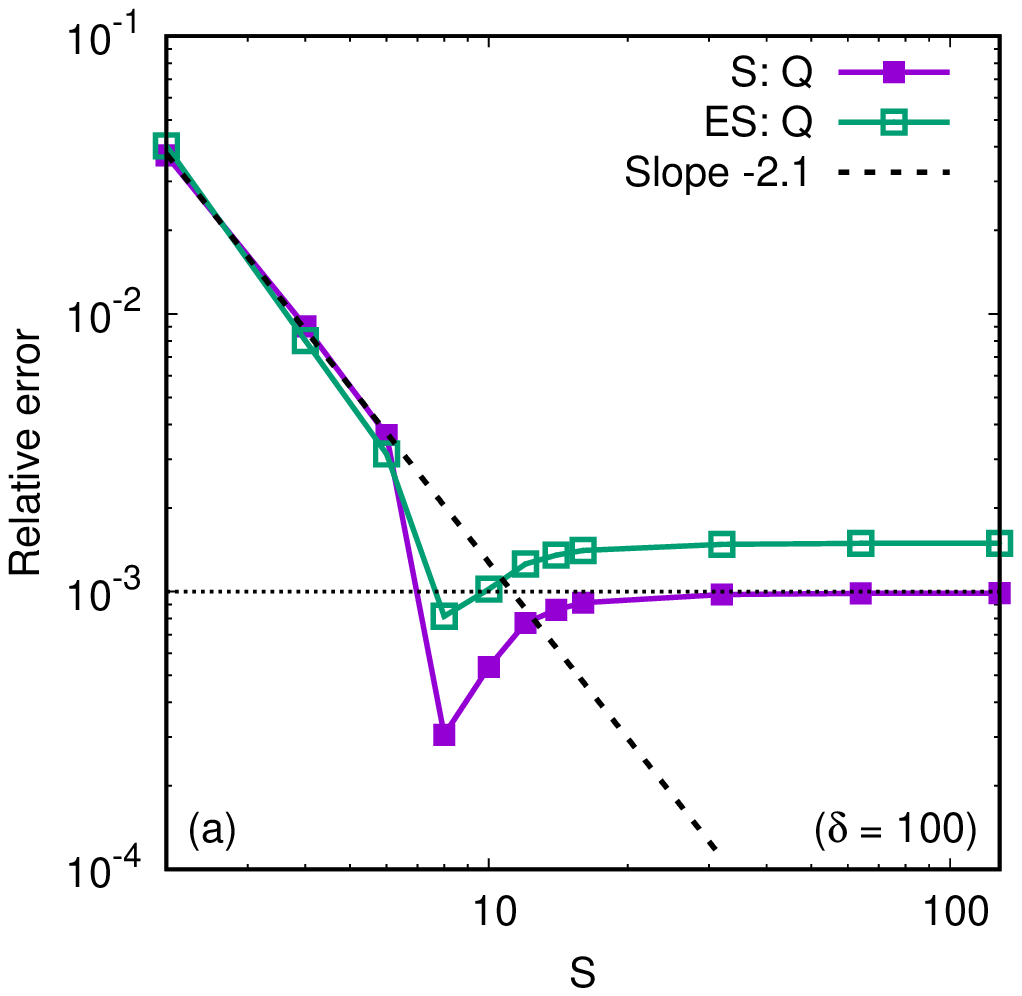} & 
 \includegraphics[width=0.48\linewidth]{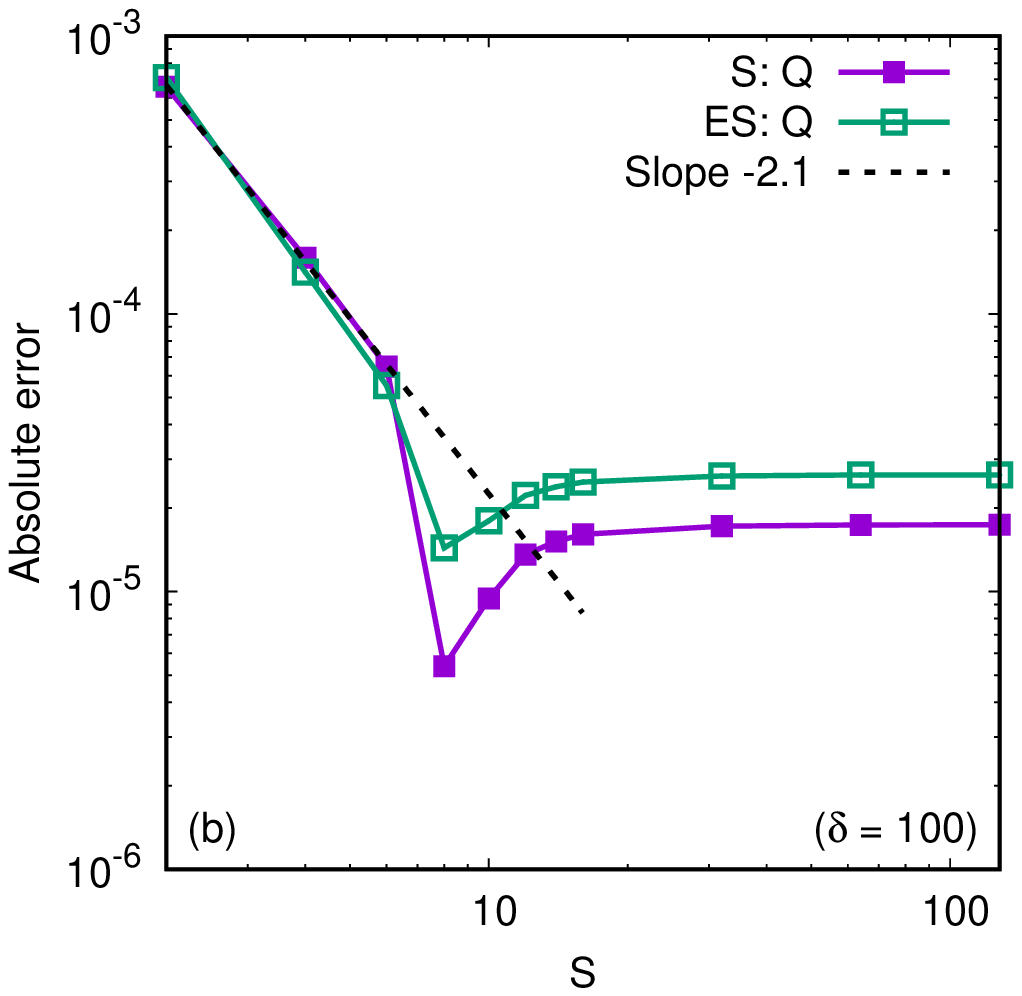} \\
 \includegraphics[width=0.48\linewidth]{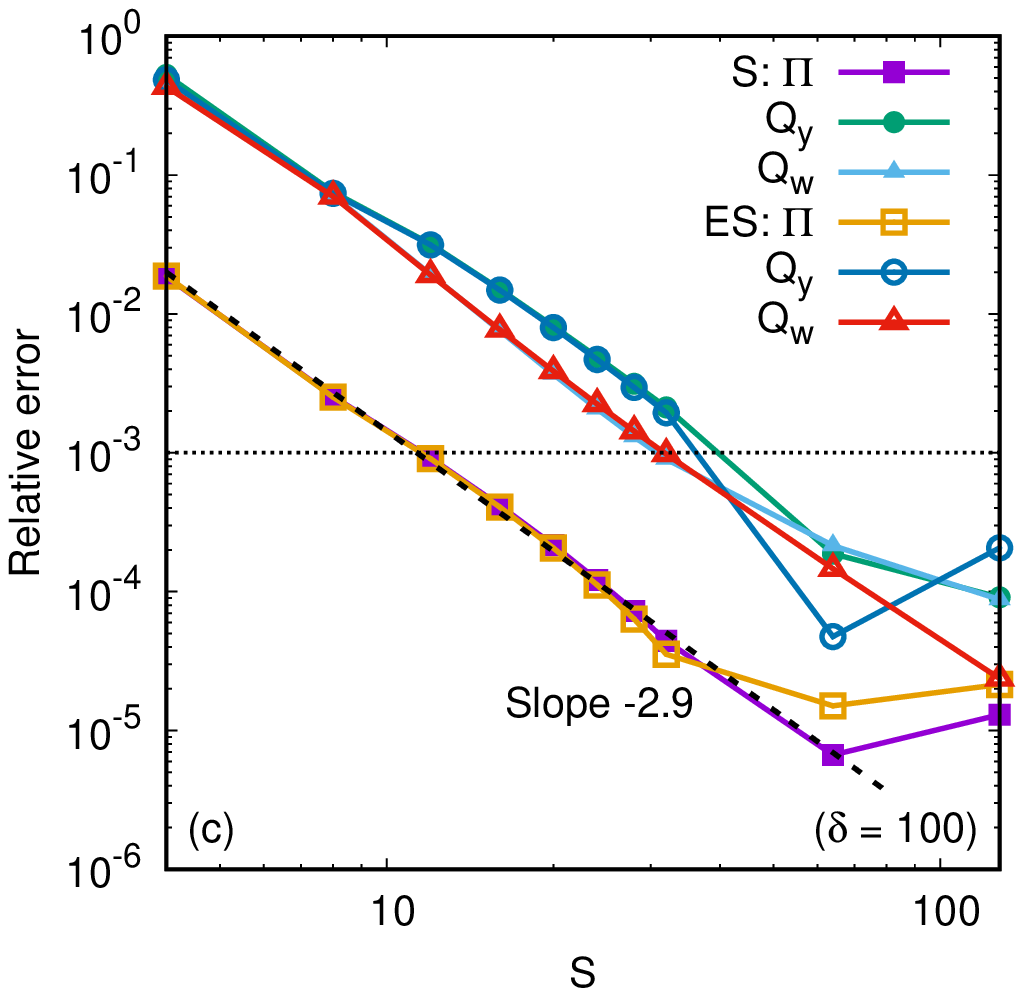} & 
 \includegraphics[width=0.48\linewidth]{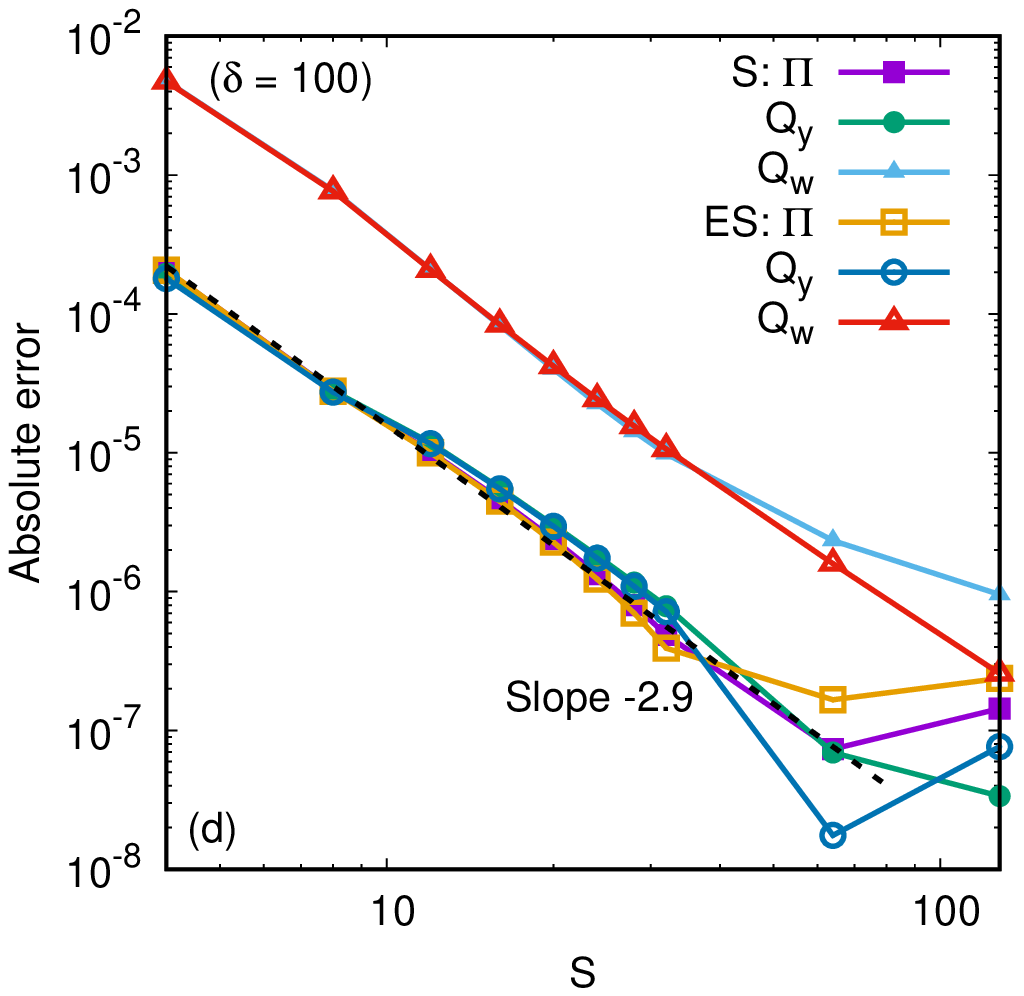} \\
 \includegraphics[width=0.48\linewidth]{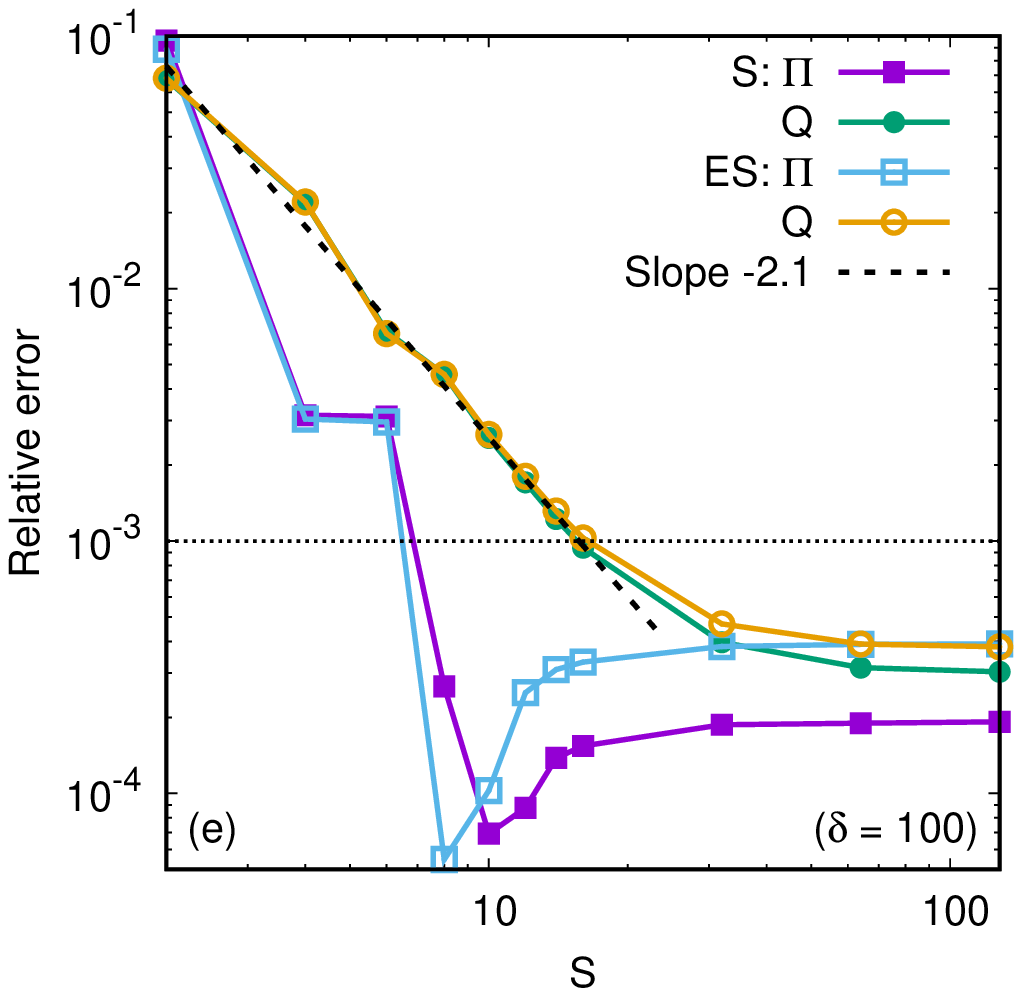} & 
 \includegraphics[width=0.48\linewidth]{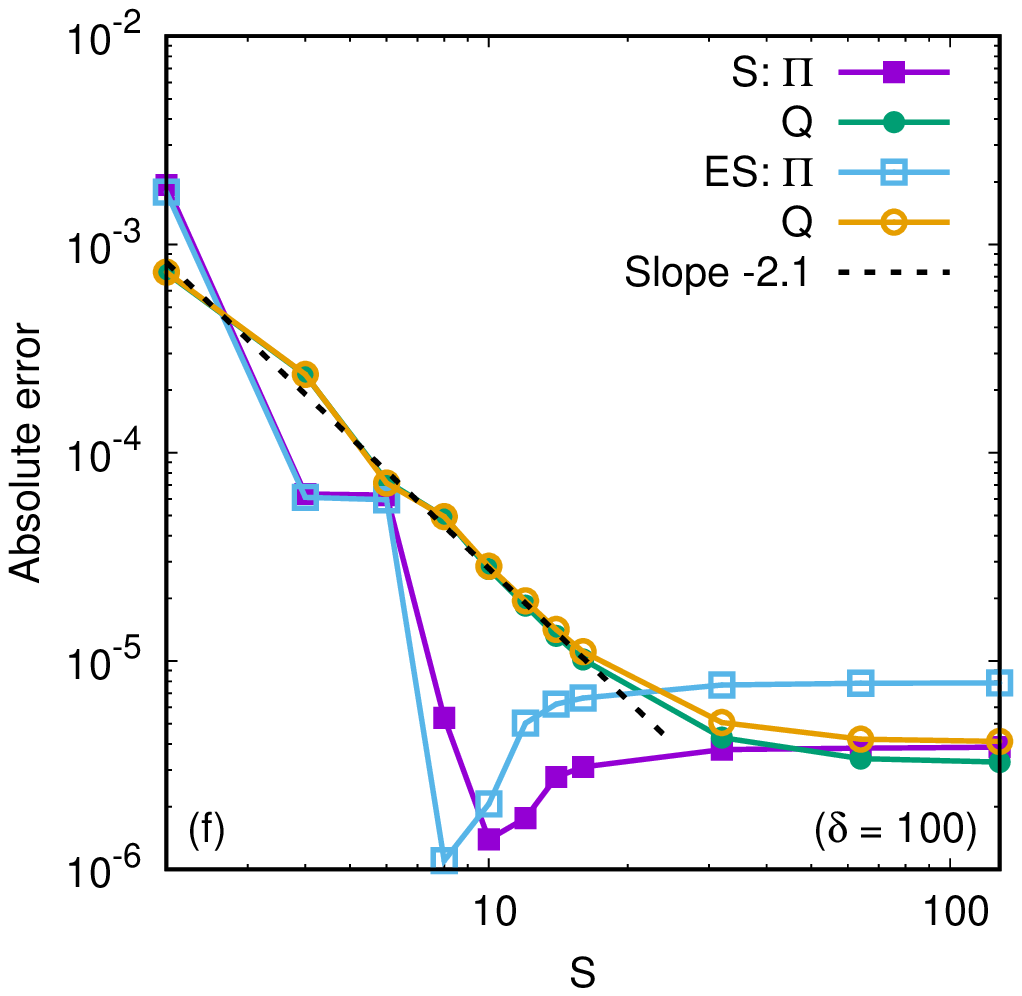}
\end{tabular}
\end{center}
\caption{Convergence test for the results obtained using the projection 
method with respect to the number $S$ of nodes in the half channel.
The heat transfer between stationary plates (top line), Couette flow 
(middle line) and heat transfer between moving plates (bottom line) 
are considered at $\delta = 100$. The results obtained using the S 
model are shown using 
filled symbols, while the ES model results are represented using 
empty symbols. The relative (left) and absolute (right) errors are computed 
at the level of the various quantities introduced in Sec.~\ref{sec:meth}
by taking the results obtained using the hybrid method with 
$Q_x^\phi = Q_x^\chi = 16$ and $S = 128$ as a reference.
The slope of the dotted line indicates the convergence order.
\label{fig:hydro:conv100}}
\end{figure}

\begin{figure}
\begin{center}
\begin{tabular}{cc}
 \includegraphics[width=0.48\linewidth]{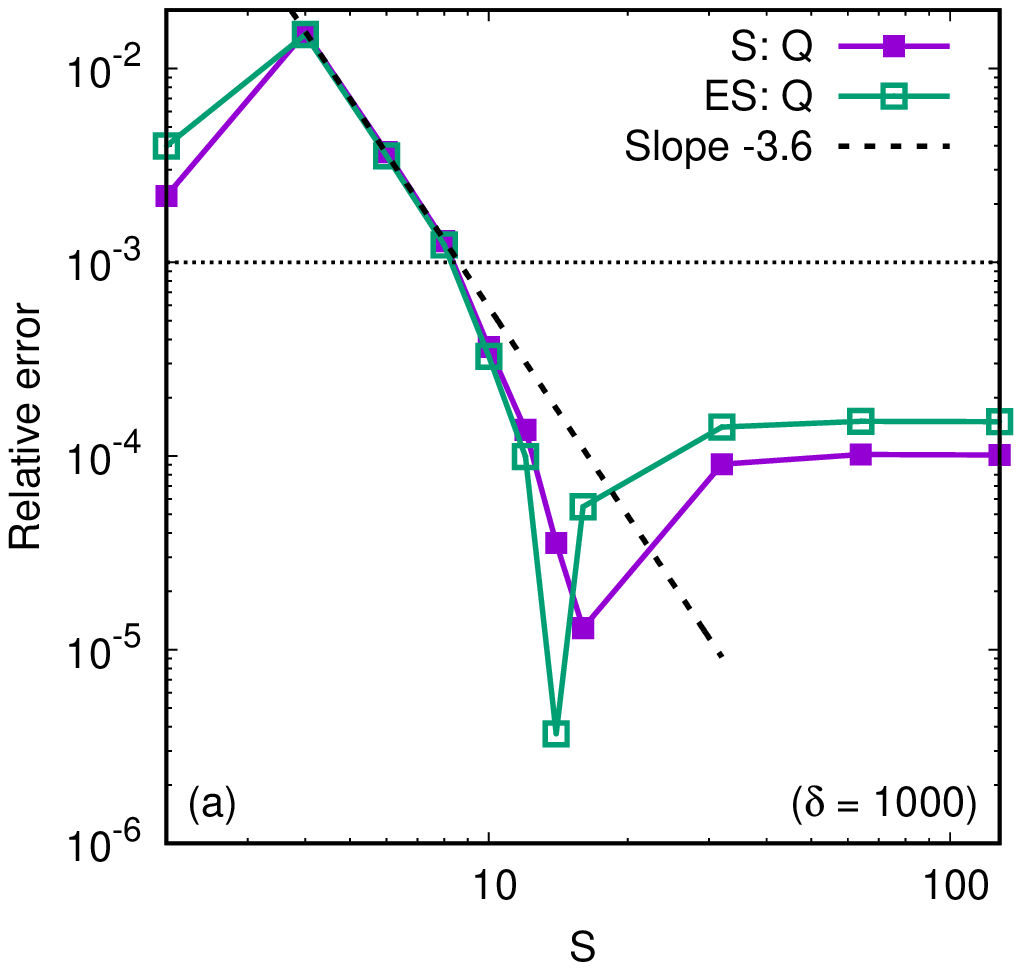} & 
 \includegraphics[width=0.48\linewidth]{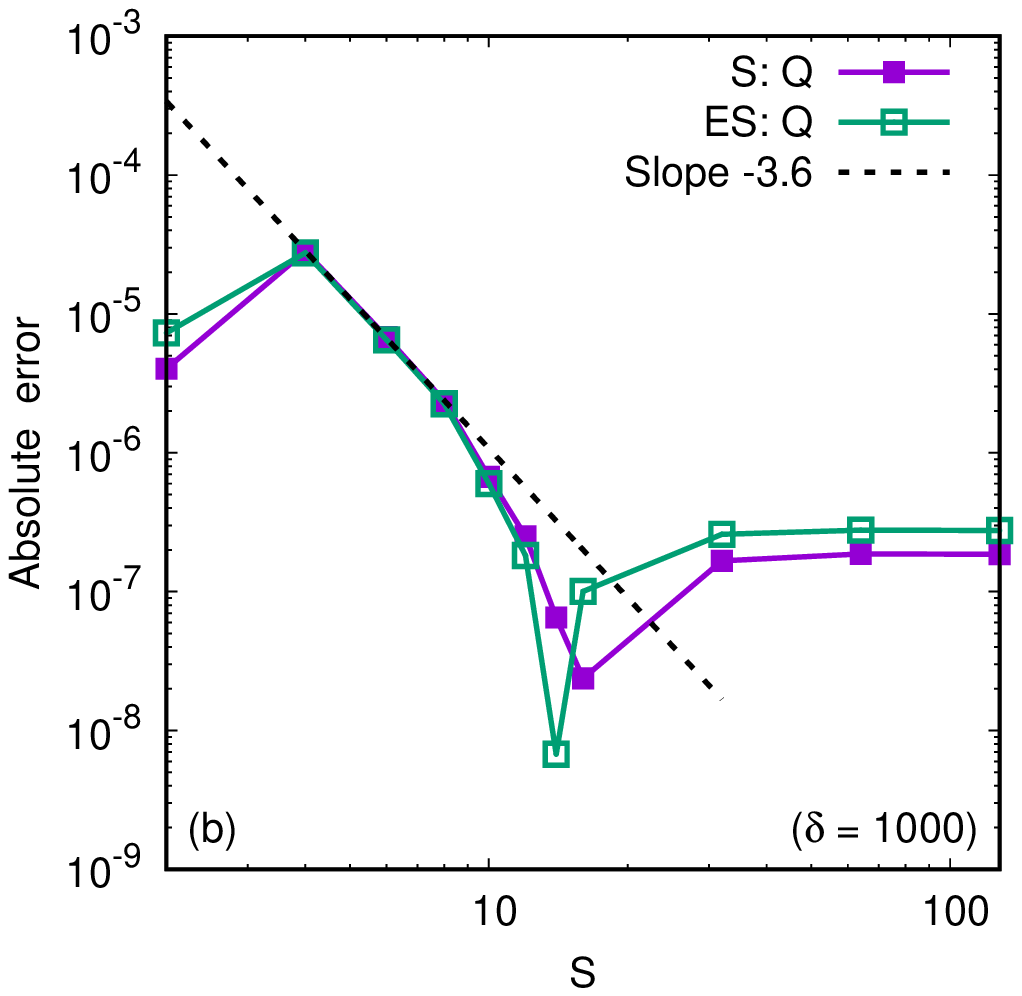} \\
 \includegraphics[width=0.48\linewidth]{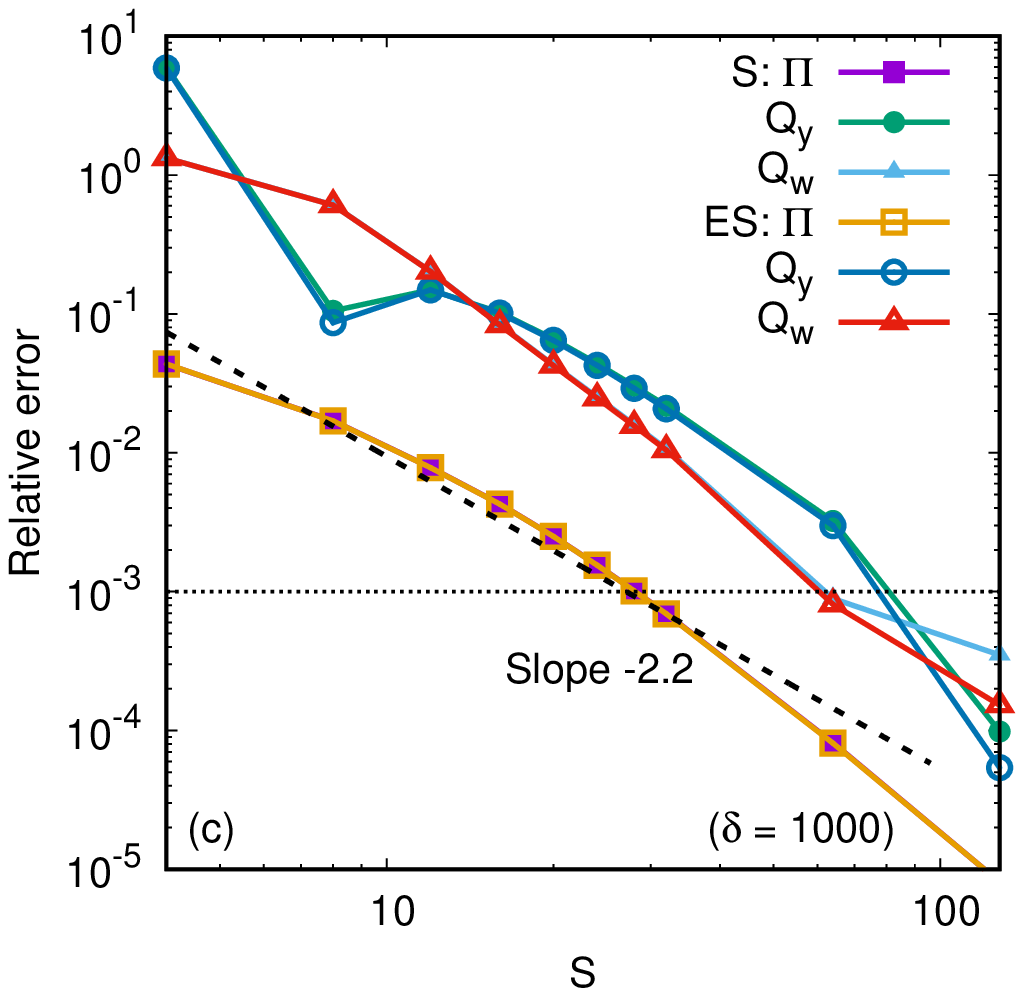} & 
 \includegraphics[width=0.48\linewidth]{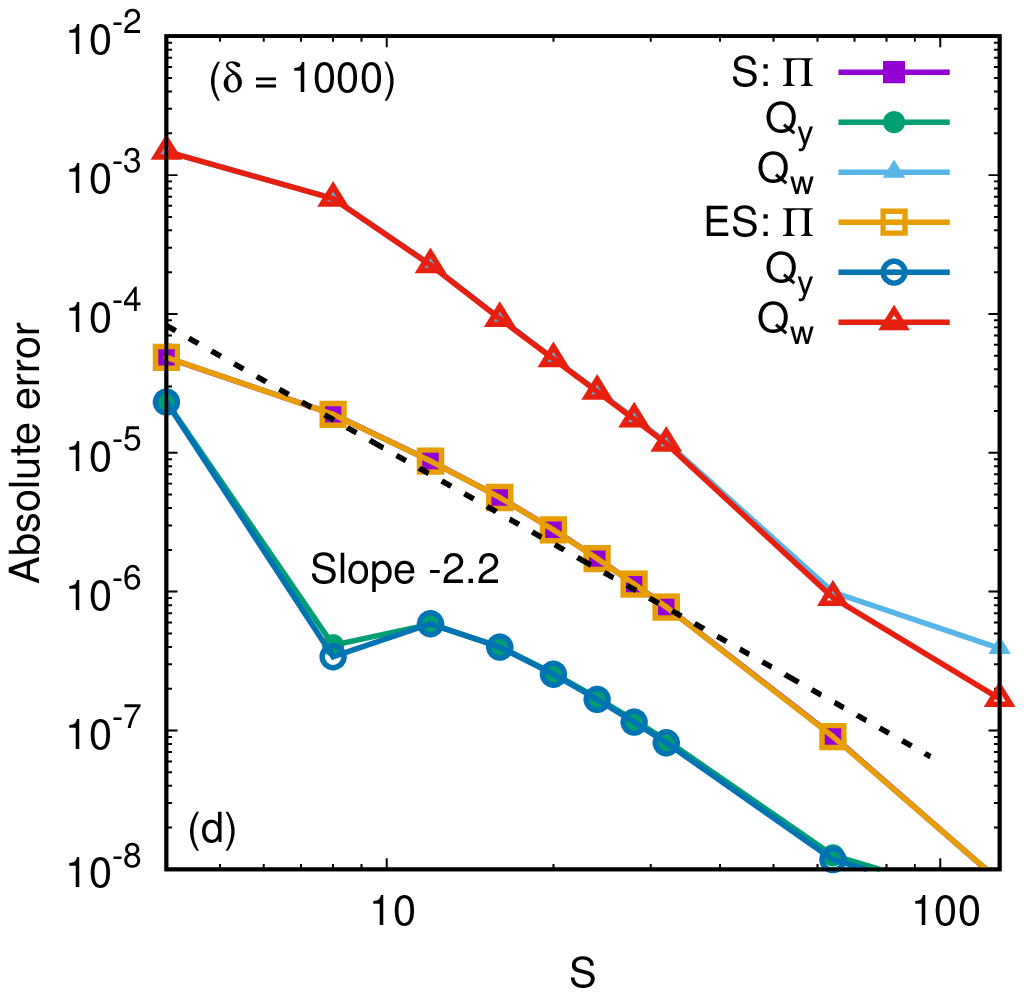} \\
 \includegraphics[width=0.48\linewidth]{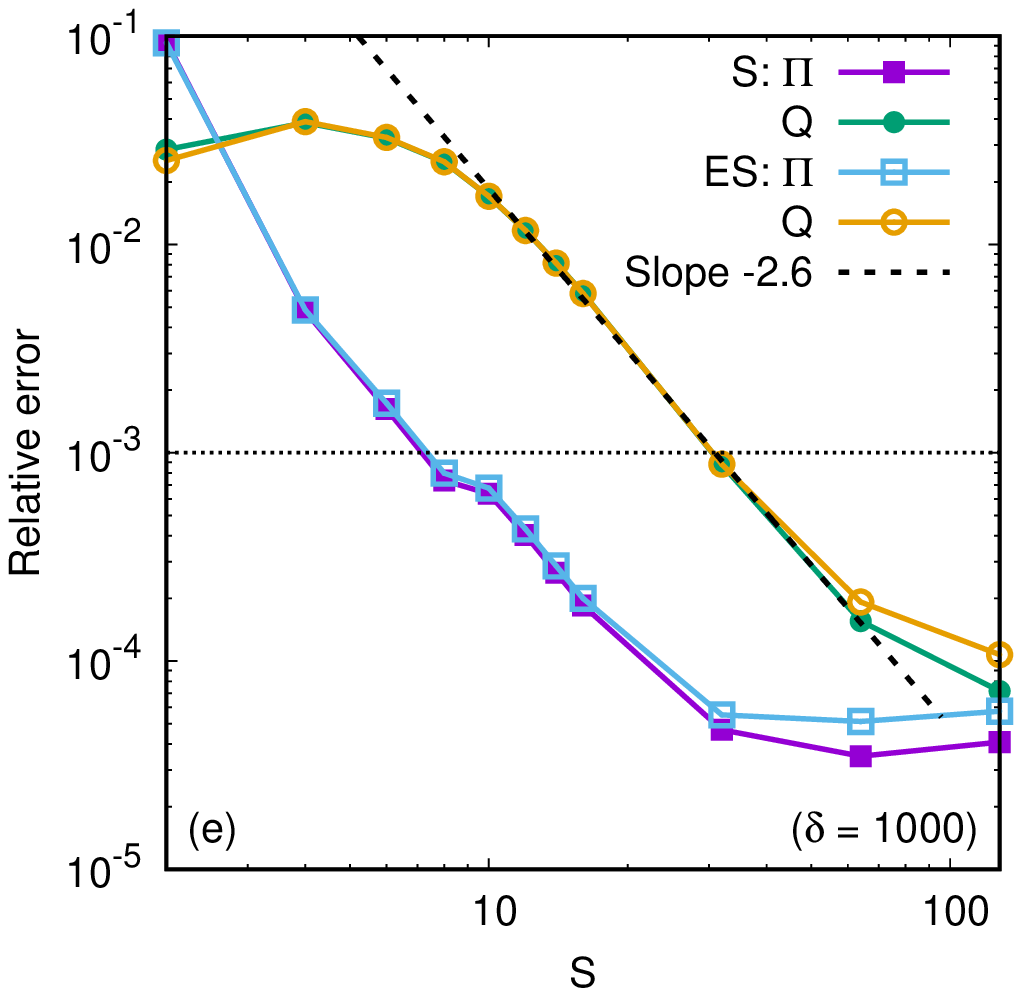} & 
 \includegraphics[width=0.48\linewidth]{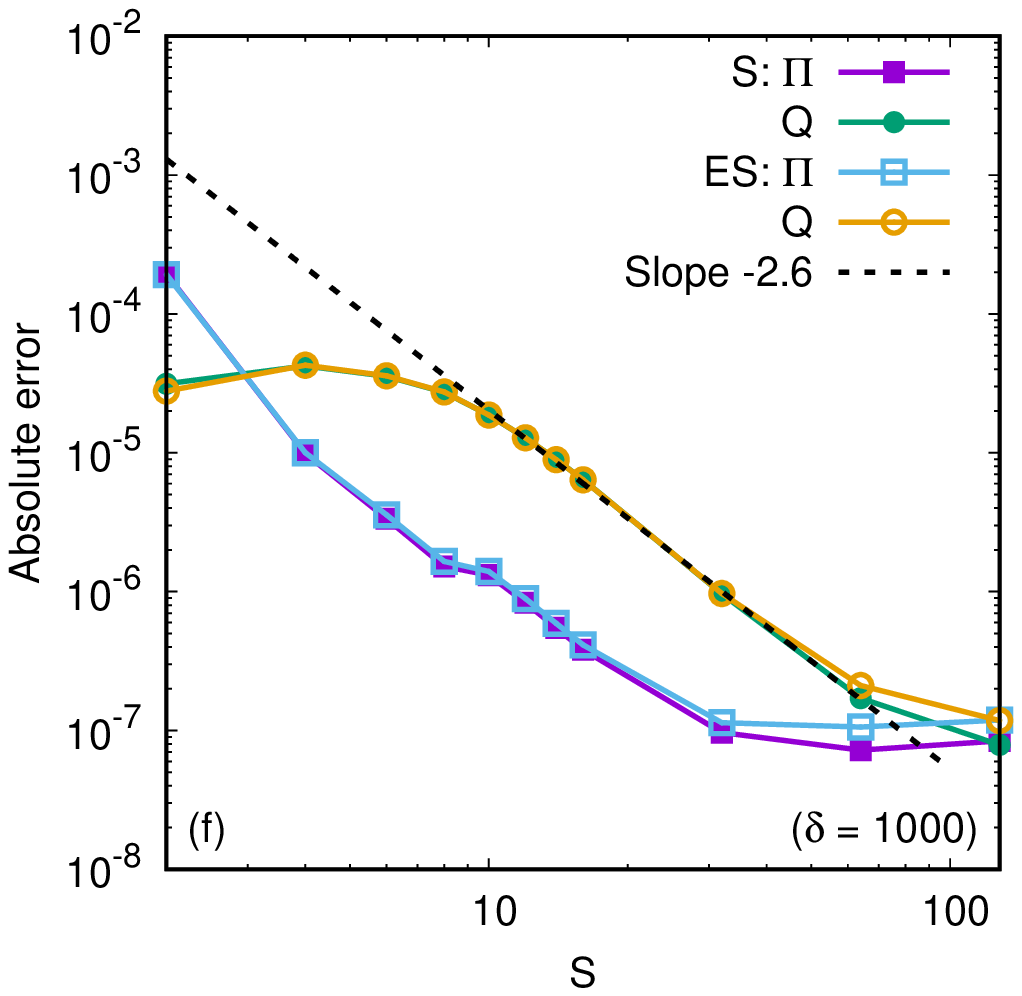}
\end{tabular}
\end{center}
\caption{Same as Fig.~\ref{fig:hydro:conv100}, for $\delta = 1000$.
\label{fig:hydro:conv1000}}
\end{figure}

\begin{table}
\begin{center}
\begin{tabular}{c|cc|cc|cc}
& \multicolumn{2}{|c|}{HT} & \multicolumn{2}{|c|}{SH} & \multicolumn{2}{|c}{HT-SH} \\
 \hline\hline
 $\delta$ & $T_{\rm S} ({\rm s})$ & $T_{\rm ES} ({\rm s})$ & $T_{\rm S} ({\rm s})$ & $T_{\rm ES} ({\rm s})$ & $T_{\rm S} ({\rm s})$ & $T_{\rm ES} ({\rm s})$ \\\hline
 $1000$ & $674$ & $320$ & $992$ & $699$ & $2106$ & $1368$ \\
 $100$ & $67$ & $32$ & $108$ & $68$ & $197$ & $135$ 
\end{tabular}
\end{center}
\caption{Computational times (in seconds) required 
to reach the steady state using the projection method 
described in \ref{app:hydro} at $\delta = 100$ and $\delta = 1000$,
in the context of the 
heat transfer between stationary plates (HT), Couette flow 
(SH) and heat transfer under shear (HT-SH) problems, considered in 
Sections~\ref{sec:ht}, \ref{sec:couette} and \ref{sec:htsh}.
\label{tab:comp2}}
\end{table}

In order to demonstrate the capabilities of the model introduced in the previous 
subsections, we performed simulations in the context of the problems introduced 
in Sections~\ref{sec:ht}, \ref{sec:couette} and \ref{sec:htsh} for 
$\delta = 100$ and $1000$. For definiteness, we considered the ${}^4{\rm He}$ gas 
at $\widetilde{T}_{\rm ref} = 300\ {\rm K}$. The simulations were performed 
using the quadrature orders $Q_x^\phi = 5$ and $Q_x^\chi = 3$ for the half-range 
Gauss-Hermite quadrature employed on the $p_x$ axis. In the case of the heat transfer 
between stationary plates (discussed in Sec.~\ref{sec:ht}), there are no other 
non-trivial degrees of freedom and the total number of distinct populations 
is $2(Q_x^\phi + Q_x^\chi)= 16$. In the Couette flow and heat transfer between moving 
plates problems, discussed in Sections~\ref{sec:couette} and \ref{sec:htsh}, 
the $p_y$ axis was discretised using the full-range Gauss-Hermite quadrature of
orders $Q_y^\phi = 4$ and $Q_y^\chi = 2$, as discussed in Sec.~\ref{sec:FDLB:quad}.
The total number of distinct population in this case is 
$2(Q_x^\phi Q_y^\phi + Q_x^\chi Q_y^\chi) = 52$. In the context of this subsection,
we refer to the method employing the models described above as the 
``projection method.''

In order to validate the results obtained with the projection method,  we also performed 
simulations using the hybrid models introduced in Sec.~\ref{sec:FDLB}, which differ from 
the former since the equilibrium distributions are replaced by truncated expansions only 
with respect to the $p_y$ axis. The quadrature orders in the hybrid approach are set to 
$Q_x^\phi = Q_x^\chi = 16$, while the quadrature orders $Q_y^\phi$ and $Q_y^\chi$ remain 
unchanged with respect to those employed within the projection method. 

Figures~\ref{fig:hydro:conv100} and \ref{fig:hydro:conv1000} show convergence tests
performed at $\delta = 100$ and $1000$, respectively,
comparing the results obtained using the projection method for various values of 
$S$ with those obtained using the hybrid method using $Q_x = 16$ and $S = 128$,
in the context of the heat transfer between stationary plates (top line), Couette flow 
(middle line) and heat transfer between moving plates (bottom line) problems. 
The left columns of Figs.~\ref{fig:hydro:conv100} and \ref{fig:hydro:conv1000}
show the relative errors, while the right columns show the corresponding absolute errors,
computed at the level of the $Q$, $\Pi$, $Q_y$ 
and $Q_w$ quantities introduced in Sec.~\ref{sec:meth}. 
The resulting convergence orders take values between $2.1$ and
$3.6$, depending on $\delta$ and on the quantity being analysed. 

While the relative errors in $Q$ and $\Pi$ quickly approach $0.1\%$ as $S$ is increased,
it can be seen from the left columns of Figs.~\ref{fig:hydro:conv100} and
\ref{fig:hydro:conv1000} that achieving the same level of relative
error for $Q_w$ and $Q_y$ in the context of the Couette flow becomes more
challenging at large $\delta$. Taking into account that all quantities decrease
in absolute value as $\delta^{-1}$ (except $Q_y$, which decreases as $\delta^{-2}$),
the relative error is correspondingly amplified and thus becomes less relevant. 
By comparison, the right columns of the same figures show that the absolute errors
are several orders of magnitude below the relative errors. It can be seen that reasonable 
results are obtained using the projection method with $S = 32$, which gives absolute 
errors that are below $10^{-4}$ for al quantities under consideration. The corresponding 
runtimes for $S = 32$ are summarised in Table~\ref{tab:comp2}.

\bibliographystyle{unsrt}

\end{document}